\documentclass{JFM-FLM_Au}

\usepackage{subcaption}
\usepackage{comment}

\lefttitle{Nikolaidou et al.}
\righttitle{Journal of Fluid Mechanics}

\title{Drag reduction regimes in air lubrication} 

\author{Lina Nikolaidou\aff{1}, Ali R Khojasteh\aff{1}, Angeliki Laskari\aff{1}, Tom van Terwisga\aff{2,3} \and Christian Poelma\aff{1}}

\affiliation{\aff{1}Process \& Energy, Faculty Mechanical Engineering, Delft University of Technology, Mekelweg 2, Delft 2628 CD, The Netherlands
\aff{2} Maritime and Transport Technology, Faculty Mechanical Engineering, Delft University of Technology, Mekelweg 2, Delft 2628 CD, The Netherlands
\aff{3}Maritime Research Institude of the Netherlands (MARIN), Haagsteeg 2, Wageningen, 6708 PM, The Netherlands}

\corresau{Lina Nikolaidou, \email{m.nikolaidou@tudelft.nl}}

\begin{document}
\maketitle

\begin{abstract}
Air lubrication regimes were studied using simultaneous drag force measurements and multi-plane imaging to characterize the regimes and identify the governing mechanisms of drag reduction. A bubbly, transitional, and air layer regime are identified over a large range of freestream velocities ($U_{\infty}$), air flow rates ($Q_{air}$), and Froude-depth numbers ($Fr_d$). For the lowest $U_{\infty}$, drag reduction lags significantly behind the non-wetted area coverage at all cases and no simple correlation exists. Within the bubbly regime, a drag increase is found for low $U_{\infty}$ with large, slow-moving bubbles forming a single layer over the plate height. For higher velocities, bubbles become smaller and disperse vertically, while the drag starts decreasing. For higher $Q_{air}$, irrespective of $U_{\infty}$, air patches start to form (transitional regime) and drag monotonically decreases, with the onset of the air layer regime at 60\% drag reduction. A new scaling of the associated critical $Q_{air}$ is proposed, combining the air exit velocity, the liquid velocity close to the air layer and $Fr_d$. For a further increase of $Q_{air}$ and low $U_{\infty}$, a thicker and smoother air layer is formed with even lower drag; for higher $U_{\infty}$, marginal differences are observed. The air layer morphology is significantly altered however, depending on $Fr_d$: for $Fr_d>0.7$, it is unbounded, extending beyond the current test section length, and for subcritical conditions (deep water regime, $Fr_d<0.61$) a closure is formed and the air layer transitions to a cavity of a specific length.
\end{abstract}



\section{Introduction}
\label{sec:intro}

Drag reduction by air lubrication is one of the techniques applied to reduce frictional resistance beneath a ship’s hull. Despite the fact that this technology has already been extensively studied and implemented on ships, the physical mechanisms responsible for drag reduction are still not fully understood. As a result, discrepancies persist between model-scale and full-scale results, and reliable scaling laws have yet to be established. This is partly due to the complex multiphase nature of these flows and the difficulty of performing experiments or numerical simulations at the high Reynolds numbers relevant to full-scale conditions. Better understanding these flows could lead to improved designs of air lubrication systems that target specific operating conditions.

In literature, these flows are mostly studied by means of injecting air into the spatially developing turbulent boundary layer (TBL) of a flat plate. In most of these studies a backward-facing step or cavitator is used upstream of the air injector \citep{zverkhovskyi2014ship, barbaca2019unsteady}. While this insert is shown to enhance the stability of the air phase, it demands an additional hull modification on existing ships. Another option is to inject air into the TBL without the use of such inserts, a configuration much less researched \citep{elbing2008bubble,jang2014experimental,nikolaidou2024effect}.

With this configuration, three air phase regimes are commonly observed. For a constant freestream velocity ($U_{\infty}$) and for small air flow rates ($Q_{air}$), a {\em{bubbly regime}} (BR) is observed. In this regime, dispersed bubbles are present in the flow. By further increasing $Q_{air}$, the bubbles coalesce to form air patches ({\em{transitional air layer regime}} - TALR). Eventually by further increasing $Q_{air}$, the air patches also coalesce and an {\em{air layer}} is formed ({\em{air layer regime}} - ALR). \cite{elbing2008bubble} performed large scale air lubrication experiments in a $13m$ long flat plate, where the shear stress was measured over several locations downstream of a slot-type injector. They defined each regime based on the resultant drag reduction (DR). More specifically, the TALR and the ALR were demarcated by 20\% and 80\% drag reduction, respectively. This was based on a single $U_{\infty}$ measurement and a single shear stress measurement location. These measurements revealed a steep increase in drag reduction with $Q_{air}$ within the TALR ($20\%<DR<80\%$) followed by a gradual convergence towards nearly 100\% drag reduction as $Q_{air}$ increased further within the ALR. In our previous study \citep{nikolaidou2024effect}, using a similar air injector as \cite{elbing2008bubble}, we also found all three regimes with a similar morphology, but no quantitative comparison with \cite{elbing2008bubble} was possible in the absence of drag force measurements. While \cite{elbing2008bubble} performed experiments in a large range of conditions spanning all three air phase regimes, their focus was not on the physical mechanisms of drag reduction. In the following sections, we discuss the drag reduction mechanisms associated with the BR, TALR, and ALR separately, based on insights from other existing literature.

Many studies have targeted the $BR$ and explored the drag reduction mechanisms following the successful results of microbubble drag reduction  \citep{mccormick1973drag}. As a result, earlier studies have focused particularly on microbubbles (e.g. \citealp{ferrante2004physical, elbing2008bubble, sanders2006bubble}). In contrast, the effects of larger, millimeter-sized bubbles have received comparatively less attention. Across these studies, drag reduction mechanisms are generally attributed to modifications of the momentum transfer in the near-wall region, such that the drag is reduced. A broad consensus highlights bubble deformability — captured by the Weber number — as the main parameter governing drag reduction \citep{verschoof2016bubble, lu2005, kim2010direct, van2013importance}. 
\cite{murai2014frictional} summarized the successful drag reduction regimes in a bubble diameter - flow speed plot. He suggested a classification of the mechanisms by looking into the bubble sizes in comparison to the coherent structures. He notes that drag reduction can be achieved either by small bubbles ($\mu m -mm$) in high speed flows ($>1$ m/s) or larger bubbles ($mm$) in low speed flows ($<0.1$ m/s). For velocities in between these two regimes, bubbles smaller than one centimeter are increasing the drag by perturbing the laminar state and triggering flow transition. Yet, results of an {\em{increase}} in drag due to bubbles are relatively scarce in literature.
\cite{savviobubble} performed air lubrication measurements on a flat plate over smooth and rough surfaces. A drag increase was reported for $U_{\infty}=2$ m/s and the lowest air flow rates tested within the BR. No information on the corresponding bubble size was reported. These drag increasing bubbles were measured to travel at approximately $0.5U_{\infty}$ while the relative bubble speed increased with increasing $U_{\infty}$ suggesting that for higher $U_{\infty}$ bubbles might migrate at the outer layer of the TBL. \cite{biswas2024effects} performed experiments in shallow channel flow, with and without salts, at relatively low Reynolds numbers. They saw that smaller bubbles (as a result of salt addition) result in drag increase. They attributed this effect in smaller bubbles being less deformable due to their more spherical size and also due to the deposition of salt ions at the bubble water interface. They noticed that smaller bubbles stay densely packed and close to the top wall, unlike the smaller salt-born bubbles which disperse near the top wall. Bubble drag increase was also found by \cite{Gabillet2002}. In that experiment, bubbles were injected from a porous layer in the bottom wall of a channel, rising to the top one.  It is shown that close to the (bottom) wall, near the injection, bubbles are of the same size as the energy containing eddies and successfully modify near-wall turbulence, while the liquid flow profile behaves like a rough TBL with the roughness parameter correlating with bubble size. \cite{hori2024numerical} performed interface-resolving DNS on a bubbly plane-Couette flow for low $Re_{\tau}$ numbers. The effect of bubbles and air layers on the drag force was investigated by varying the Weber and Froude numbers for moderate volume fractions of air (1\% - 5\%). They found that in small Weber and high Froude cases, bubbles increase drag with respect to the single-phase flow. By decomposing the stresses in contributions of advection, diffusion and interfacial tension, they found that in these drag-increasing cases, there is a large contribution from the interfacial tension term, which enhances momentum transfer.

Within the TALR, larger air patches are formed (TALR). This regime is largely unexplored with only a few instances in literature \citep{jang2014experimental,nikolaidou2024effect}. No conclusive information on the drag reduction mechanisms is reported, but it is hypothesized that drag reduction comes directly from the reduction of wetted area. Regarding the ALR, the same hypothesis holds. At the same time, it is shown that this regime is sensitive to upstream disturbances from the incoming TBL \citep{elbing2013scaling}. Similarly, in \cite{nikolaidou2024effect} it was found that the incoming TBL conditions affect both the onset of the ALR and also the air layer topology: bigger air layer thickness to incoming TBL thickness ratios result in an earlier onset of ALR and a larger air layer length. In addition, local, instantaneous incoming TBL organization also governs break up of the air layer close to the injector \citep{laskari2025effects}.

Since the $ALR$ is the desired regime in terms of drag reduction, special attention has been given in literature to the critical air flow rate $Q_{crit}$, the one needed for an air layer to be formed. While it has been shown that $Q_{crit}$ increases with $U_{\infty}$, only a few scaling attempts have been reported. In \cite{elbing2013scaling} a potential scaling of the $Q_{crit}$ is proposed by analyzing a single bubble, subjected to shear flow. This scaling indicates that the transition to ALR depends on the ratio of buoyancy to turbulent shear forces, with buoyancy promoting phase separation and turbulent fluctuations enhancing phase dispersion. While a good collapse of most results was exhibited, the authors suggest it is not conclusive and should be regarded as a starting point for further work on the onset of ALR \citep{elbing2013scaling}. Scaling of the $Q_{crit}$ is also discussed in \cite{peifer2020air}, where the authors adopt the following approach: they assume that an air layer is formed and hypothesize that interfacial instabilities — such as Kelvin–Helmholtz waves and turbulent fluctuations — act to wet the surface. They therefore propose that the frequency of wetting events depends on both the outer liquid flow characteristics and the thickness of the air layer. Despite these efforts, there is still no scaling of the critical air flow rate that captures all phenomena and further tests are needed.

While the aforementioned experiments span a wide range of conditions, a parameter not highlighted enough is the depth of the experimental facility and its potential effect on the air phase topology and dynamics. In the case of the ALR, a free surface is present between the air layer and the surrounding liquid. Using the liquid freestream velocity $U_{\infty}$ and the facility depth $d$, a Froude-depth number, $Fr_d$, can be defined. Only a limited number of studies explicitly address the role of $Fr_d$.  \cite{barbaca2019unsteady} reported a change in cavity closure behavior and hypothesized that it was related to variations in $Fr_d$. In contrast, \cite{makiharju2013scaling}, who conducted measurements exclusively in the supercritical regime ($Fr_d > 1.4$), suggested that the cavity closure mechanism should remain similar at lower Froude numbers, since the dominant parameters ($U_{\infty}$, fluid properties, and closure configuration) remain unaltered with $Fr_d$. In most studies, due to the limited depth of facilities, $Fr_d>1$ such that the flow is supercritical, analogous to the supercritical open channel flow regime \citep{elbing2008bubble, sanders2006bubble,makiharju2013scaling, elbing2013scaling}. In these studies, although imaging of the air phase is not performed, it is stated that once the air layer is formed it covers the total length of the flat plate. Only few studies exists within the subcritical regime yielding  $Fr_d<1$ \citep{zverkhovskyi2014ship, nikolaidou2024effect}. 
For this regime, it has been shown, using potential flow theory, that a stable cavity length exists and can be predicted by the dispersion relation \citep{matveev2003limiting,butuzov1966limiting,butuzov1967artificial}. More specifically, for deep water conditions (approximately $Fr_d<0.61$), it has been shown that the air cavity length is in good agreement with half a gravity wave predicted by the dispersion relation \citep{Qin2019, nikolaidou2024effect}. However, no systematic analysis of the influence of $Fr_d$ has been conducted, and this constitutes one of the main objectives of the present work.

In order to investigate the potential effect of $Fr_d$ on the air phase topology, imaging data are required. Past experimental studies have typically focused on measuring drag forces \citep{elbing2008bubble,sanders2006bubble}, while imaging data are often limited to the BR (for bubble sizes) in few conditions, or targeted in the closure region \citep{makiharju2013scaling, elbing2013scaling} of the air layer. The global morphology of the air phase regime is rarely captured. On the other hand, imaging of the air phase topology by means of high speed imaging or x-ray densitometry \citep{barbaca2019unsteady,gawandalkar2025examination,Qin2019} have been used to gain insight to the air phase topology of similar flows. However no studies exists that combine both drag force measurements and imagining of the global air phase topology. 

In this paper, simultaneous measurements of the drag force acting on a flat plate TBL and imaging of the air phase regimes are performed. The latter allows us to probe the air phase topology and perform a quantitative analysis of the air phase characteristics. By combining it with drag measurements, we aim to gain physical insight into the mechanisms responsible for drag reduction. To that end, the air flow rate, freestream velocity and Froude-depth number are systematically varied, allowing a wide range of air phase regimes to be investigated and general trends to be identified. These are discussed and compared for different $Fr_d$ conditions (subcritical and supercritical) to fill in the aforementioned knowledge gap in this context. 

The remainder of this paper is organized as follows. In \S \ref{sec:exp_setup} the experimental setup and methods are given. In \S \ref{onephaseflow} the single phase flow parameters of a developing TBL along a flat plate are given which are then used to evaluate the influence of air. Sections \ref{supercritical_results} and \ref{subcrtitical_results} both describe the multiphase flow results associated with $Fr_d > 1$ and $Fr_d < 1$ respectively. In \S \ref{regimeMAPsec} all results are combined in a single air phase regime map and scaling considerations of the critical air flow rate are presented.

\section{Experimental Setup \& methods}
\label{sec:exp_setup}

\subsection{Facility}

\begin{figure}
\centering
\includegraphics[width= \linewidth]{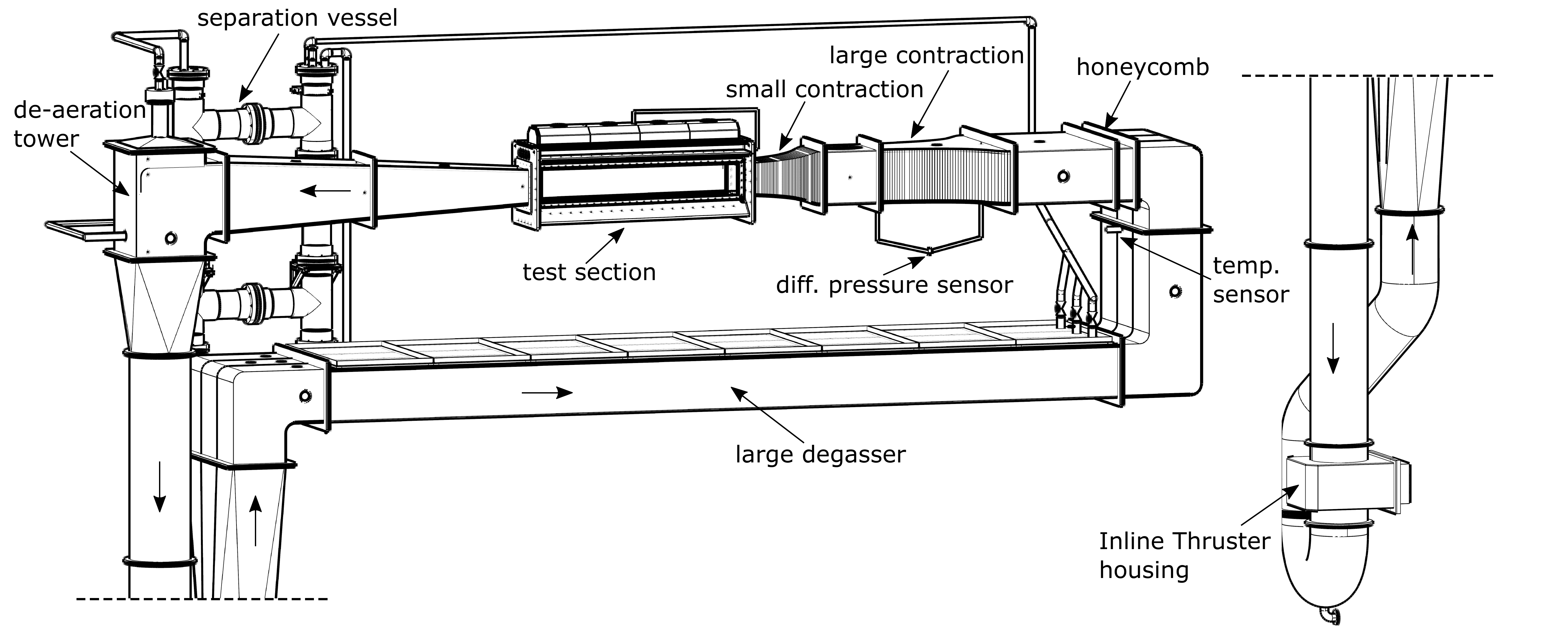}
\caption{Sketch of the multiphase flow tunnel. Arrows indicate the liquid flow direction. For an indication of the scale: the test section is 2.1 m long. The sketch is adapted by the original one by P. Poot.}
\label{MPFT_overview}
\end{figure}

The experiments were performed in the new multiphase flow tunnel (MPFT) at the Ship Hydromechanics laboratory of Delft University of Technology. The flow is driven by a Voith Inline Thruster (VIT; Voith, Heidenheim, Germany) resulting in low noise and vibrations. The operational range of the impeller leads to freestream velocities $U_{\infty}$ = 2.13 m/s to 13 m/s and with the addition of an orifice plate downstream of the rim-driven thruster, a minimum $U_{\infty}$ = 0.5 m/s can be reached. To facilitate air lubrication experiments, the MPFT features two kind of degassing mechanisms to prevent air circulation: a de-aeration tower and a large horizontal degasser (Figure~\ref{MPFT_overview}). In the de-aeration tower, the air-water mixture impacts a sloped plate that breaks air patches, allowing air to escape via a separation vessel, while water flows downstream. The horizontal degasser features a specially designed top lid that collects rising bubbles and directs them to the separation vessel. The former degassing mechanism is mainly effective for lower velocities ($<3$ m/s) and the latter for higher ones. Before entering the test section the flow is conditioned with a honeycomb and two contractions (Figure~\ref{MPFT_overview}) to ensure a uniform inflow and low turbulence intensity ($< 0.6\%$  as measured by laser doppler anemometry \citep{nikolaidou2026flowcharacterizationdelftmultiphase}).

The test section is 2.136 m long and the cross section varies from 0.3 $\times$ 0.3 $m^2$ at the inlet to $0.3 \times 0.32$ $m^2$ at the outlet due to a sloping bottom. The sloping bottom was designed to compensate for the TBL growth for 2 m/s. The test section is tightly sealed with a lid, and it can be pressurized up to 2 bar overpressure. Transparent windows are present on the sides of the tunnel and the bottom, while the top has limited access with 4 viewing windows. The water level in the test section can be monitored with a transparent tube and is kept constant throughout the experiments.

\subsection{Test plates}

Two different test plates were used during the experiments, one for single phase flow and one for air lubrication measurements. The former was a simple flat plate without inserts, used to measure the TBL flow for validation of our measurement system (see section \ref{onephaseflow}). The air lubrication plate was more complex and was subsequently used for multiphase flow measurements (see sections \ref{supercritical_results} and \ref{subcrtitical_results}). The test plates were positioned flush with the ceiling of the test section and were submerged in water (depth $d=0.3$ m). Both were made out of transparent polycarbonate, with length $L=1.95$ m and width $B=0.295$ m. The thickness was 11 mm and 36 mm (in 3 layers of 12 mm) for the single phase and air lubrication plates, respectively. A zig-zag strip of 0.8 mm thickness was placed 10 cm upstream of the leading edge of the test plates in the tunnel contraction to ensure a turbulent boundary layer.

\begin{figure}
\begin{center}
\begin{subfigure}[t]{.48\textwidth}
  \centering
    \includegraphics[width=\textwidth]{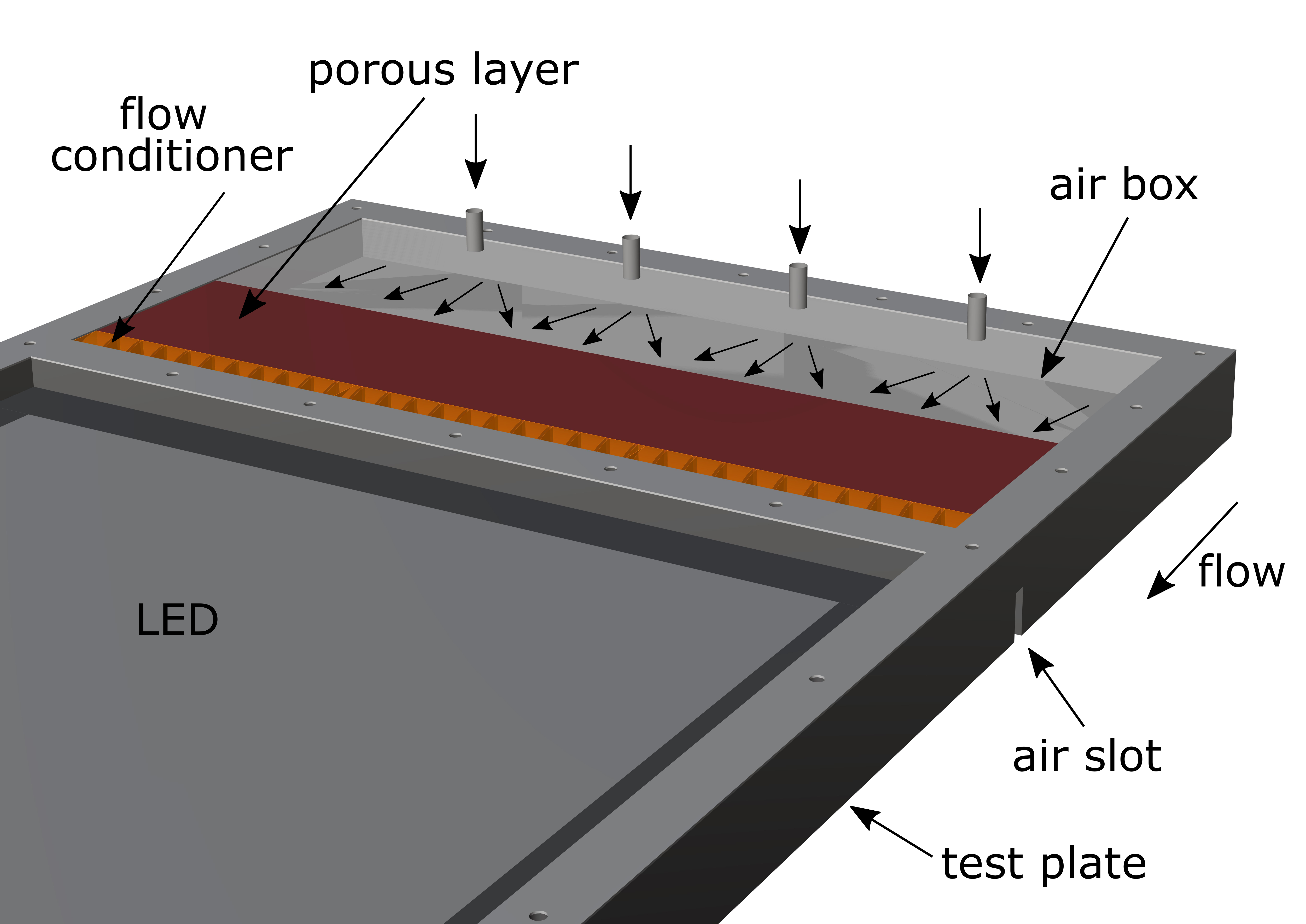}
  \caption{}
  \label{airslot}
  \end{subfigure}%
    \hspace{0.001\textwidth}
\begin{subfigure}[t]{.48\textwidth}
  \centering
\includegraphics[width=\textwidth]{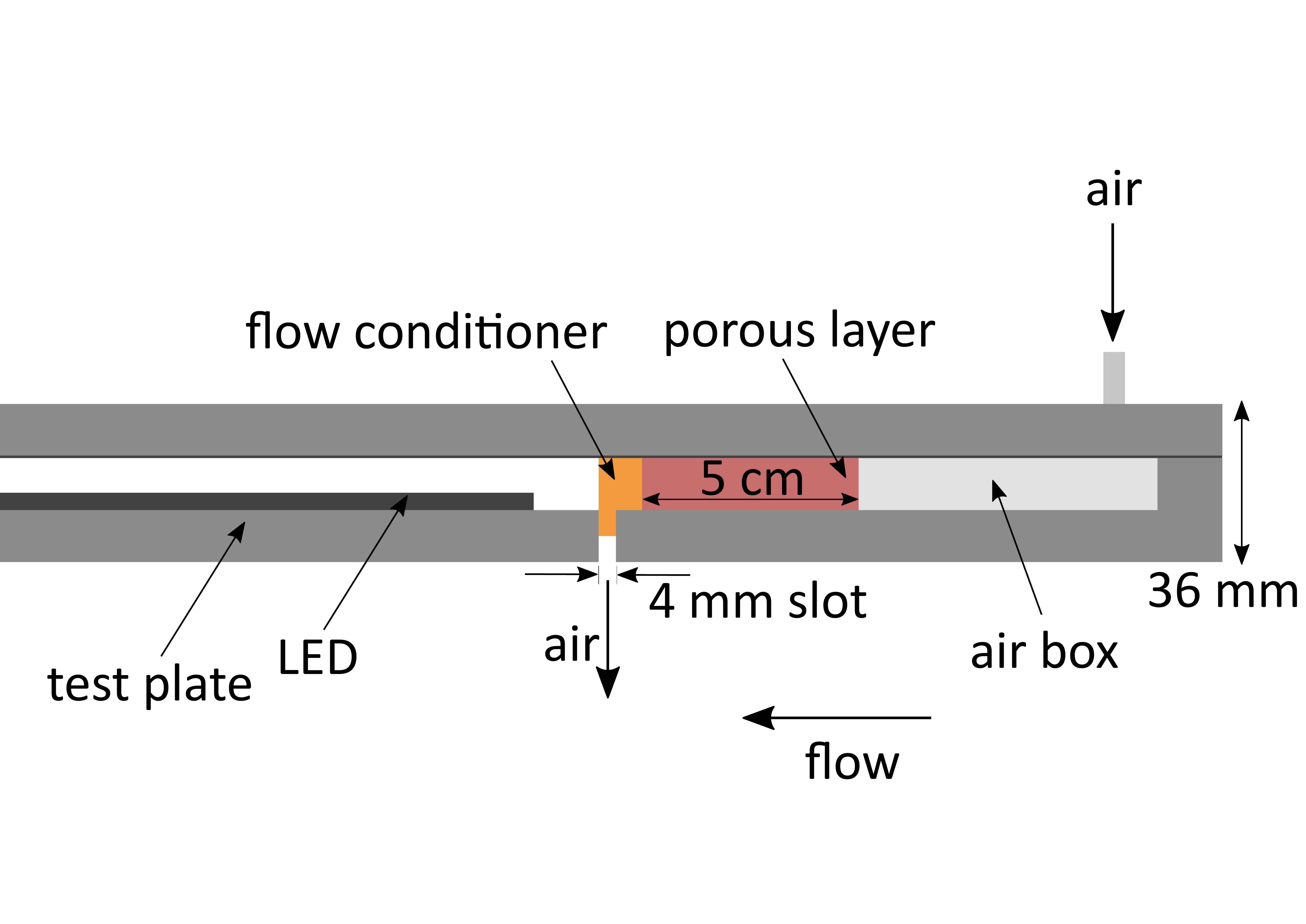}
  \caption{}
  \label{skegs}
  \end{subfigure}
  \caption{Sketch of the air injector. (a) 3D view from the top and (b) side view. Liquid flow is from right to left.}
    \label{CSinj}
\end{center}
\end{figure}

In the case of the air lubrication plate, air was introduced vertically from the top layer through four manifolds equally spaced along the plate's width (see Figure~\ref{CSinj}). It was then dispersed in an air box within the second layer, whose downstream end was fitted with a porous material and flow conditioner, ensuring homogeneous distribution. At the last layer, air was injected vertically into the liquid turbulent boundary layer (TBL) via a $t=4$ mm wide slot-type injector (same as in our previous work \cite{nikolaidou2024effect}), spanning the width of the plate and located 140 mm downstream of its leading edge. The last layer of the plate was also fitted with side fences along its length (2 mm thick and 49 mm in height) to restrict air escaping from the sides (see Figure~\ref{drag_force_system}). Within the plate's middle layer an in-house made LED panel was placed downstream of the injection slot (in a watertight housing) to illuminate the entire plate length and width downstream of the injector, as required for the imaging system (see section \ref{imagingsection}).

The air injection was controlled via a mass flow controller (Bronkhorst EL-FLOW Select series) located several meters upstream of the injector location along with an air filter. The flow meter was factory calibrated for a nominal air flow range of 4 - 500 $l_{n}$/min with an accuracy of $\pm$ 0.5 \% error in the reading value and $\pm$ 0.1 \% of the full range while the repeatability is 0.2\% of the measured value.

\subsection{Drag force measurement system}

Both test plates were fitted to a drag force measurement system during their respective uses. This custom-made force balance was designed and built to assess the total friction force of the lower surface of the test plate. The design was based on previously established ones \citep{zverkhovskyi2014ship}. A load cell (type BM6A by Zemic) with a maximum capacity of 6 kg ($60 N$) was used. The cell was made from IP68 stainless steel and was suitable for underwater applications both in fresh and salt water.

\begin{figure}
\centering
\includegraphics[width= 0.75\linewidth]{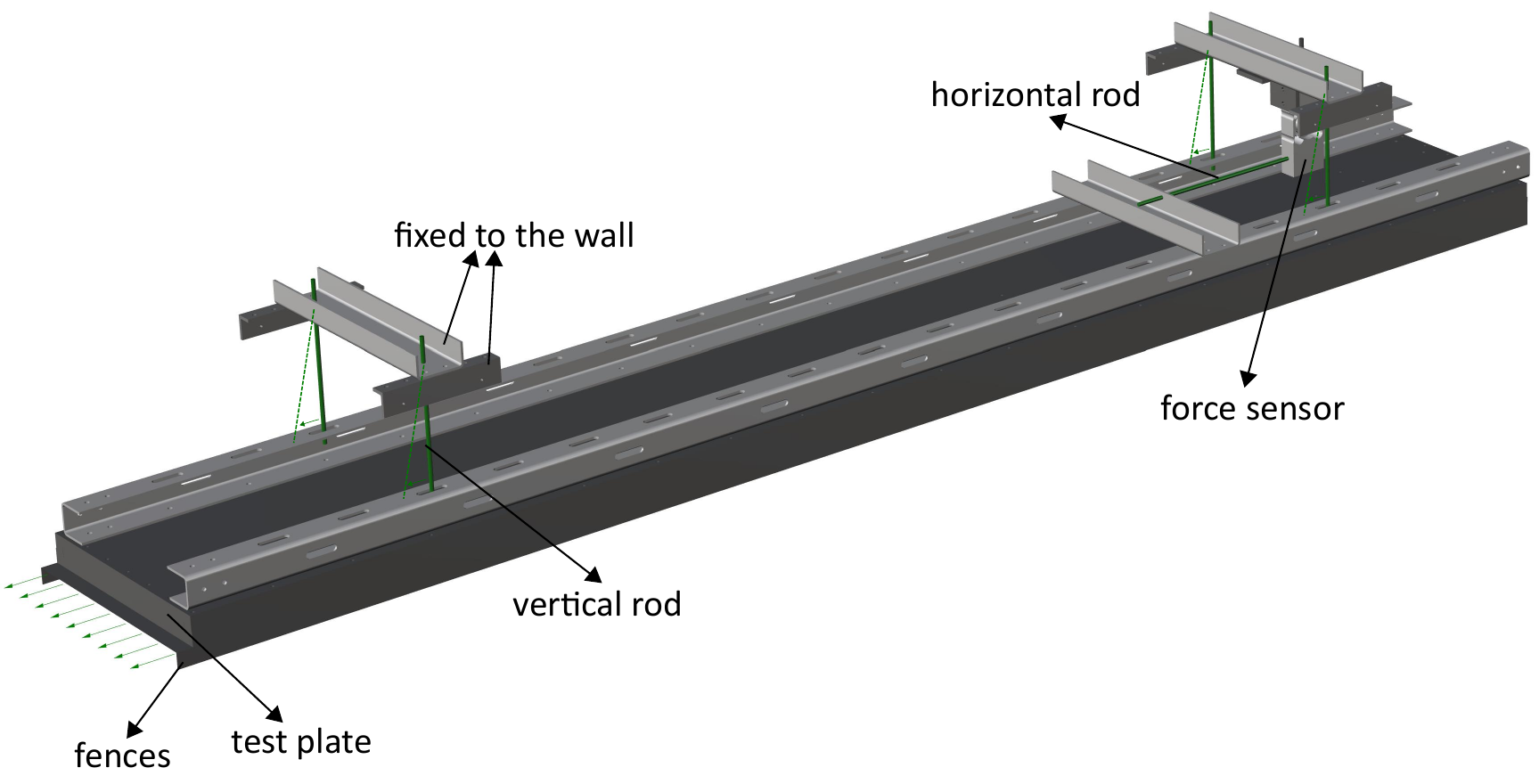}
\caption{Sketch of drag force measurement system attached to the air lubrication plate.}
\label{drag_force_system}
\end{figure}

The test plate was suspended by four vertical rods, with their top end fixed to the tunnel side walls and their bottom attached to longitudinal bars supporting the plate over its length (see Figure~\ref{drag_force_system}). This allowed plate motion only in the flow direction with alignment screws enabling precise leveling (angle of attack $\pm 1^\circ$). The streamwise displacement of the plate was measured by the load cell, mounted 370 mm from its leading edge. The plate's width and length were adjusted to maintain a 1.5 mm gap between the plate and the tunnel walls to allow movement and avoid interference. For the air lubrication plate, apart from the side fences, a plastic film at the trailing edge was also used to prevent air from escaping through that gap.

The force balance had to be calibrated in situ. In operating conditions the test section was closed from the top (see Figure~\ref{MPFT_overview}) and filled with water to the brim, resulting in the sensor being fully submerged in water. The calibration was performed without the lid, but with water on top of the plate and the sensor being underwater. Once the lid was closed, the experiments took place and in the end the calibration was repeated to ensure that nothing changed in the system in between. The calibration was performed using standard forces (weights) and a pulley arrangement in accordance to the guidelines given by the \cite{ittc2017}. The calibration involved 16 steps in total of loading and unloading conditions raging from zero to 55 N with a preload as well. A linear relation was found between the output signal in volt and the force in Newton, with a small standard error ($\sigma = 6.9  \times10^{-3}$ $V$, equivalent to 0.0018 $N$).

\subsection{Imaging system}
\label{imagingsection}

\begin{figure}[htp]
    \centering
    \begin{subfigure}{0.45\linewidth}
        \centering
        \includegraphics[width=\linewidth]{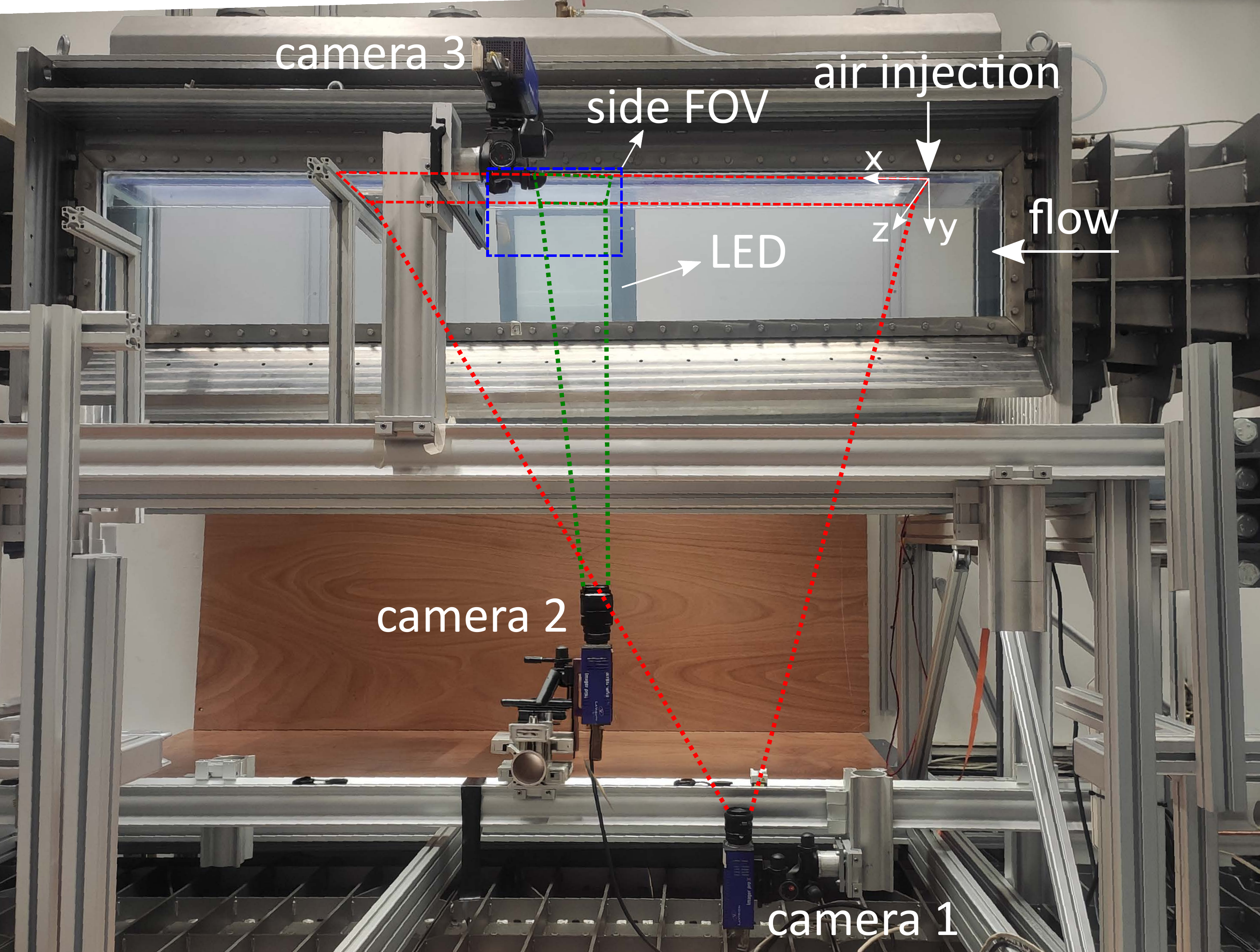}
        \caption{Camera setup 1.}
        \label{camera_setup_1}
    \end{subfigure}
    \hspace{0.01mm}  
    \begin{subfigure}{0.45\linewidth}
        \centering
        \includegraphics[width=\linewidth]{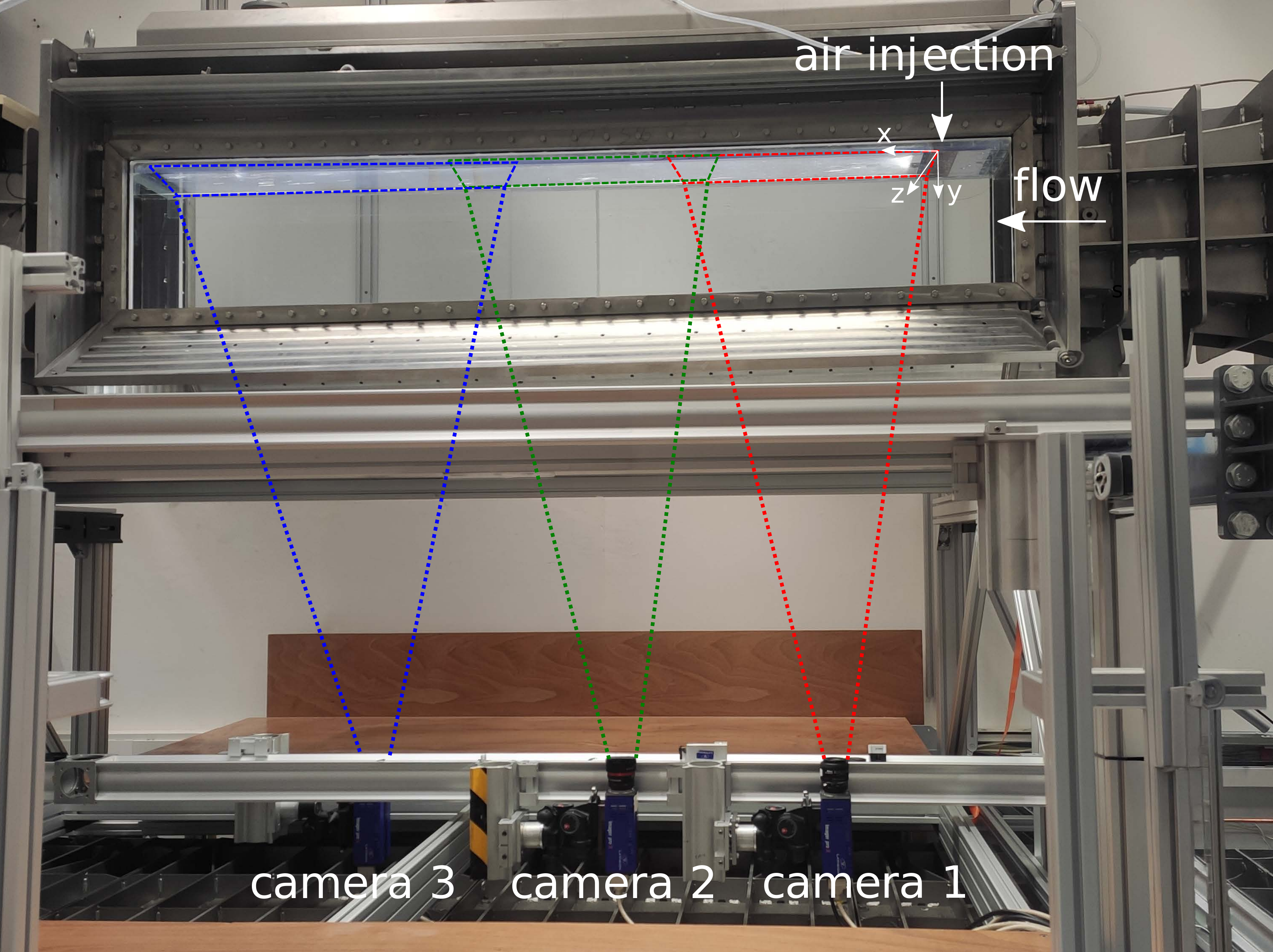}
        \caption{Camera setup 2.}
        \label{camera_setup_2}
    \end{subfigure}
    \caption{Imaging systems used for the multiphase flow campaign.}
    \label{fig:main}
\end{figure}

For the air lubrication measurements, two different imaging systems were used, to accommodate two different research objectives. First, we wanted to capture the three-dimensional air phase topology synchronously with the measured drag force, in order to gain insights into the different drag reduction mechanisms. This necessitated multiple plane views ($x-y$, $x-z$; see Figure \ref{camera_setup_1} for the coordinate system used), high resolution and thus smaller fields of view (FOVs). Second, we wanted to associate the measured drag force with the \textit{total} air covered plate area, and probe their correlation. This required large FOVs on the $x-z$ plane (see Figure \ref{camera_setup_2}).

Three Imager Pro X 4M cameras (10-bit, LaVision) with a pixel size of 7.4 $\mu m$ and a resolution of 2048 $\times$ 2048 pixels were used for both setups in different configurations and with different lenses, as described below. For wall-parallel imaging planes, illumination was provided by the aforementioned LED panel built within the test plate, while wall-normal imaging was back-lit by another LED panel resting on the tunnel side wall (see Figures \ref{CSinj} and \ref{camera_setup_1}, respectively). Davis 10.2 (LaVision GmbH) was used for data acquisition.

\begin{table}
  \begin{center}
\def~{\hphantom{0}}
  \begin{tabular}{ccccc}
      camera No. & FOV (mm $\times$ mm) & focal length (mm) & $f^{\#}$ & resolution (px/mm) \\[3pt]
       1&740$\times$740 &24 &5.6 &2.75\\
       2& 110$\times$110& 105&5.6 & 18.48\\
       3& 70$\times$70&105 &5.6 & 30.16\\
  \end{tabular}
  \caption{Camera setup 1 details.}
  \label{tab:setup1}
  \end{center}
\end{table}

For the first setup, camera 1 was positioned in a down-up configuration imaging the $x-z$ plane downstream of the injection slot. This camera provided information on the global air phase characteristics. Camera 2 captured a smaller FOV of the same plane with an increased spatial resolution. Images from this camera were used for estimating smaller features (e.g. bubble sizes). Lastly, a third camera (3) capturing the $x-y$ plane was used to estimate the bubbly/air layer thickness. This camera was placed in the middle of the test section, except for some tests where it was traversed close to the air injector. All three cameras were synchronously capturing independent snapshots with an acquisition frequency of 0.8 to 7.4 Hz depending on the $U_{\infty}$. For this campaign $U_{\infty}$ varied from 0.5 to 5.5 m/s, while $Q_{air}$ from 5 to 400 l/min. Details of this camera setup are given in Table~\ref{tab:setup1}.

For the second setup, the same three Imager Pro X 4M cameras were used, but in a different configuration. They were positioned below the test section in a down-up configuration to collectively capture the total $x-z$ extent of the test plate downstream of the injection slot. A small overlap between the three field-of-views (FOVs) was present to allow stitching. The acquisition frequency varied from 1.11 to 1.94 Hz depending on $U_{\infty}$, to ensure statistically independent snapshots, whose total number varied from 1000-1500. In this campaign $U_{\infty}$ was 2 m/s and $Q_{air}$ varied from 5 - 140 l/min. The exposure time was set to 200 $\mu s $ to achieve less than one pixel motion blur. Details of this camera setup are given in Table~\ref{tab:setup2}.

\begin{table}
  \begin{center}
\def~{\hphantom{0}}
  \begin{tabular}{ccccc}
      camera No. & FOV (mm $\times$ mm) & focal length (mm) & $f^{\#}$ & resolution (px/mm) \\[3pt]
       1 &510$\times$510 &35 &4 &4\\
       2 & 490$\times$490& 35&4 & 4.22\\
       3 & 780$\times$780&24 &4 & 2.61\\
  \end{tabular}
  \caption{Camera setup 2 details.}
  \label{tab:setup2}
  \end{center}
\end{table}

\subsection{Experimental program \& procedure}

Two experimental campaigns were performed. First, single phase flow measurements of the drag force acting on the flat plate were performed for a freestream velocity range of 2 to 7 m/s. These included drag force measurements of the flat plate without imaging. Secondly, single and multiphase flow measurements of the drag force acting on the air lubrication plate were performed with synchronous imaging of the air phase. For this campaign the two imaging systems described above were used. First, a short campaign was performed using camera setup 2. Secondly, measurements using camera setup 1 were performed. A summary of the experimental conditions for the latter is given in Table~\ref{tab:expcond}.

\begin{table}
  \begin{center}
\def~{\hphantom{0}}
  \begin{tabular}{lccc}
      \textbf{Parameters} & \textbf{Definition} & \textbf{Value} & \textbf{Units} \\[3pt]
       {Freestream velocity} & $U_{\infty}$ & $0.5-7$ & m/s\\
       {Air flow rate} & $Q_{air}$ & $5-450$ & l/min\\
       Reynolds number & $Re_L=U_{\infty}L/\nu$ & $1 - 14 \times 10^6$ & -\\
       Froude-depth number & $Fr_d=U_{\infty}/\sqrt{gd}$ & $0.29 - 4$ & -\\ 
       non-dimensional air flux & $C_Q=Q_{air}/BtU_{\infty}$ & $0.012 - 1.4$ & -\\
  \end{tabular}
  \caption{Experimental parameters of the multiphase flow campaign using camera setup 1.}
  \label{tab:expcond}
  \end{center}
\end{table}

For a given rotational frequency of the VIT, the mean drag force was recorded at 100 Hz and averaged over 10 minutes, which was sufficient for convergence. Simultaneously, the differential pressure across the contraction and the water temperature (as shown in Figure~\ref{MPFT_overview}) were recorded and averaged. The differential pressure was used to estimate $U_{\infty}$, with a correction applied, as set by earlier LDA measurements \citep{nikolaidou2026flowcharacterizationdelftmultiphase}. The signals were continuously recorded via LabVIEW. For the multiphase cases, measurements were first performed without air followed by those with air, the mass flow rate of which was also recorded and averaged. 

\section{Single phase flow}
\label{onephaseflow}

In this section, single phase flow results are presented in the absence of air injection. Firstly, the characteristics of the single phase TBL developing along the flat plate are given. Secondly, the drag force measurement system is validated against theoretical predictions. These will set the baseline for the multiphase flow measurements introduced later.

\subsection{Flat plate TBL characterization}

The flat plate TBL thickness was measured with LDA \citep{nikolaidou2026flowcharacterizationdelftmultiphase} in three streamwise locations along the plate and multiple velocities (see Table~\ref{tab:top_BL_thickness} for one $U_{\infty}$). Due to the sloping bottom of the test section, the freestream velocity was found to slightly decrease at the most downstream position ($\approx$ 2\%). Due to the position of the flat plate at the tunnel's ceiling, the TBL was found to originate approximately 30 cm prior to the plate's leading edge and with a growth was not strictly canonical.

\begin{table}
  \begin{center}
\def~{\hphantom{0}}
  \begin{tabular}{cc c c c}
      x (m) & $Re_{x}\times10^6$ & $\delta_{99}$ (mm) & $U_{\infty}$ (m/s) & $\nu/u_{\tau}$ ($\mu$m)\\[3pt]
       0.1 & 2.4 & 16   & 5.08 & 4.38\\
0.975 & 7.6 & 51 & 5.06 & 4.77\\ 
1.85 & 12.8 & 77 & 4.98 & 5.19\\
  \end{tabular}
  \caption{Experimental parameters of the single phase flow campaign.}
  \label{tab:top_BL_thickness}
  \end{center}
\end{table}

\subsection{Drag force measurement validation}

In order to validate the drag force measurement system, measurements of the drag force acting on both the flat plate and the air lubrication plate were performed over a large range of freestream velocities. The total drag force coefficient was calculated as:

\begin{equation}
    C_{D}=\frac{2D_{no air}}{\rho U_{\infty}^2A}
    \label{formulaCD}
\end{equation}
where $\rho$ is the calculated water density based on the measured temperature, $D_{no air}$ is the measured drag force acting on the flat plate, $U_{\infty}$ is the freestream velocity and $A$ is the test plate (bottom) surface area.

\begin{figure}[!t]
\centering
\includegraphics[width=0.95\linewidth]{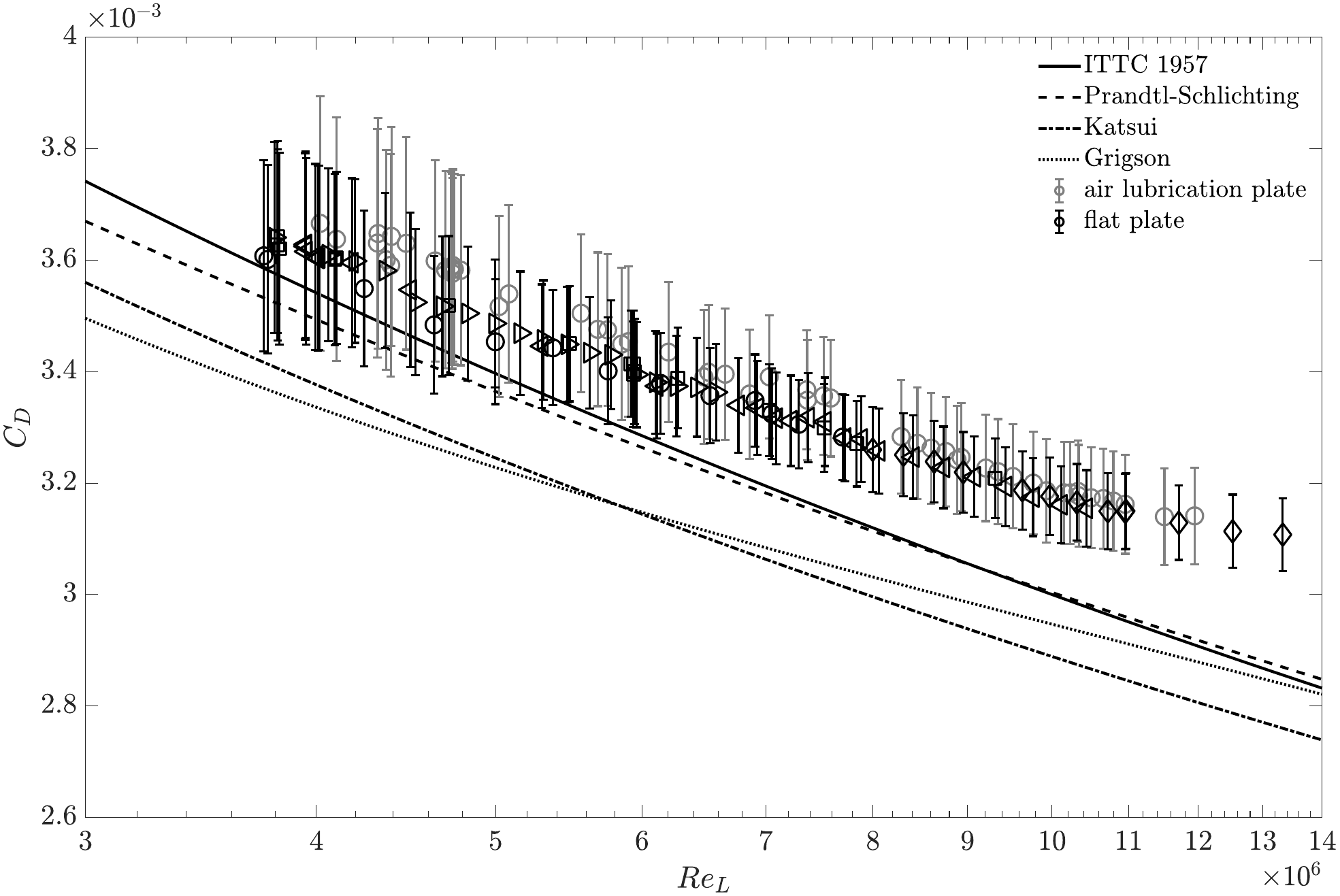}
\caption{Flat plate total skin friction coefficient of the current measurements along with friction lines from literature \citep{ittc1957,schlichting1979boundary,katsui2005proposal,cwb1999planar}. Black markers are from measurements with a flat plate. Different marker shapes are from different measurement days. Grey markers are from measurements with the air lubrication plate. Error bars represent the bias and precision errors following \cite{ittc2002}.}
\label{CD}
\end{figure}

The flat plate measurements of $C_{D}$ are compared with reference friction lines from literature (Figure~\ref{CD}). More specifically, ITTC's \citep{ittc1957} and Prandtl-Schlichting's \citep{schlichting1979boundary} semi-empirical formulas are plotted. Following \cite{bose2005specialist}, Grigson's \citep{cwb1999planar} and Katsui's \citep{katsui2005proposal} lines are also obtained through numerical integration of the local friction coefficient in the boundary layer. The Reynolds number is defined as $Re_{L}=U_{\infty}L/\nu$. Use of the aforementioned TBL origin did not significantly  alter the results presented here and as such this correction was not applied. The trend is consistent, and the results show good overall agreement. The experiments appear to be biased towards higher values compared to all the theoretical friction lines. Tests performed without a lid over the test section, revealed systematically lower values of the drag coefficient (approximately 5-6\%) hinting to possible added resistance from water motion \textit{above} the plate (supported also by visual observations). Since we are interested in changes in drag however, further investigation of such effects was considered out of the scope of the present study. Negligible differences were also found between the drag coefficients of the two test plates (with area differences due to the fences and sealing strip fitted on the air lubrication plate taken into account); thus, any potential form drag effects due to the different geometry of the air lubrication plate were marginal.

An uncertainty analysis of $C_{D}$ was performed following the recommendations by \cite{ittc2002}. For both test plates, the largest source of uncertainty is the bias error from the calibration curve fit, due to the sensor's large force range. A small precision error is found, as seen in the repeatability of flat plate measurements performed on different days, shown with different (black) markers in Figure~\ref{CD}. The air lubrication plate has slightly higher uncertainty due to additional bias from the manufacturer error (material tolerances), caused by the system's complexity.

Following the aforementioned baseline checks in absence of air (injection), results of multiphase flow conditions are discussed next.

\section {Air injection - supercritical conditions}
\label{supercritical_results}

In this section, multiphase flow results pertaining to {\em{supercritical}} flow conditions are discussed, i.e. all cases $Fr_d>1$. In section~\ref{subcrtitical_results} {\em{subcritical}} flow results are presented separately. This classification will be more clear once both conditions are presented and discussed.

\subsection{Baseline drag reduction and air regime topology}

\begin{figure}
\centering
\includegraphics[width=0.95\linewidth]{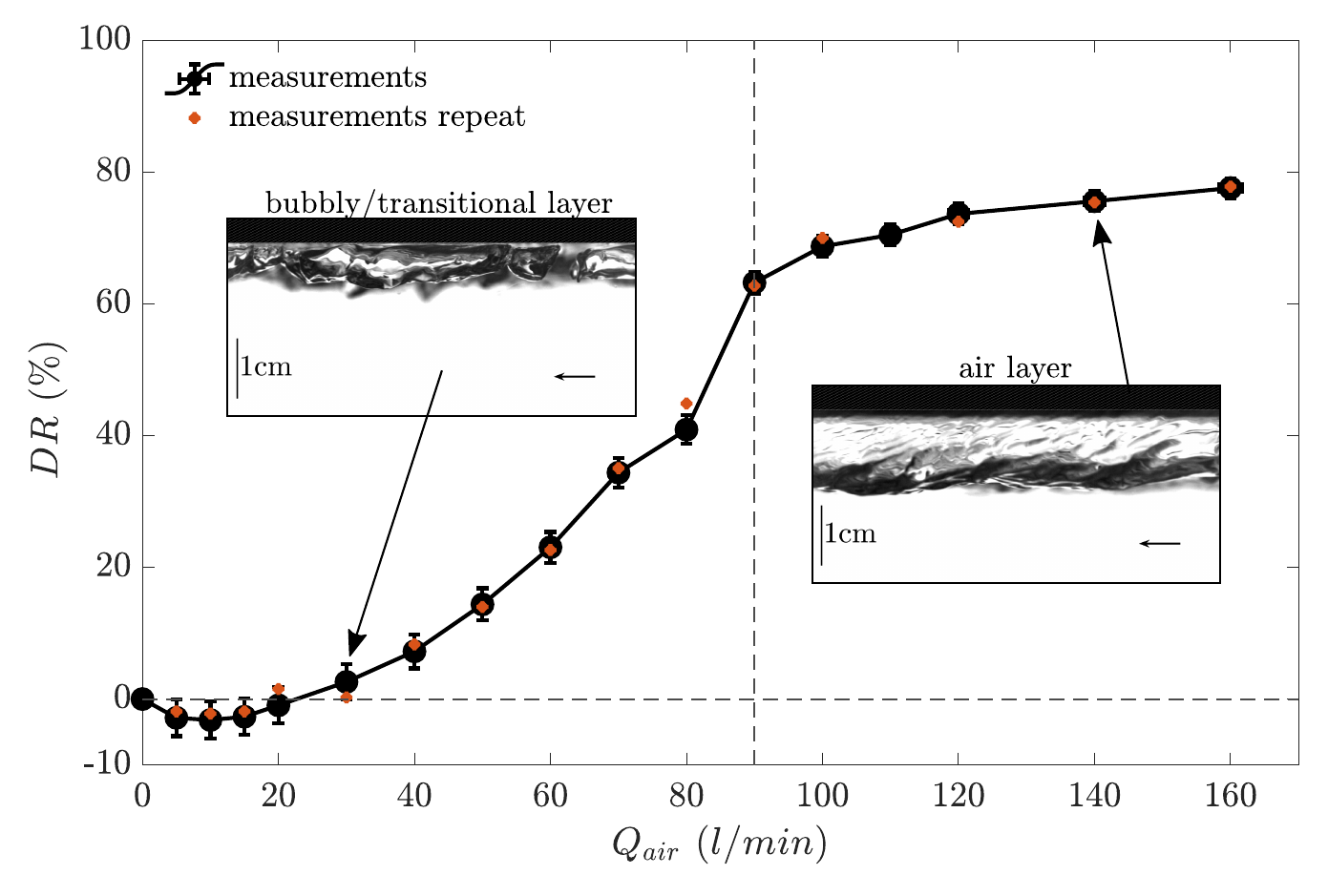}
    \caption{Drag reduction with increasing air flow rate for a nominal freestream velocity of 2 m/s ($Fr_d=1.24$). The vertical dashed line demarcates the transition to the air layer regime at $Q_{air}=Q_{crit}$. Right insert: side view ($x-y$ plane) of the air layer at $Q_{air}=140$ l/min. Left insert: side view ($x-y$ plane) of the bubbly regime at $Q_{air}=30$ l/min.}
\label{intro_DR}
\end{figure}

We here first introduce the general air phase regime topology, along with the corresponding drag forces. A typical drag reduction curve obtained during the experiments can be seen in Figure~\ref{intro_DR}.\\

In what follows, drag reduction is calculated with:

\begin{equation}
    DR = c \times \frac{D_{no air}-D_{air}}{D_{no air}}
    \label{DR}
\end{equation}
where $D_{no air}$ in the drag force of the air lubrication plate without air ($Q_{air}=0$), $D_{air}$ is the drag force of the test plate with air ($Q=Q_{air}$) and $c$ is a constant. The constant $c$ is introduced because $D_{air}$ is the force experienced by the test plate in its entirety (see Figure 3). In reality, only a part of the test plate can be lubricated (downstream of the injector and up to a height equal to the air layer thickness on the side fences). Using the measured maximum air thickness (see section \ref{transReg}) and assuming a linear increase in $DR$ with the covered test plate area, we obtain $c=1.34-1.35$. Although introducing $c$ adds an additional bias error to  $D_{air}$, this correction is applied since we are interested in drag reduction over the area that can be covered by air: downstream of the air injector and along a part of the side fences area. This also enables us to compare our results with other studies where typically \textit{local} shear sensors are used to estimate drag reduction. These sensors only consider the local effect of air by definition \citep{elbing2008bubble,makiharju2013scaling}. 

\begin{figure}
    \centering
    \begin{subfigure}[t]{0.9\textwidth}
        \centering
       \includegraphics[width=\textwidth]{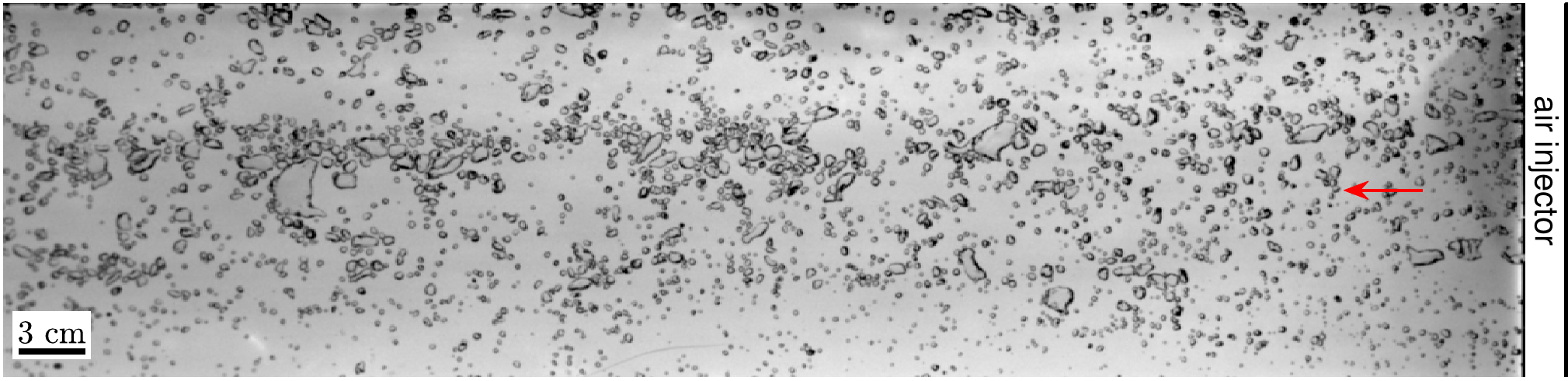} 
        \caption{bubble regime}
        \label{charaBUBBLY}
    \end{subfigure}
    \hfill
    \begin{subfigure}[t]{0.9\textwidth}
        \centering
        \includegraphics[width=\textwidth]{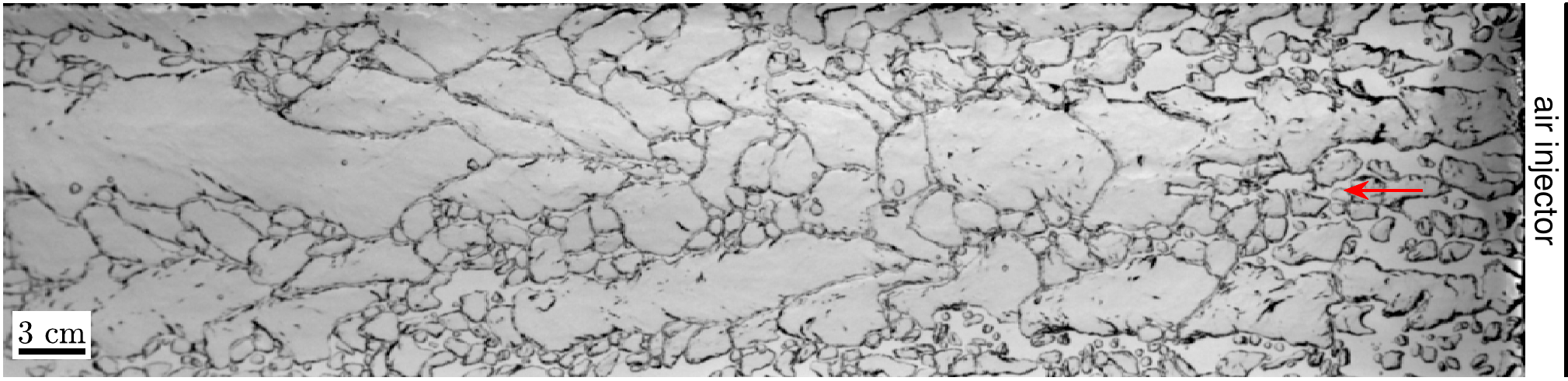} 
        \caption{transitional regime}
        \label{charTRANS}
    \end{subfigure}
    \hfill
    \begin{subfigure}[t]{0.9\textwidth}
        \centering
        \includegraphics[width=\textwidth]{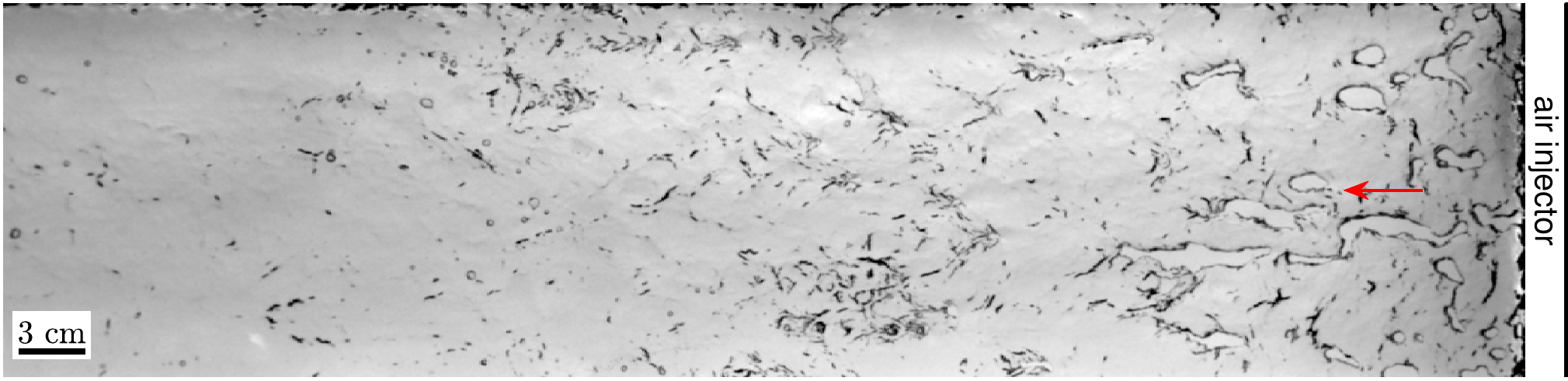} 
        \caption{air layer regime}
        \label{charLAYER}
    \end{subfigure}
    \caption{Characteristics images of the multiphase flow topology: (a) bubbly, (b) transitional, and (c) air layer regime, for $U_{\infty}=2$ m/s. Air flow rate increases from top to bottom. Black rectangle indicates the injector slot location, while the white region could not be imaged.}
    \label{CIs}
\end{figure}

For small air flow rates, a \textit{bubbly} regime is observed (Figure~\ref{intro_DR}), where dispersed bubbles are present in the flow (Figure~\ref{charaBUBBLY}). In our configuration, bubbles are not formed due to the injector geometry (e.g. orifices plate or porous layer as used in \cite{sanders2006bubble} and \cite{Madavan1985}), but they are formed at the injection slot due to the impact of the air flow with the incoming TBL, which causes the air to break into bubbles. A similar mechanism was presented by \cite{shen2006influence} and \cite{park2015drag}. Interestingly enough, from Q = 0 to Q = 20 l/min a drag {\em{increase}} ($DR<0$) is measured, which peaks at Q = 10 l/min. Drag increase cases are rarely reported in literature. The drag increase changes to a decrease by further increasing Q. This drag behavior within the bubbly regime will be addressed in detail at section \ref{bubbly}. 

By further increasing the air flow rate, the bubbles coalesce and air patches are present in the flow (Figure~\ref{charTRANS}), as also reported in \cite{nikolaidou2024effect} and \cite{elbing2008bubble}. While drag is further decreased in this transitional regime, it happens gradually, unlike the steep change for $DR>20\%$, previously reported by \cite{elbing2008bubble} for a similar configuration. It is therefore hard to demarcate the bubbly and transitional regimes through the corresponding drag reduction curve.  \cite{nikolaidou2024effect} found a steep increase in non-wetted area in this region. The availability of synchronous force measurements and imaging data in the current experiments enables the comparison of the two. This will be further elaborated in section \ref{NWA}.

For $Q_{air}=Q_{crit}$, the individual air patches coalesce and an air layer (Figure~\ref{charLAYER}) of a certain thickness is formed (see right side insert in \ref{intro_DR}). This jump in morphology is not gradual but sharp (here observed within the 5 l/min resolution of the air flow). It is correlated with $\approx60\%$ drag reduction and a clear change in the slope of the $DR-Q_{air}$ curve. On average, the air layer de-wets the top wall across the total length of the test section except for the region close to the air injector, where wetted patches persist. The air layer extends beyond the field-of-view of  Figure~\ref{charLAYER}, and the closure can thus not be observed in this image. The integrity of the air layer is disturbed by various factors observed during the experiments. This will be further elaborated in section \ref{transReg}. Due to the supercriticality of the flow, the air layer is unbounded \citep{matveev2003limiting}, meaning that the closure of this air layer extents outside of the test section length (until somewhere in the diffusor). In that sense the morphology of the air layer is different from air cavity cases formed behind a backward facing step or a cavitator \citep{zverkhovskyi2014ship} or in the case of subcritical conditions ($Fr_d<1$) \citep{nikolaidou2024effect}. The latter will be further explored in section \ref{subcrtitical_results}.

\subsubsection{Hysteresis effects \& repeatability}

\begin{figure}[!ht]
\centering
\includegraphics[width=0.75\linewidth]{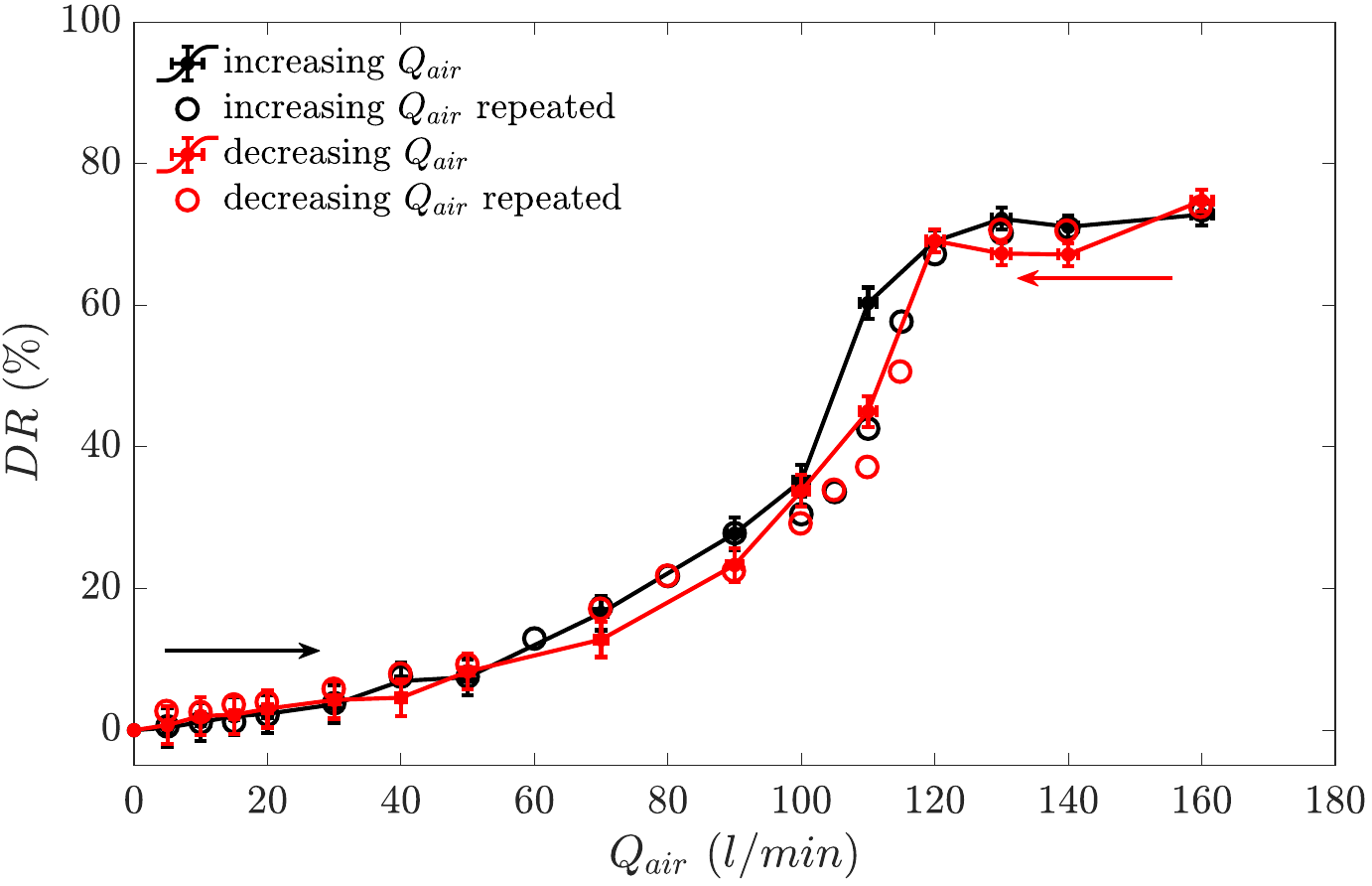}
\caption{Drag reduction measurements in the case of increasing air flow rate from (bubbly to air layer regime) and decreasing air flow rate (air layer to bubbly regime) for $U_{\infty}=3$ m/s ($Fr_d=1.75$). Repeat measurements are also shown (without error bars).}
\label{hysterisis}
\end{figure}

For all the experiments performed, the air flow rate was increased from $Q_{air}=0$ to $Q_{air}=Q_{max}$ and all three regimes (bubbly, transitional and an air layer) were observed. In order to investigate any hysteresis effects the opposite sequence was also tested: starting from the highest air flow rate where an air layer regime was present, drag was measured as air flow rate was decreased to zero. Both scenarios can be seen in Figure~\ref{hysterisis}. The error bars depict the experimental uncertainty due to the bias error. Repeated measurements for each scenario are shown separately.

In the bubbly regime ($Q<60$ l/min), differences fall within experimental uncertainty. In the transitional regime (and especially between 100 l/min to 120 l/min), larger differences arise, exceeding uncertainty, but repeated measurements also show high variability. This suggests that the nature of the transitional regime itself drives $DR$ variations rather than hysteresis effects. The importance of repeated measurements is highlighted here. Conversely, the air layer regime onset is consistent for all measurements, with drag reduction within experimental uncertainty.

Based on the above, we have demonstrated that there is no difference between the air flux required for creation and maintenance of the air layer, which is the desired regime for drag reduction. That means that the supplied air flow rate should always be at least equal to the critical one ($Q_{air}=Q_{crit}$), to be able to maintain the air layer. Lowering the air flow rate would mean that the air layer would break and drag would be increased. That is one of the main disadvantages of this air injection strategy compared to partial cavities reported by (among others) \cite{zverkhovskyi2014ship}, \cite{makiharju2010ventilated}, \cite{Arndt2009}, and \cite{gawandalkar2025examination}, who measured a difference in formation and maintenance air flow rates.

\subsubsection{Non-wetted area and drag reduction correlation}
\label{NWA}

Next, for the same drag reduction results that have been presented in Figure \ref{intro_DR}, the $total$ non-wetted area coverage $A_{nw}$ downstream of the air injector is calculated (see Table \ref{tab:setup2} and section \ref{sec:exp_setup} for details on the experimental setup). To the best of the authors knowledge, this correlation has not been explored in this context before, only qualitative discussions have linked the percentage of non-wetted area with drag reduction. 

For each $Q_{air}$ ranging from 5 l/min to 100 l/min, bubbly/ air layer coverage was quantified by bubble segmentation using Artificial Intelligence (AI). More specifically, a zero-shot Segment Anything Model (SAM) was implemented as an extension of the flow segmentation framework and modified for bubbly flow images \citep{khojasteh2024practical}. Although SAM can be fine-tuned for bubbly flows \citep{khojasteh2024practical} and related two phase flow studies have reported retraining the model for complex interfaces \citep{kuccuk2025segmenting}, the zero-shot configuration in the present work provided sufficient performance to proceed. For each image, the automatic mask generator sampled a uniform grid of point prompts to generate candidate bubble masks; candidates were filtered using the model predicted Intersection over Union (IoU) stability score criterion, and box based non maximum suppression, with optional small region cleanup via min mask region area \citep{kirillov2023segment}. To isolate dispersed bubbles and ignore the large connected patches that appear at higher air flow rates ($\geq70$ l/min), masks were post processed by projected pixel area $A$ and only masks within a prescribed range were used. For higher air flow rates ($>80$ l/min), where bubble patches became large and image contrast decreased, the automatic segmentation was refined manually using an interactive add/remove step with the SAM predictor: spurious regions were removed by selecting masks, and missing/partial bubbles were added using point prompts and, when needed, bounding box prompts to trim oversized predictions prior to analysis \citep{kirillov2023segment}. Following the calculation of $A_{nw}$ of all the images, an average value is calculated.

\begin{figure}[!t]
\centering
\includegraphics[width= 0.75\linewidth]{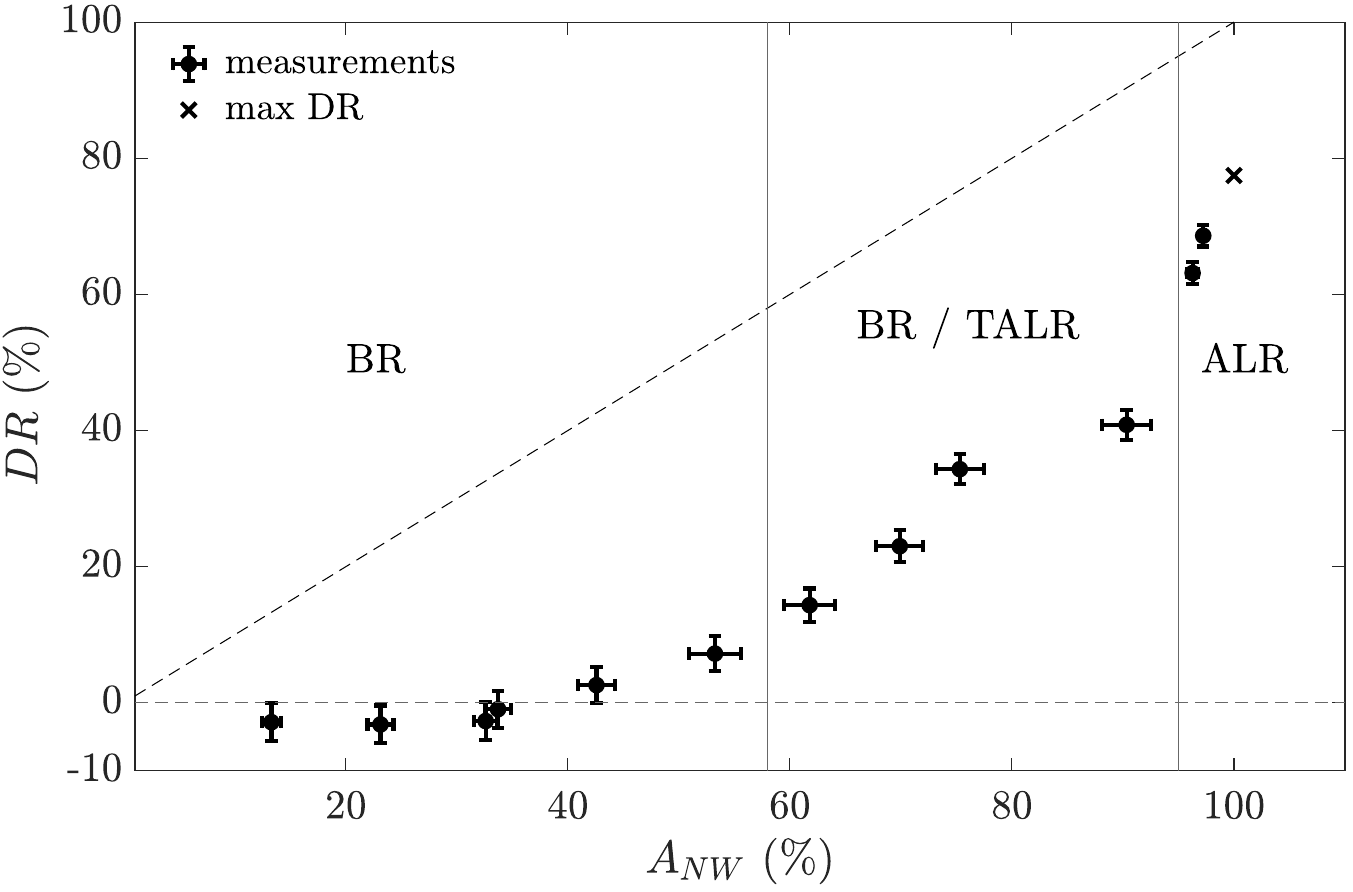}
\caption{Non-wetted area, calculated from down-up images versus drag reduction measurements for $U_{\infty}=2$ m/s ($Fr_d=1.24$). Air flow rate increases from left (5 l/min) to right (100 l/min).  See Figure \ref{intro_DR} for the air flow rates specification. Dashed line indicates $DR = A_{nw}$ for reference.}
\label{ANWvsDR}
\end{figure}

Results show that the non-wetted area coverage is not equal to the resultant drag reduction in any of the three air phase regimes. On the contrary, the $DR$ lags behind the $A_{nw}$. Overall, the $DR-A_{nw}$ trend is similar with the  $DR-Q_{air}$ one (see Figure \ref{ANWvsDR} and Figure \ref{intro_DR}) for the BR and TALR, while a different trend is seen for the ALR. Although a correlation can be seen between $DR$ and $A_{nw}$ in each regime separately, no universal correlation exists. 

Starting with the BR, it can be seen that for low $Q_{air}$, increasing $A_{nw}$ (via increasing $Q_{air}$) results in a drag {\em{increase}}. That persists up to a 35\% coverage of the flat plate. Further increasing $A_{nw}$ up to 53\% results in a low drag reduction of around 7\%. It must be noted that the non-wetted area within the bubbly regime does not correspond to the amount of air de-wetting the top wall: it is just the projection of the bubbles to the top wall which may or may not be touching the wall. 

By further increasing $A_{nw}$ within TALR, $DR$ increases at a higher rate. As a result, the lag between the parameters is reduced, and a better correlation is seen. In this regime, bubbles/ air patches are ``flatter'' and most likely the calculated $A_{nw}$ is closer to the actual amount of air de-wetting the wall. However, the lag remains considerable: a 75\% non-wetted area corresponds to merely 32\% $DR$. Comparing the last two points within the TALR, a big increase in $A_{nw}$ is calculated (17\%). However, this increase is not proportionally translated to a $DR$ increase (6\%). 

A clear jump in $DR$ is observed when further increasing $A_{nw}$, which coincides with the onset of ALR ($Q_{air}=Q_{crit}$): a moderate increase in $A_{nw}$ (4.5\%) results in a large DR increase (22\%) owing to the TALR air patches coalescing into an air layer. At this point, the $A_{nw}$ is already 92\% with only a few wetted patches close to the air injector (see Figure \ref{wettedpatches}, discussed in detail later). Interestingly, an even steeper increase of $DR$ to 69\% is observed when $A_{nw}$ reaches 97\%. Finally, a maximum DR of 77\% (marked by $\times$ in Figure \ref{ANWvsDR}) is measured for an almost 100\% air coverage. Within the ALR, for a further increase of $Q_{air}$ above $Q_{crit}$, the wetted patches near the wall never cease to exist but the air layer gets thicker, which could possibly contribute to the DR increase (Figure \ref{tairQ}). In the next section, the reasons for this DR increase within $ALR$ will be discussed. 

From the above, it can be concluded that the non-wetted area coverage alone is not enough to characterize the drag reduction regimes or provide clear insights regarding the governing mechanisms. As a result, drag force measurements and multi-plane images of the air phase are needed for each condition.

\subsection{Effect of freestream velocity on drag reduction curves}
\label{DRcurves}

\begin{figure}[!t]
\centering
  \includegraphics[width=\textwidth]{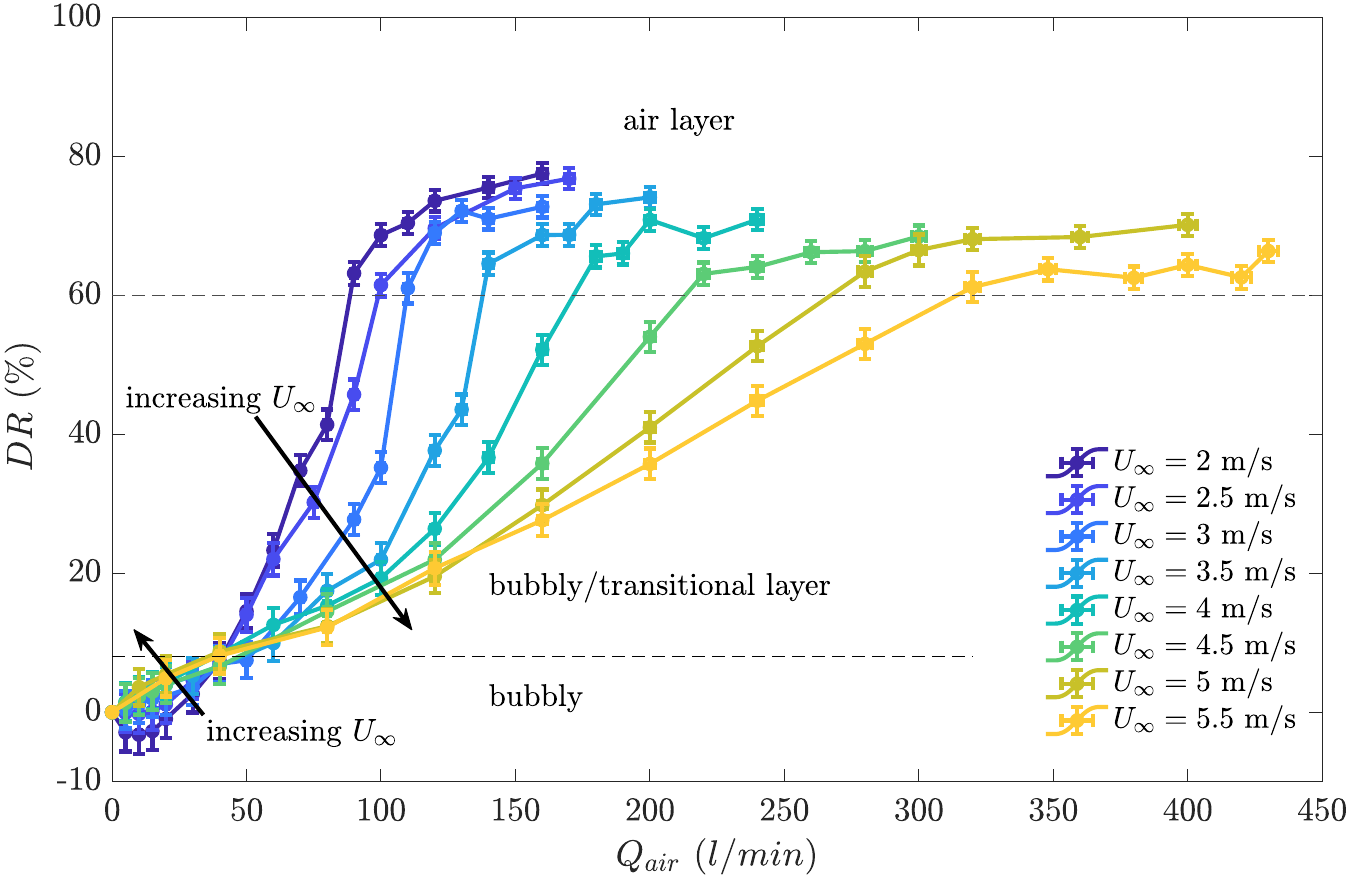}
\caption{Drag reduction curves, for different $U_{\infty}$. The dashed lines demarcate the regime transitions.}
\label{DRcurvesall}
\end{figure}

The effect of $U_{\infty}$ on drag reduction is systematically examined across the different air phase regimes. Figure~\ref{DRcurvesall} presents drag reduction as defined in equation \ref{DR}, plotted against air flow rate $Q_{air}$ for varying $U_{\infty}$. For clarity, repeated measurements are omitted, although good repeatability is achieved (see appendix \ref{more_dr}). Across all $U_{\infty}$, the general trend is consistent: a bubbly, a transitional, and an air layer regime emerge for increasing $Q_{air}$ with similar morphological features. 

In the current experiments, the effect of $U_{\infty}$ is different for low ($Q_{air}<40$ l/min) and higher air flow rates: for the former, increasing $U_{\infty}$ results in a monotonic increase of the drag reduction (when keeping $Q_{air}$ constant), while the opposite is true for higher air flow rates ($Q_{air}>40$ l/min). This enables a distinction between the bubbly and transitional regimes, which is not possible when analyzing a single $U_{\infty}$ curve. Further investigation of the physical meaning of this ``pivot point'' (here at $DR=8\%$), potentially via non-dimensionalizing $Q_{air}$ in Figure \ref{DRcurvesall} could provide a more formal distinction between the regimes. The transition from TALR to ALR on the other hand is defined as the first occurrence of $DR = 60\%$, coinciding with a clear reduction in the $DR-Q_{air}$ slope and supported by flow visualization (for all $U_{\infty}$). In \cite{elbing2008bubble}, a $DR$ 20\% and a 80\% classification was made for the bubbly-transitional and transitional-air layer regime limits. However, these limits and the drag reduction levels in general are not directly comparable to those of present study due to differences in the measurement techniques---specifically, the use of a \textit{local} versus a \textit{global} drag measurement. Better drag reduction agreement is expected in the bubbly and transitional regimes, where the air distribution is more homogeneous over the width and length of the domain. In contrast, in the air layer regime, strong spatial non-uniformity - e.g. wetted patches near the injector and an intact layer downstream - likely lead to lower \textit{global} DR estimates.

For all $U_{\infty}$, after transitioning to ALR, further increasing $Q_{air}$ has only a small effect on DR. For example, for $U_{\infty}=5$ m/s at $Q_{air}=Q_{crit}$, $DR$ is 63\%. To achieve a 11\% increase in $DR$, the air flow rate should be increased by $43\%$. This is especially relevant for real ship applications, where a cost-benefit analysis is used to determine the working conditions of the air lubrication system. The intermediate region ($40<Q<Q_{crit}$) is characterized by drag reduction lines of various slopes. For increasing $U_{\infty}$, the slope of the curves becomes lower. That means that for higher $U_{\infty}$, relatively more injected air is needed for an increase in $DR$.

To further explore these different effects of increasing $U_{\infty}$ across the three regimes, in the next sections the air phase characteristics are quantified using the available imaging data and the results are discussed in tandem with the corresponding drag reduction behavior.  

\subsubsection{Characteristics of the bubbly regime}
\label{bubbly}

In this section, we will focus on the early stages of the bubbly regime ($Q_{air}<40$ l/min in Figure~\ref{DRcurvesall}). For these low air flow rates, the multiphase flow exhibits a different behavior than in the rest of the regimes: for constant $Q_{air}$, increasing $U_{\infty}$ leads to a decrease in drag, while for all other air flow rates the opposite is true. This points to a potentially different drag reduction (or augmentation) mechanism than that of the transitional and air layer regimes.

\begin{figure}[!t]
  \centering
  \begin{minipage}[b]{0.6\textwidth}
    \centering
    \begin{subfigure}{\textwidth}
      \centering
      \includegraphics[width=\linewidth]{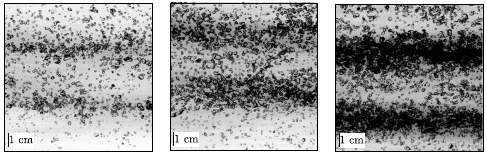}
      \caption{}
      \label{bubblyINSTA5}
    \end{subfigure}

    \vspace{0.5em}

    \begin{subfigure}{\textwidth}
      \centering
      \includegraphics[width=\linewidth]{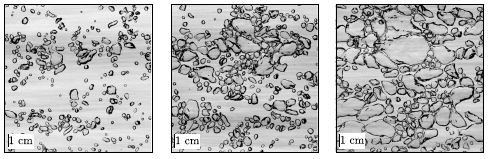}
      \caption{}
      \label{bubblyINSTA2}
    \end{subfigure}
  \end{minipage}
  \vspace{0.1em}
  \raisebox{1.7\height}{
  \begin{minipage}[c]{0.35\textwidth}
    \centering
    \begin{subfigure}{\textwidth}
      \centering
      \includegraphics[width=\linewidth]{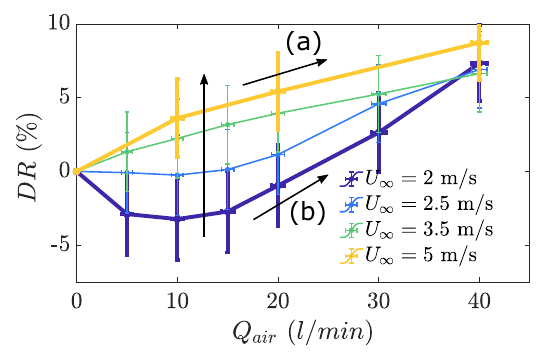}
      \caption{}
      \label{Dr_all_small_general}
    \end{subfigure}
  \end{minipage}
}
  \caption{Instantaneous images of the bubbly regime in the $x-z$ plane for (a) $U_{\infty}=5$ m/s and (b) $U_{\infty}=2$ m/s. $Q_{air}$ increases from left to right (10 l/min, 20 l/min and 40 l/min). (c) The corresponding drag reduction curves. The black vertical arrow indicates that drag reduction increases with $U_{\infty}$ for low air injection rates. Flow is from right to left.}
  \label{Dr_all_small}
\end{figure}

To explore this, we zoom in the region of $Q_{air}<40$ l/min in the DR profiles and assess it together with corresponding instantaneous bubbly images in the $x-y$ and $x-z$ planes (Figure~\ref{Dr_all_small}). For the highest liquid velocity ($U_{\infty}=5$ m/s), increasing $Q_{air}$ has a negligible effect on bubble size (Figure \ref{bubblyINSTA5}), while a monotonic increase of drag reduction is measured (Figure  \ref{Dr_all_small_general}). The most notable effect is an apparent increase in the {\em{number}} of bubbles, with a higher concentration along two streaks at the centerline of the plate. Such inhomogeneities can be often traced back to upstream disturbances and small geometry variations yet none could be identified definitively in our facility. For the lowest velocity ($U_{\infty}=2~\mathrm{m/s}$), increasing $Q_{air}$ results in qualitatively larger bubbles (Figure~\ref{bubblyINSTA2}) but the corresponding $DR$ here does not increase monotonically: it initially decreases from zero to negative values (denoting a drag increase) and reaches a local minimum, before increasing again (Figure \ref{Dr_all_small_general}). A similar (albeit slightly lower) drag increase was also observed for $U_{\infty}=2.5~\mathrm{m/s}$. Further tests with lower $U_{\infty}$, revealed that the local minimum in $DR$ increases in magnitude with decreasing freestream velocity (all the way down to approximately 1.2 m/s, see chapter \ref{subcrtitical_results} for more details). Earlier studies on the topic have associated the drag reducing abilities of bubbles to both their size and deformability \citep{verschoof2016bubble}, as well as the general topology (closely packed vs dispersed, \cite{biswas2024effects}). In this study we intentionally focused on acquiring synchronous imaging of both the wall-projected bubble area (Figure \ref{Dr_all_small}) and of the vertical bubble organization (Figure \ref{vertical_pos}) which we will leverage in what follows in order to try and interpret these drag changes.

\begin{figure}
\begin{center}
    \begin{subfigure}[t]{\textwidth}
        \centering
    \includegraphics[width=0.45\textwidth]{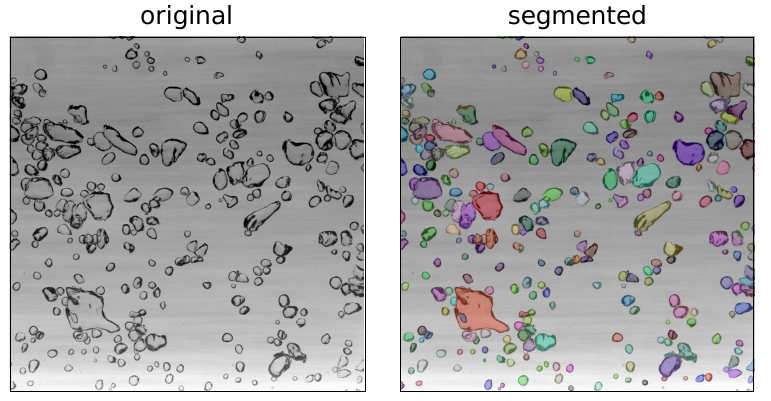}
        \hspace{0.01cm}
\includegraphics[width=0.45\textwidth]{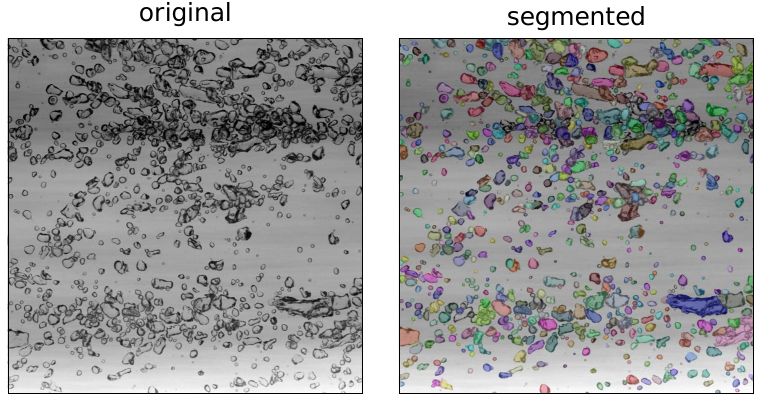}      
  \caption{}
  \label{bubble_seg}
    \end{subfigure}
        \vspace{0.01cm}
\begin{subfigure}[t]{0.45\textwidth}
  \centering
    \includegraphics[width=\textwidth]{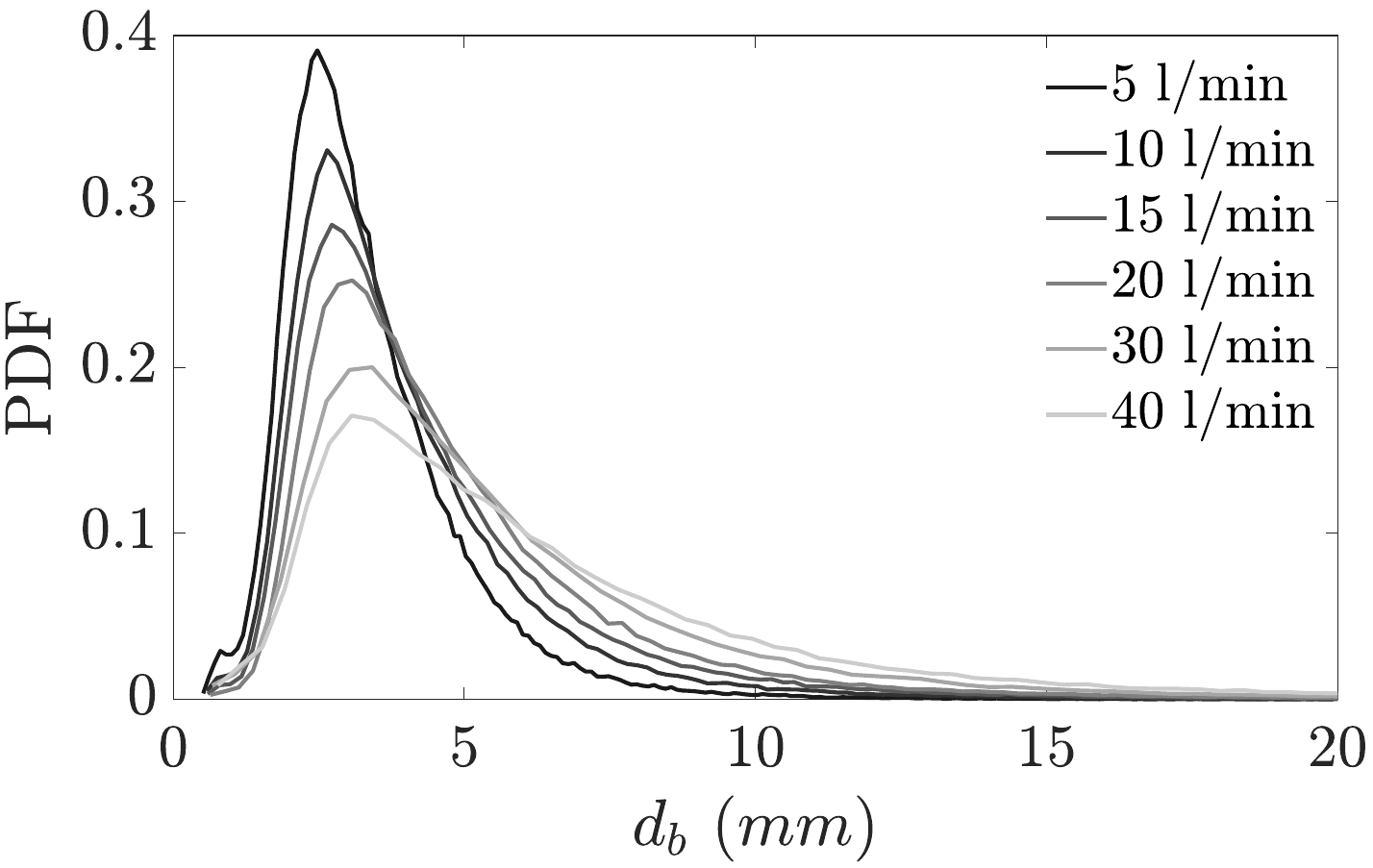}
  \caption{}
  \label{2mps_PDF}
  \end{subfigure}%
  \hspace{0.01cm}
\begin{subfigure}[t]{0.45\textwidth}
  \centering
\includegraphics[width=\textwidth]{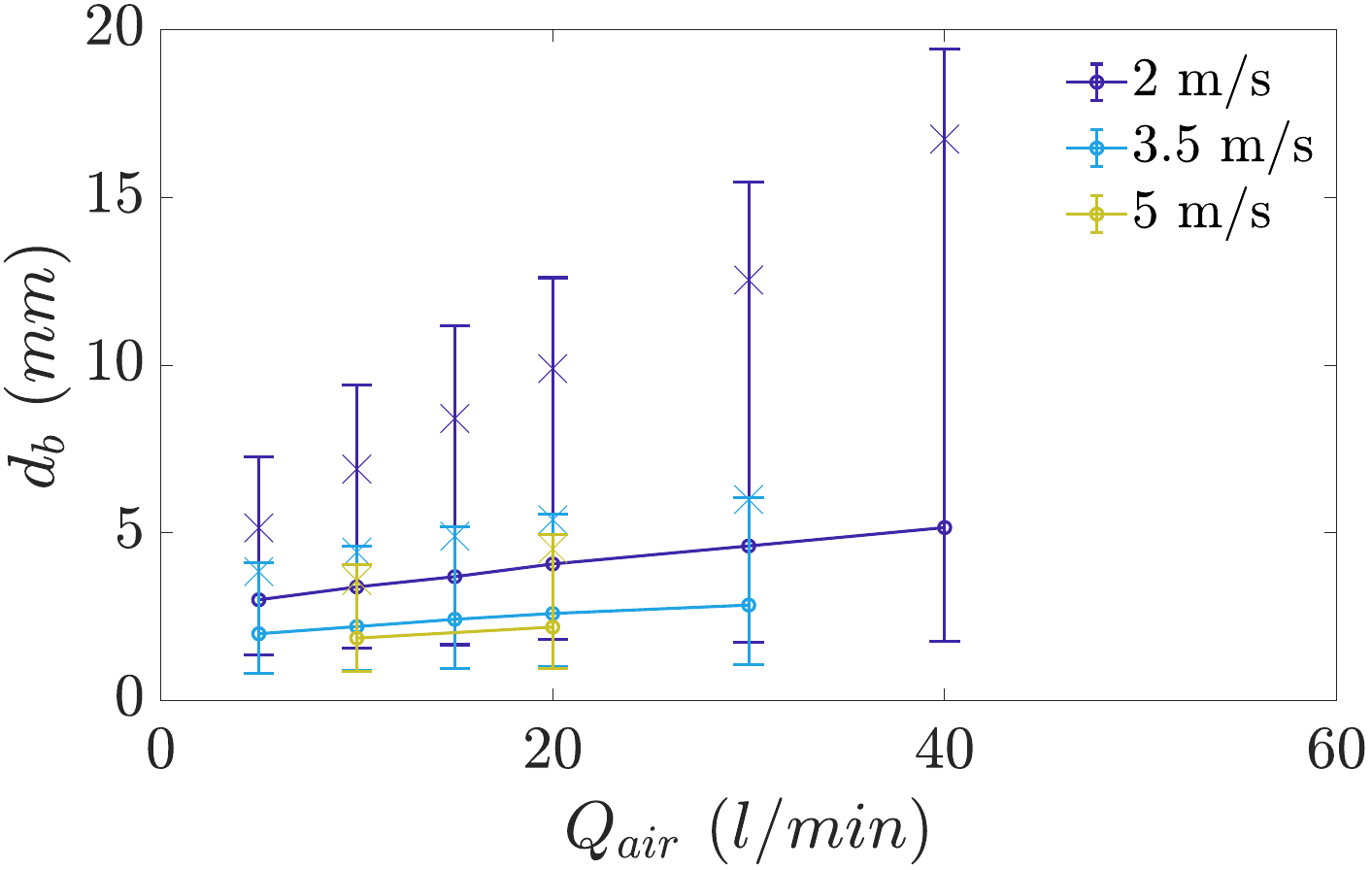}
  \caption{}
  \label{qair_db}
  \end{subfigure}
  \caption{Bubble size statistics. (a) original and segmented images for $U_{\infty}=2$ m/s (left) and $U_{\infty}=3.5$ m/s (right). (b) Probability distribution of the bubble diameter (\( d_b \)) for air flow rates ranging from \( Q_{\text{air}} = 5 \) to \( 40 \) l/min and \( U_{\infty}=2 \) m/s. Vertical dashed line indicates the thickness of the bubbly layer \(y_b\). (c) Bubble size range for three \( U_{\infty} \) and various air flow rates. The error bars represent the 2.5th and  97.5th percentile at the lower and upper end respectively. The area-weighted mean bubble size is indicated by ``x'', while the median is denoted by ``o''.}
    \label{bubble_stats}
\end{center}
\end{figure}

In order to identify any other links between bubble topology and $DR$, we first quantify the bubble sizes from our imaging data. Towards this end, we detect the bubble boundaries in the $x-z$ plane using the AI-based image segmentation introduced in section \ref{NWA}. An example of the bubble detection can be seen in Figure~\ref{bubble_seg}. For every $U_{\infty}-Q_{air}$ pair, 600 images are processed (corresponding to 150-300 bubbles per image for the lowest velocity and more than $1000$ for the highest one). The equivalent bubble diameter $d_b$ is then calculated from the 2D bubble area as $d_b=2\sqrt{A_b/ \pi{}}$. In Figure~\ref{2mps_PDF}, the probability density function of all bubble diameters is shown for $U_{\infty}=2$ m/s. The bubbles range in size from a few $\mathrm{mm}$ to a maximum of 2 $\mathrm{cm}$, with a wider $d_b$ distribution and larger values for increasing $Q_{air}$; for all cases the distribution exhibits a lognormal behavior, as has been observed for bubbles and droplets in turbulence in a large range of configurations due to the underlying physics of coalescence and breakup (\citealp[see][among many others]{Mouza2005,Jacob2010}). For this $U_{\infty}$, the mode is mostly insensitive to $Q_{air}$ ($\approx3-4$ mm) and approximately equal to the injector slot width ($t=4$ mm). For higher $U_{\infty}$  (Figure \ref{qair_db}), the range and mean values of $d_b$ both decrease, indicating that bubble sizes become more uniform and thus their distributions narrower (see also Figure~\ref{bubble_stats_5_35} in Appendix \ref{ML}); yet, this velocity effect becomes less noticeable for $U_{\infty}>3.5$ m/s. Finally, from a quantification perspective, it should also be noted that, for $U_{\infty}=5$ m/s (for which the single phase $u_{\tau}$ was accurately estimated, see table \ref{tab:top_BL_thickness}), the extracted $d_b$ range ($1.86-2.19~\mathrm{mm}$) agrees remarkably well with the maximum bubble diameter predicted by the Kolmogorov-Hinze theory, as adapted for bubbles within the high shear region of wall-bounded flows by \cite{sanders2006bubble}: $d_{max}=(\sigma/2\rho)^{3/5}(\kappa y/u_{\tau}^3)^{2/5}$. Here, $y=y_b$ is the outermost bubble wall-normal location (defined in the next paragraphs) and results in a $d_{max}$ between $1.9-2.2~\mathrm{mm}$. 
 
 Connecting these quantitative results for $d_b$ with our drag results discussed above and with earlier studies is far from straightforward. Increasing the freestream velocity (for a constant $Q_{air}$) results in decreasing bubble size (Figure \ref{bubblyINSTA5} \& \ref{bubblyINSTA2}, \ref{qair_db}) and also drag reduction (vertical line in  Figure~\ref{Dr_all_small_general}). This suggests that drag reduction increases with decreasing $d_b$. This is in line with flat plate results from \cite{savviobubble}, who observed a drag increase in bubbly regimes for the lowest velocity they tested ($U_{\infty}=2\mathrm{m/s}$), a trend that was reversed in higher velocities. Yet, the opposite effect was reported by \cite{biswas2024effects} in a lower Reynolds number experiment, where smaller bubbles were shown to increase drag. This was attributed to smaller bubble deformability, a notion introduced and supported by various other earlier studies \citep{lu2005,verschoof2016bubble,van2013importance}.  Following the friction velocity Weber number definition by \cite{lu2005}, here we find $0.5<We<15$, where $We=\rho u_{\tau}^2d_b/\sigma$. This indicates that for all our cases, bubbles can be considered deformable. Those present at the lowest velocities are the most deformable ($We=15$) and yet they lead to the highest measured drag, indicating that deformability notions are not sufficient here to predict drag performance. For bubbles of size similar to ours (larger than coherent structures, but smaller than the TBL thickness, $10\times l^{+}<d_{b}<\delta$), a different drag reduction mechanism was proposed by \cite{murai2014frictional}: bubble motion. Due to the various forces acting on the bubbles in such regimes, bubbles move in a cyclic manner similar to ejections and sweeps \citep{murai2006turbulent}, effectively reducing shear stress. While in our study we did not measure bubble motion, as mentioned before, only our $DR$ results for $U_{\infty}\geq3\mathrm{m/s}$ (drag decrease) are in line with the regime map from \cite{murai2014frictional}; for lower velocities we measure drag increase instead. What becomes clear from the above is that information from $x-z$ planes (either in terms of total non-wetted area as was shown in section \ref{NWA}, or in terms of $d_b$ and deformability as was shown here), even when it is quantitative, is not sufficient to explain the drag reduction effect of the bubbles.

\begin{figure}
    \centering
    \includegraphics[width=0.8\textwidth]{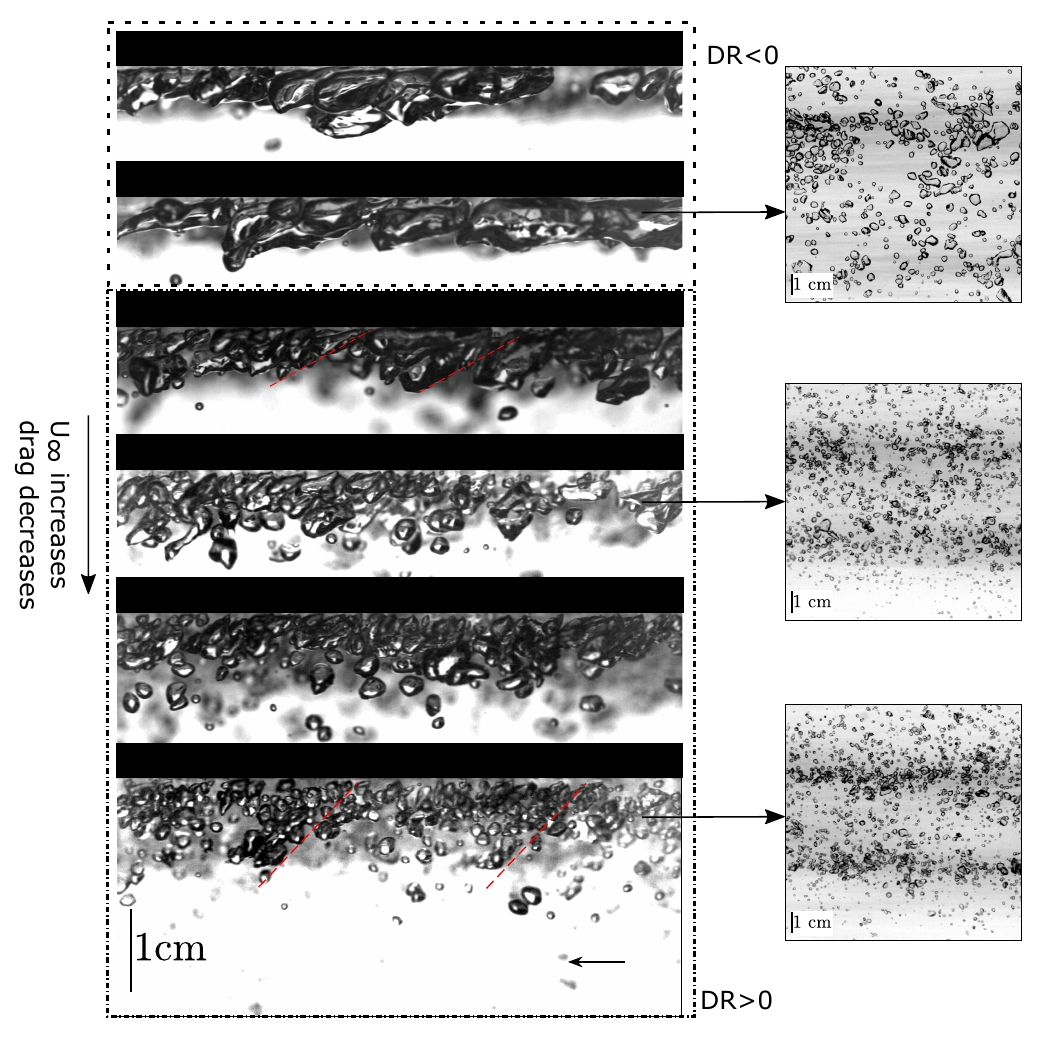}
    \caption{Side view images of the bubbly regime ($Q_{air}=10$ l/min). $U_{\infty}$ increases from top to bottom from 2 to 5 m/s. The drag decreases from top to bottom respectively.}
    \label{vertical_pos}
\end{figure}

However, aside from wall-parallel bubble imaging, we also have synchronous imaging in a wall-normal plane ($x-y$), which allows us to additionally analyze the vertical organization of the air phase and further probe its connection to drag characteristics. This is something that, to the best of our knowledge, has not been explored before in the current context. Figure~\ref{vertical_pos} showcases side view snapshots from this vertical plane for $Q_{air}=10$ l/min and increasing $U_{\infty}$ (from top to bottom). In the case of the two lower velocities, bubbles are present in a single layer below the wall, and are severely flattened due to shear/buoyancy, while for $U_{\infty}\geq3\mathrm{m/s}$, smaller, more spherical bubbles are dispersed over multiple wall-normal locations. It is also readily observable that, while the individual bubble size $d_b$ decreases for increasing $U_{\infty}$ (Figure \ref{qair_db}), the overall bubbly layer thickness, $y_b$ increases. More specifically, for the lowest velocity, $y_b\approx6\mathrm{mm}$, showing a small increase as we increase $Q_{air}$. For $U_{\infty}=3.5 \mathrm{m/s}$, $y_b\approx6.5\mathrm{mm}$, again showing a slight increase with $Q_{air}$. Lastly, for $U_{\infty}=5 \mathrm{m/s}$, a clear increase in the bubbly thickness is measured: $9\leq y_b\leq13\mathrm{mm}$. It must be noted that in the case of the lowest $U_{\infty}$, $y_b$ is representative of the height of a single bubble, but for higher ones where dispersion is at play, $y_b$ corresponds to multiple bubbles. In all cases, $y_b^{+}$ is much larger than 300 and up to $0.25\delta$ indicating that bubbles reach the outer layer of the TBL. For both single and multi-bubble layers, it is also interesting to note the clear shear induced slope (see dashed red lines in Figure \ref{vertical_pos} – flow is from right to left), reflecting the well-established downstream angle of coherent structures in wall-bounded turbulence. 

Returning to the topological disparity of the bubble layer for different $U_{\infty}$, it allows us to draw a more straightforward connection with the measured drag.  Large bubbles present for $U_{\infty}< 3$ m/s, sliding over the flat plate in single-bubble layers increase the drag. A possible mechanism could be that this single bubble layer acts similar to roughness and thus initially increase drag. It can be assumed that, as $Q_{air}$ increases further, the increase in non-wetted area eventually becomes sufficient for drag reduction, especially when a transitional regime of large air patches is reached. The latter happens for a much lower $Q_{air}$ compared to higher freestream velocities (see also \cite{nikolaidou2024effect}), and this is why DR for $U_{\infty}< 3$ m/s catches up ($Q_{air}=40$ l/min, Figure \ref{Dr_all_small_general}) and eventually outperforms the one for $U_{\infty}> 3$ m/s ($Q_{air}>40$ l/min, Figure \ref{DRcurvesall} and see also next section). On the other hand, for $U_{\infty}> 3$ m/s, much smaller bubbles are present in the TBL and they are vertically stacked, creating a multi-bubble layer, whose thickness increases with $U_{\infty}$. Their drag reduction performance is initially far superior to the single bubble layer, providing drag decrease for even the smallest $Q_{air}$ tested here. As $Q_{air}$ increases further, although the DR keeps increasing, it is at a much slower pace than for lower $U_{\infty}$ (see Figure \ref{DRcurvesall}), reflecting the much higher needs for $Q_{air}$ to reach the transitional and air layer regimes (see sections \ref{transReg} and \ref{regimeMAPsec}).  

From an application perspective, it must be noted that the single layer bubble, present at low $U_{\infty}$, will not be relevant for full-scale ship applications, since ship service speeds far exceed $3\mathrm{m/s}$, so smaller bubbles dispersed in the vertical would be expected in most conditions. 

\subsubsection{Characteristics of the transitional and air layer regime}
\label{transReg}

Following the discussion for the bubbly regime in the previous section ($Q_{air}<40$ l/min and $DR<8$\%), we now focus on higher air flow rates. As $Q_{air}$ increases, a transitional regime is reached first, in which bubbles have coalesced to larger air patches (see \cite{nikolaidou2024effect} and Figure \ref{charTRANS}) and DR increases monotonically with $Q_{air}$ (see also Figure \ref{DRcurvesall}). Subsequently, as $Q_{air}$ increases further, DR starts to plateau around DR=60\% where the ALR is achieved. With respect to drag reduction for these cases ($Q_{air}>40$ l/min), the trend with increasing $Q_{air}$ is overall very similar for all velocities tested here and follows the monotonic increase and plateauing discussed above (Figure \ref{DRcurvesall}). There are however two main differences observed, as freestream velocity increases from 2 to 5 m/s. First, a decreasing slope for the DR curves (for $8\%<DR<60\%$ in Figure \ref{DRcurvesall}), and second, a progressive reduction in the rate of DR increase to almost zero (constant DR), for $DR>60\%$. Since these two refer to the transitional and ALR regimes, respectively, we will discuss each of them in succession, in conjunction with the corresponding imaging data and image processing results.
\newline
\newline
\textbf{Bubbly/Transitional regime ($8\%\leq DR\leq60\%$)}

\begin{figure*}[!ht]
\begin{center}
  \centering
\includegraphics[width=0.8\textwidth]{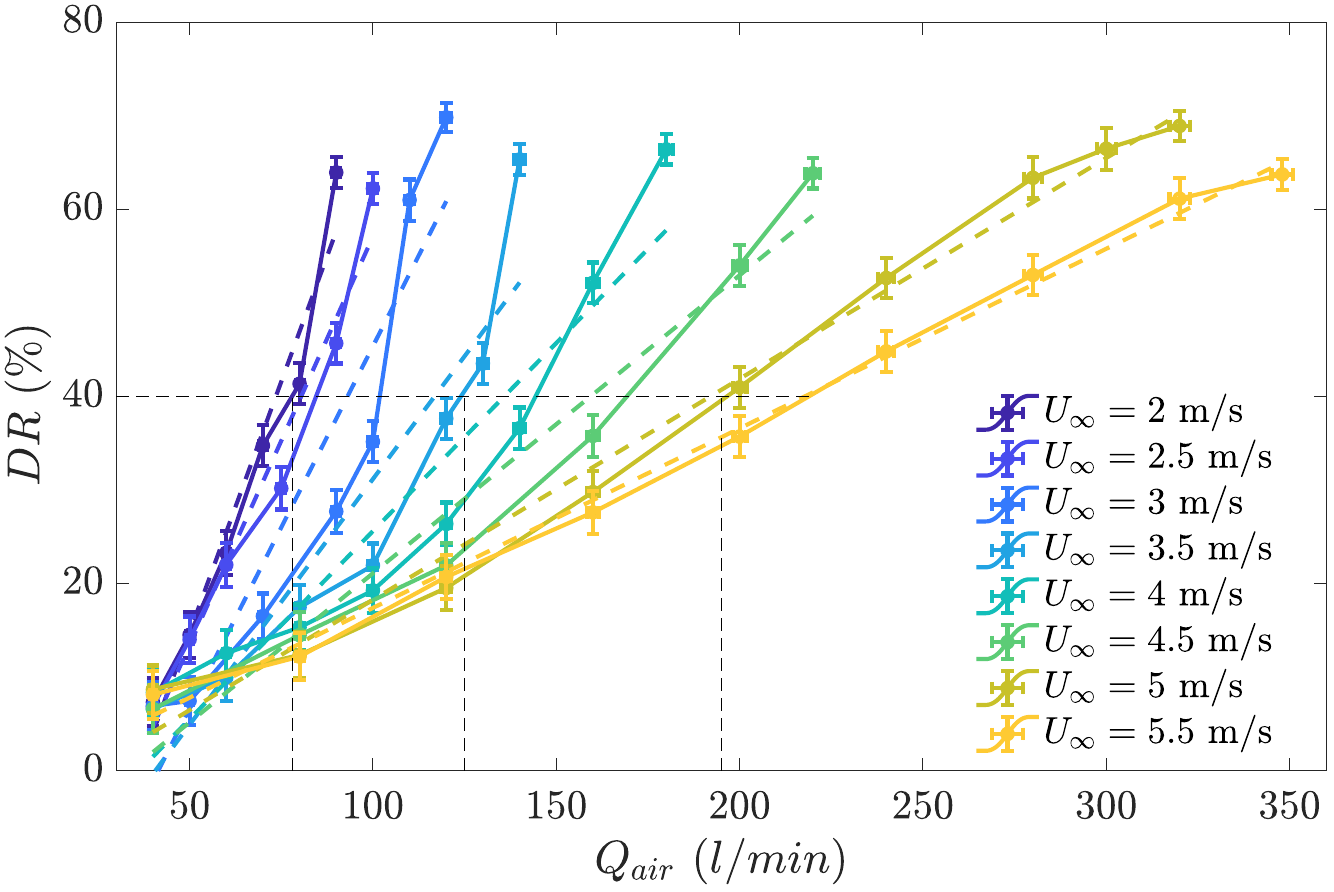}
  \caption{Slopes of the drag reduction curves within the transitional regime. Horizontal dashed line corresponds to $DR=40\%$.}
    \label{slopedecrease}
\end{center}
\end{figure*}

\begin{figure}[htbp]
    \centering
    \begin{subfigure}[b]{0.3\textwidth}
        \centering
        \includegraphics[width=\textwidth]{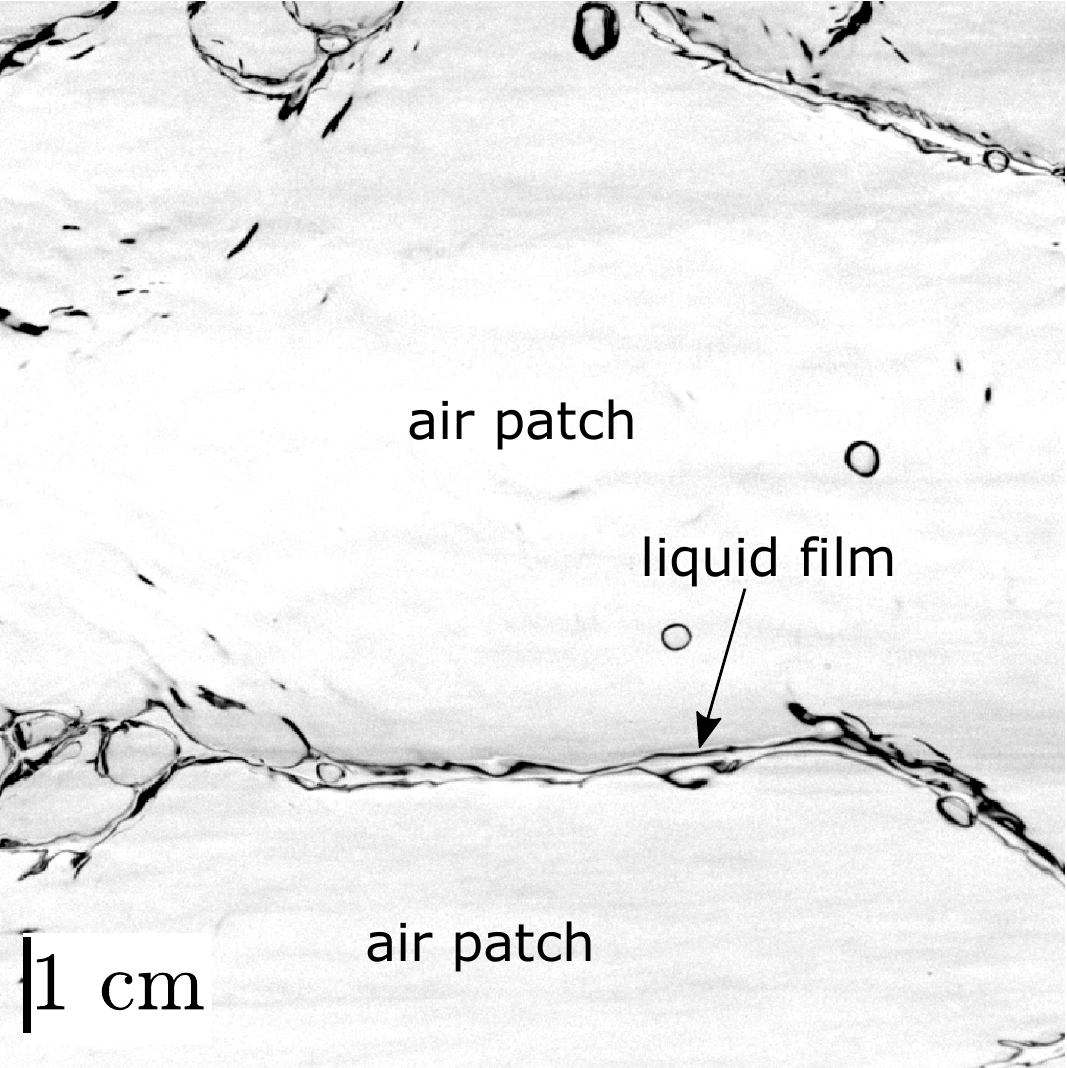} \\[1ex]
         \includegraphics[width=\textwidth]{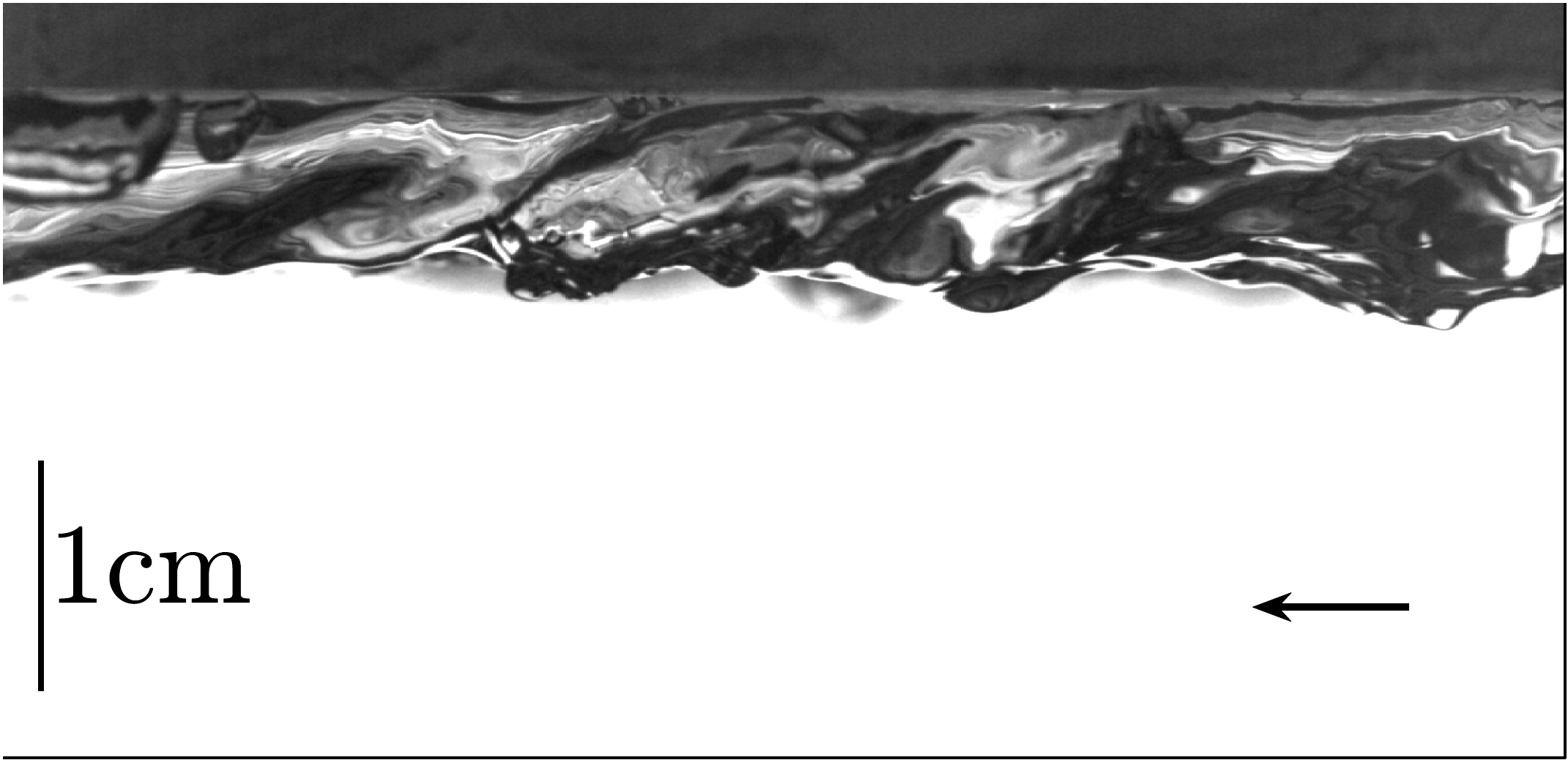}
         \caption{$U_{\infty}=2$ m/s \& $Q_{air}=80$ l/min}
         \label{U2Q80}
    \end{subfigure}%
    \hspace{0.4cm}
    \begin{subfigure}[b]{0.3\textwidth}
        \centering
    \includegraphics[width=\textwidth]{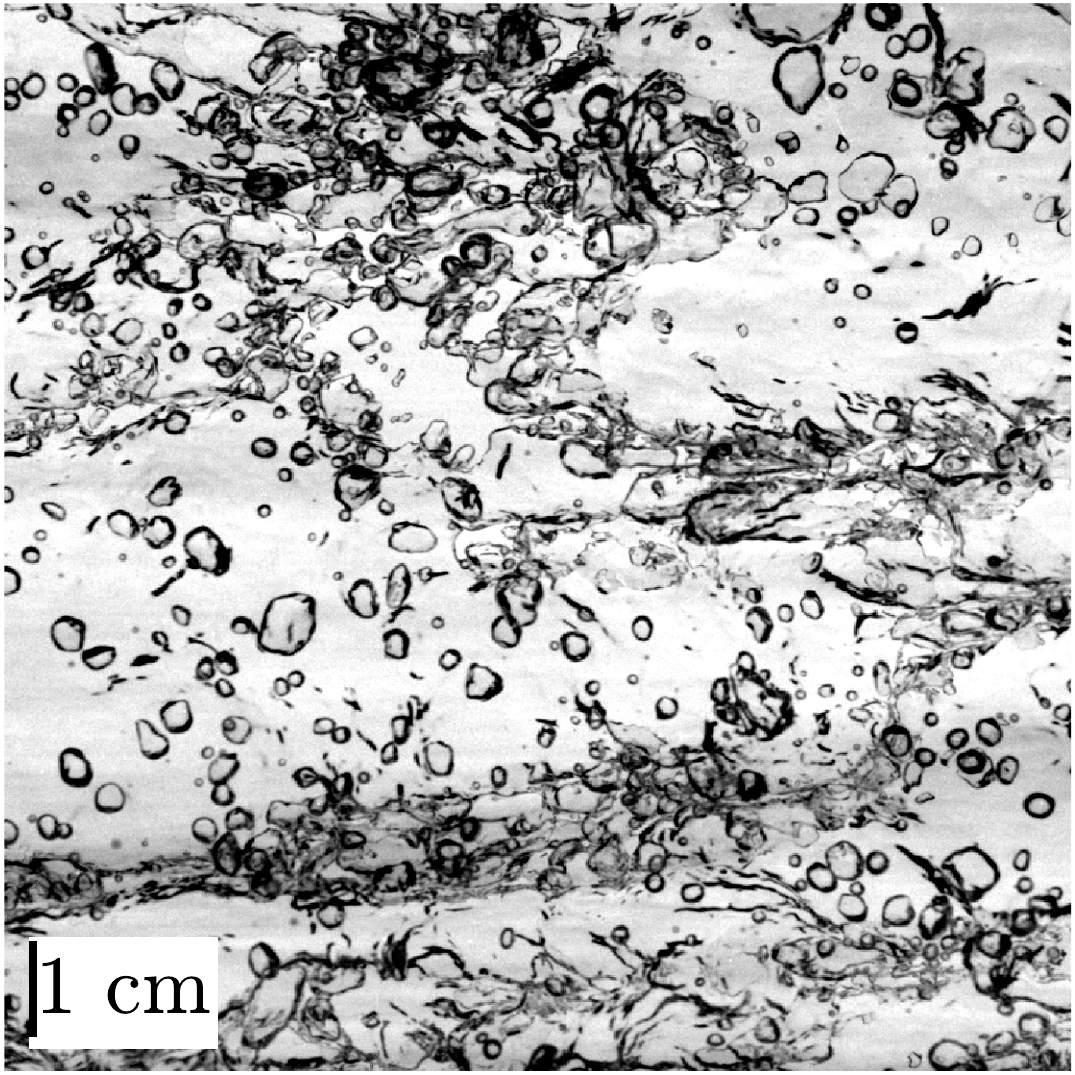} \\[1ex]
     \includegraphics[width=\textwidth]{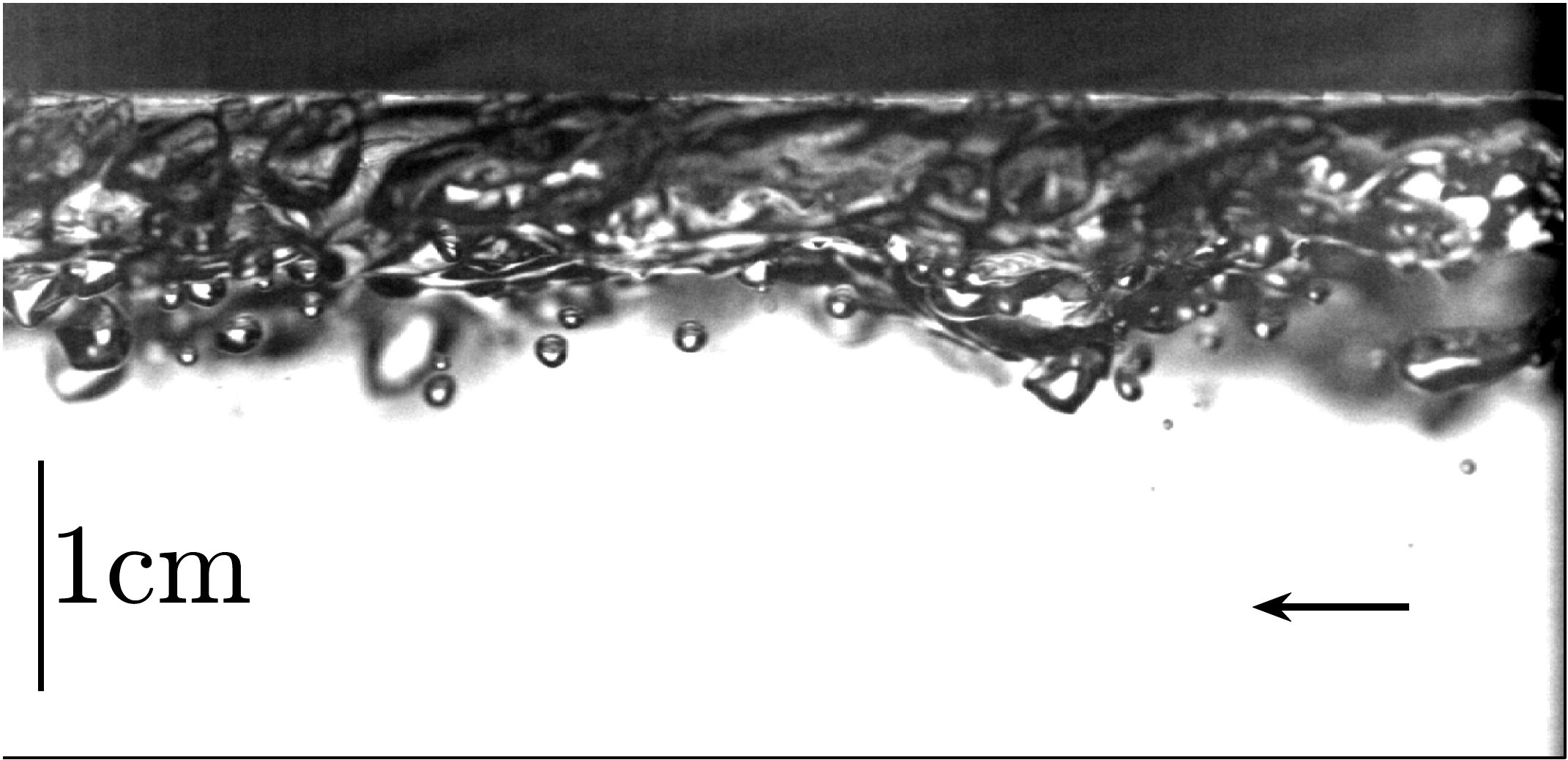}
        \caption{$U_{\infty}=3.5$ m/s \& $Q_{air}=130$ l/min}
        \label{U35Q130}
    \end{subfigure}%
        \hspace{0.4cm}
    \begin{subfigure}[b]{0.3\textwidth}
        \centering
    \includegraphics[width=\textwidth]{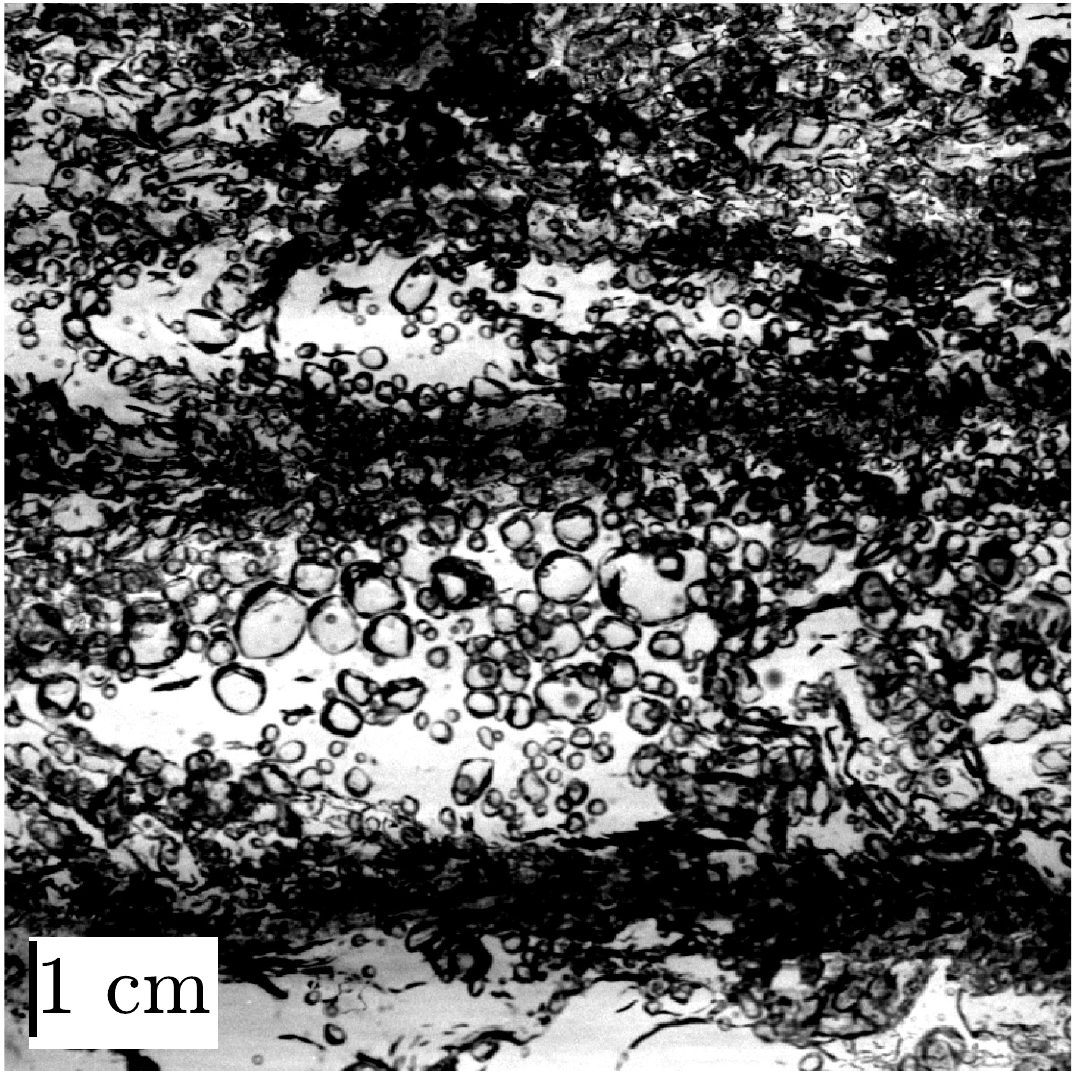} \\[1ex]
     \includegraphics[width=\textwidth]{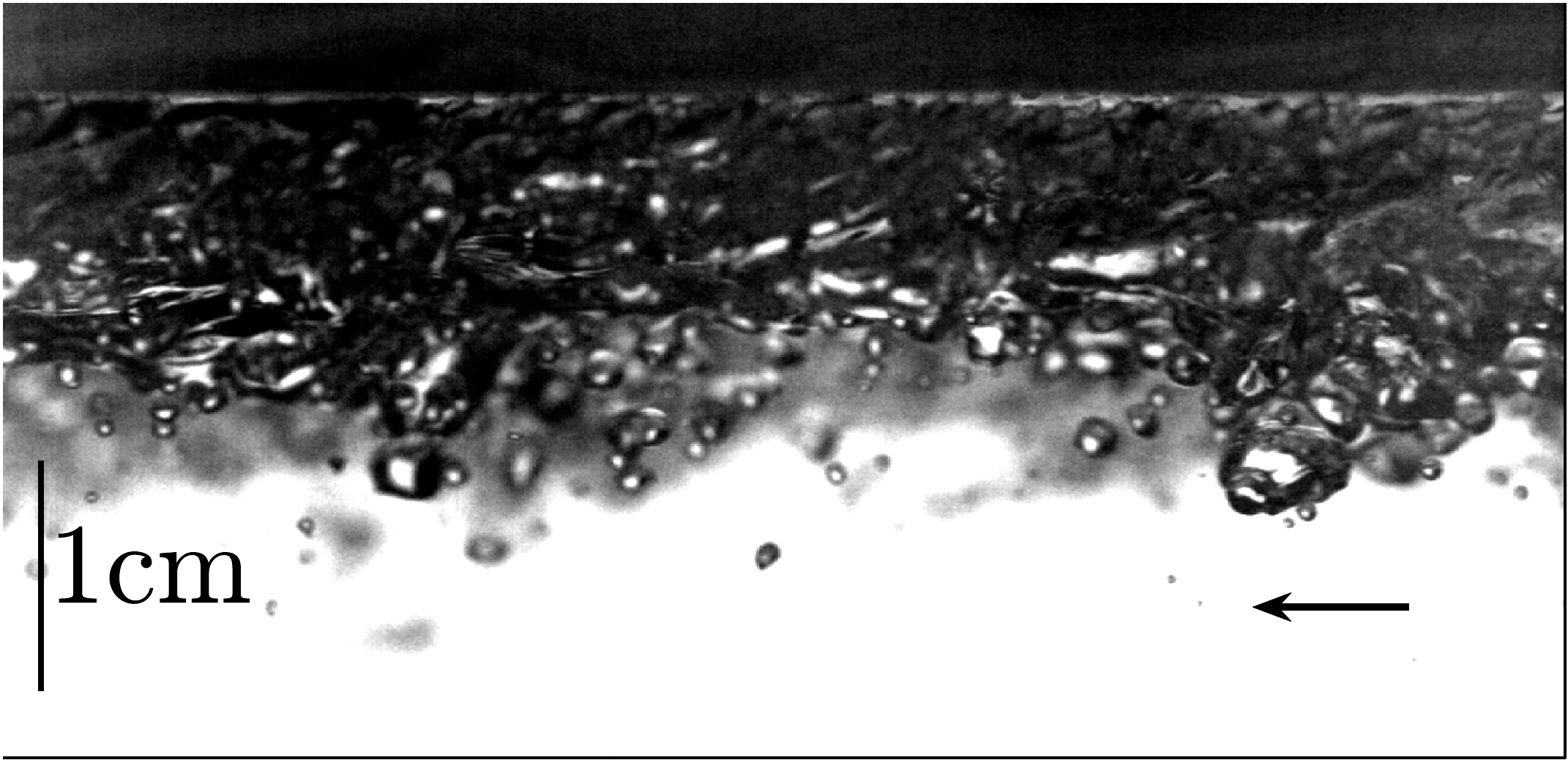}
    \caption{$U_{\infty}=5$ m/s \& $Q_{air}=200$ l/min}
    \label{U5Q200}
    \end{subfigure}
    \caption{Characteristic transitional regime images ($x-z$ \& $x-y$ plane) corresponding to  $DR\approx40\%$ for three representative $U_{\infty}$. Flow is from right to left in all panels.}
    \label{transREGIME}
\end{figure}

Despite the similar, monotonic increase in DR for all $U_{\infty}$, there is a distinct decrease in slope, indicating a lower rate of DR increase with $Q_{air}$ for higher velocities. For instance, $Q_{air}=90$ l/min is needed for $U_{\infty}=2$ m/s to reach a $DR=40\%$, while for $U_{\infty}=3.5$ m/s and $U_{\infty}=5$ m/s these become 130 l/min and 200 l/min, respectively (Figure \ref{slopedecrease}). As mentioned above, an increase in $Q_{air}$ also leads to a coalescence of bubbles and the appearance of larger air patches, which then contribute to DR by an increase in non-wetted area (Figure \ref{ANWvsDR}). From the above (and also our observations in \cite{nikolaidou2024effect}) it is then to be expected, that such a transition will happen at a slower rate as $U_{\infty}$ increases. From the available imaging data, both from the wall-parallel as well as the wall-normal imaging planes, this can indeed be seen (Figure \ref{transREGIME}). For the same DR percentage (40\%), a transitional topology is particularly clear for the lowest velocity (see Figure \ref{U2Q80}), where air patches, separated by liquid films, but forming a single wall-normal layer, are seen very close to the wall. The situation shifts as $U_{\infty}$ increases to 3.5 m/s, with the outer boundary of the air phase becoming more deformed, and bubbles appearing around it (Figure \ref{U35Q130}). For the highest velocity, only a dense layer of bubbles is present (5 m/s, Figure \ref{U5Q200}), indicating that a transitional regime might not have been reached despite the very high $Q_{air}$ involved. Distinguishing between a bubbly and a transitional regime therefore becomes very challenging. Thus, when also taking into account the topology from Figure \ref{vertical_pos}, it becomes clear that for the low velocities, the initially drag increasing bubbles flatten due to shear, coalesce and lead to a fast rate of DR for $Q_{air}>40$ l/min, while for the very high velocities, the (slower) monotonic increase in DR is linked instead to an increasingly denser packing and minimal coalescence of much smaller bubbles. 
\\
\begin{figure}
\centering
\includegraphics[width= 1\linewidth]{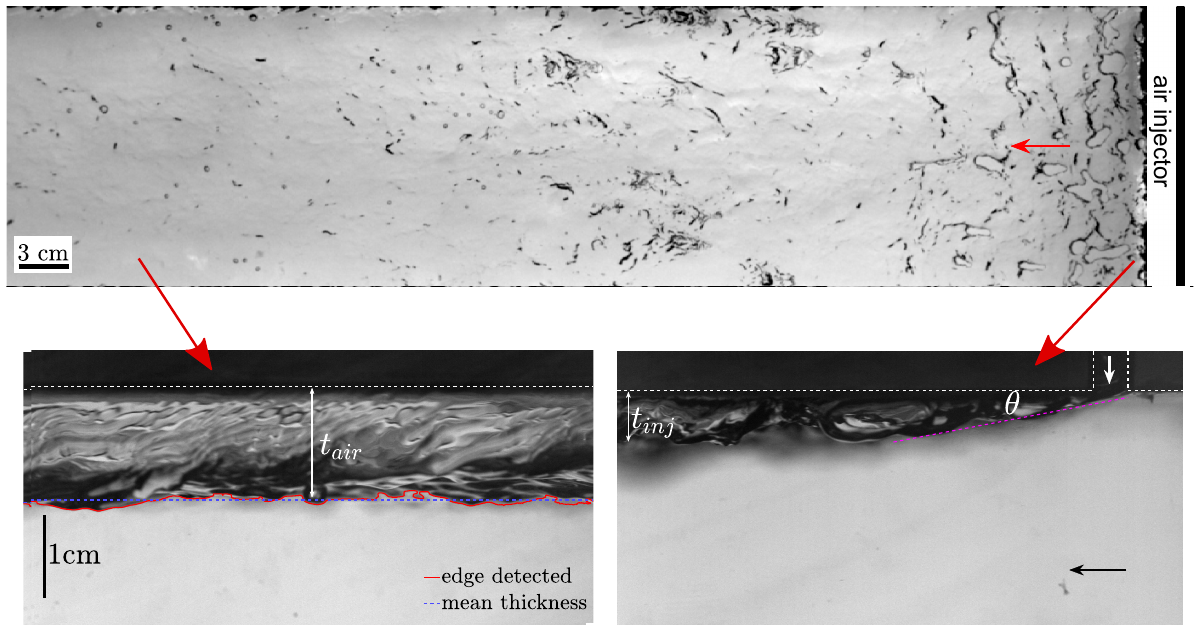}
\caption{Instantaneous image of the air layer regime for $U_{\infty}=2$ m/s. The mean air layer thickness $t_{air}$ is measured in the middle of the test section (84 cm downstream of the injection location) and initial air layer thickness $t_{inj}$ is measured close to the injector (6.2 cm downstream).}
\label{closeToInjPlot}
\end{figure}

\begin{figure*}[!ht]
\begin{center}
\begin{subfigure}[t]{0.45\textwidth}
  \centering
\includegraphics[width=\textwidth]{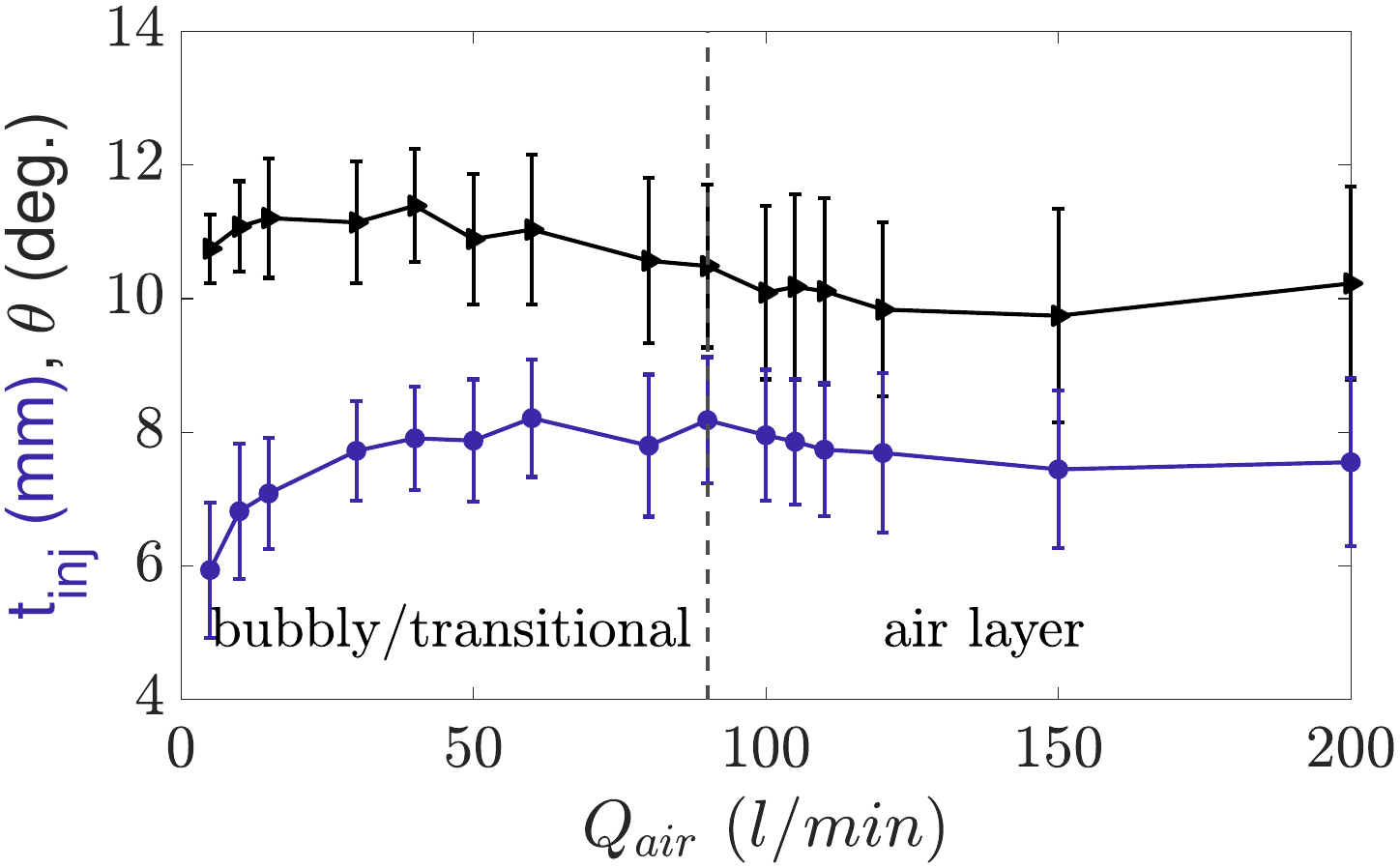}
  \caption{}
  \label{tinjthetaQ}
  \end{subfigure}%
  \hspace{0.01cm}
\begin{subfigure}[t]{0.45\textwidth}
  \centering
 \includegraphics[width=\textwidth]{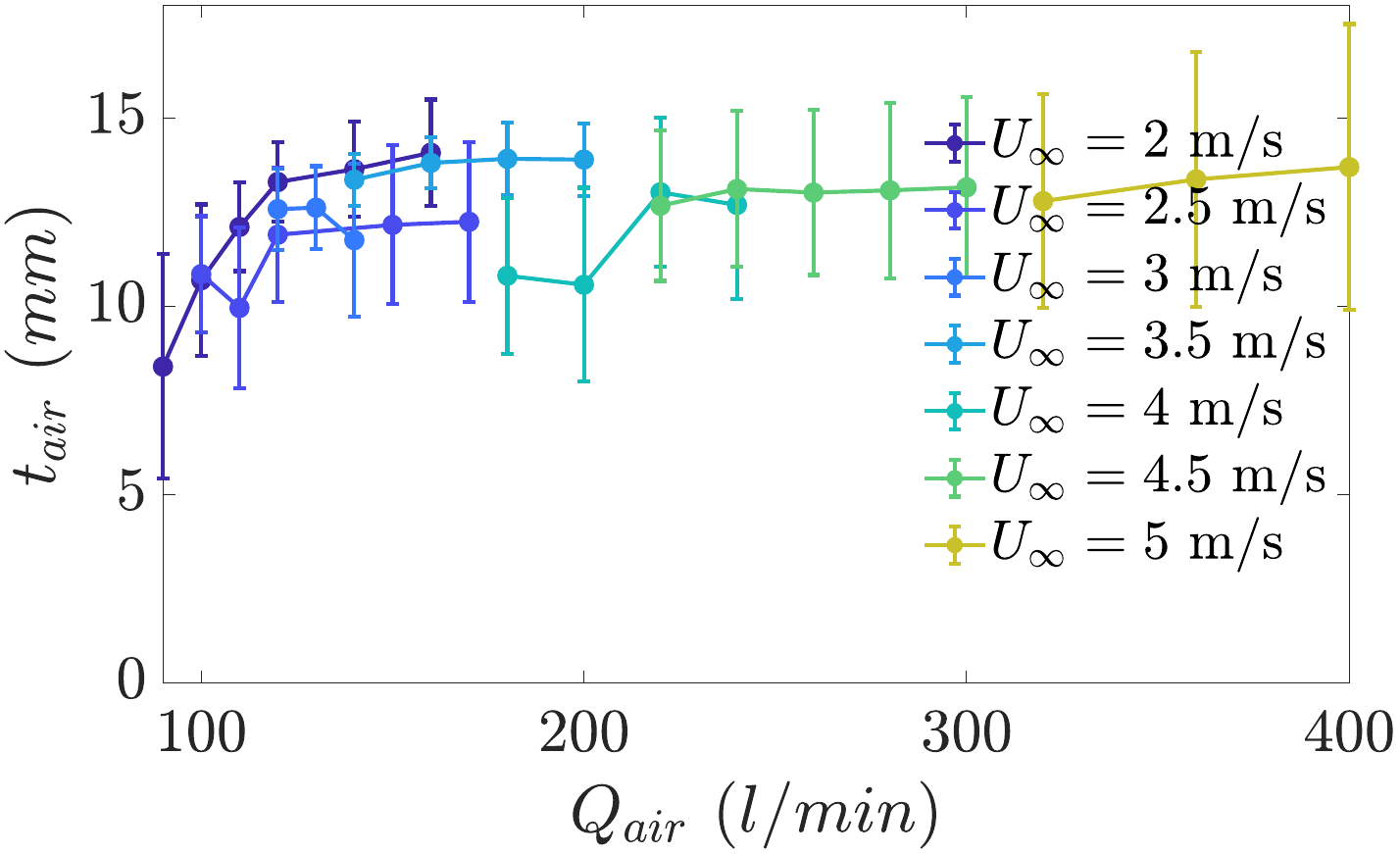}
  \caption{}
  \label{tairQ}
  \end{subfigure}
  \caption{(a) Air layer thickness close to the injector $t_{inj}$ and exit angle $\theta$ for $U_{\infty}$ = 2 m/s (see also Figure \ref{closeToInjPlot} for definitions). The vertical dashed line indicates $Q_{crit}$. (b) Mean air layer thickness $t_{air}$ development for increasing $Q_{air}$ for multiple $U_{\infty}$. Only $Q_{air}>Q_{crit}$ is shown, meaning that the air layer is formed. Error bars are the standard deviation due to the physical variation in thickness.}
\label{allthickness}
\end{center}
\end{figure*}

\newpage 

\textbf{Air Layer regime ($DR\geq60\%$)}

As mentioned above, for the last regime of drag reduction, all cases reach a plateau starting at $DR=60\%$. The main difference across the velocities tested, is that for the highest ones, drag reduction shows a marginal increase with $Q_{air}$ (absolute change approximately 5\%), while for low velocities, a much larger increase is observed (approximately 15\%, Figure \ref{DRcurvesall}). Here we leverage wall-parallel imaging data downstream of the injector and wall-normal ones both in the middle of the plate as well as close to the injector for some more quantitative discussions (Figure \ref{closeToInjPlot}). The goal is twofold: first to characterize the topology of the ALR and how it varies with $U_{\infty}$, something that is lacking in current literature. Secondly, to evaluate whether there are any links between that topology and DR.
Specifically, we examine the air layer thickness and its coherence. As both affect the non-wetted area, they could lead to changes in DR.

Towards that end, we first employ an edge detection algorithm on the wall-normal imaging planes to identify the outer boundary of the air layer. This allows us to estimate the air layer thickness at the middle of the test section, $t_{air}$ (Figure \ref{closeToInjPlot}), and both the initial air layer thickness, $t_{inj}$, and its exit angle, $\theta$ near the injector, where the air jet bends towards the wall due to buoyancy and the liquid momentum (see Figure \ref{closeToInjPlot}).%
\begin{figure}[!ht]
\begin{center}
\begin{subfigure}[t]{0.45\textwidth}
  \centering
\includegraphics[width=\textwidth]{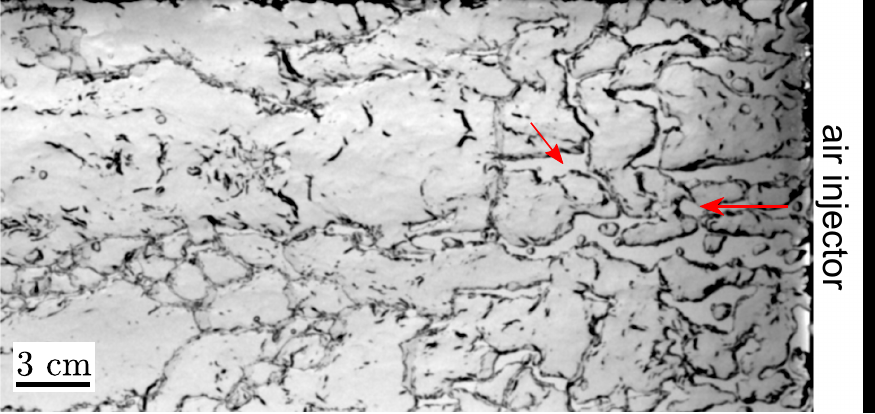}
  \caption{$Q_{air}=Q_{crit}=100$ l/min}
  \label{wet1}
  \end{subfigure}%
  \hspace{0.01cm}
\begin{subfigure}[t]{0.45\textwidth}
  \centering
\includegraphics[width=\textwidth]{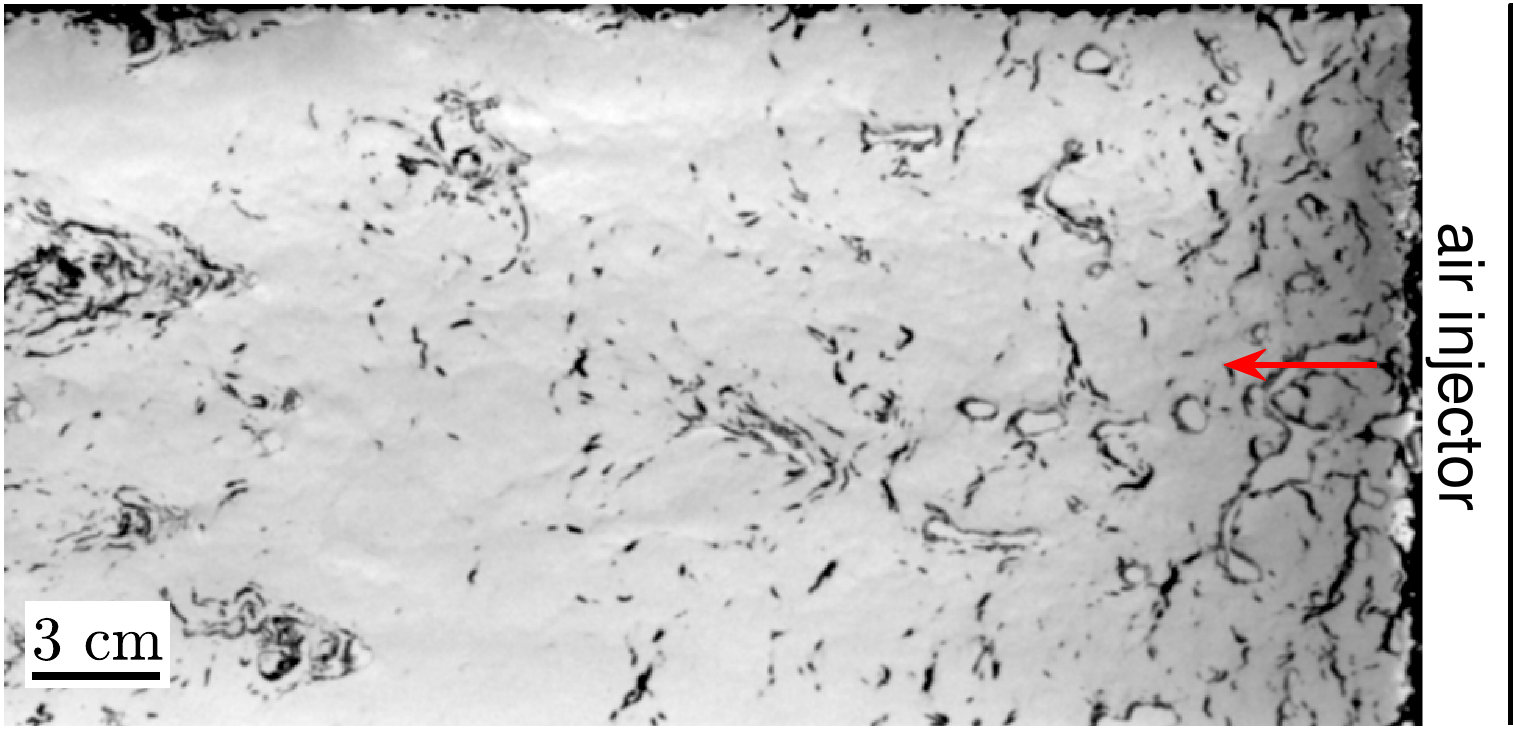}
  \caption{$Q_{air}=170$ l/min $>Q_{crit}$}
  \label{wet2}
  \end{subfigure}
\vspace{0.5em}
\begin{subfigure}[t]{0.45\textwidth}
  \centering
\includegraphics[width=\textwidth]{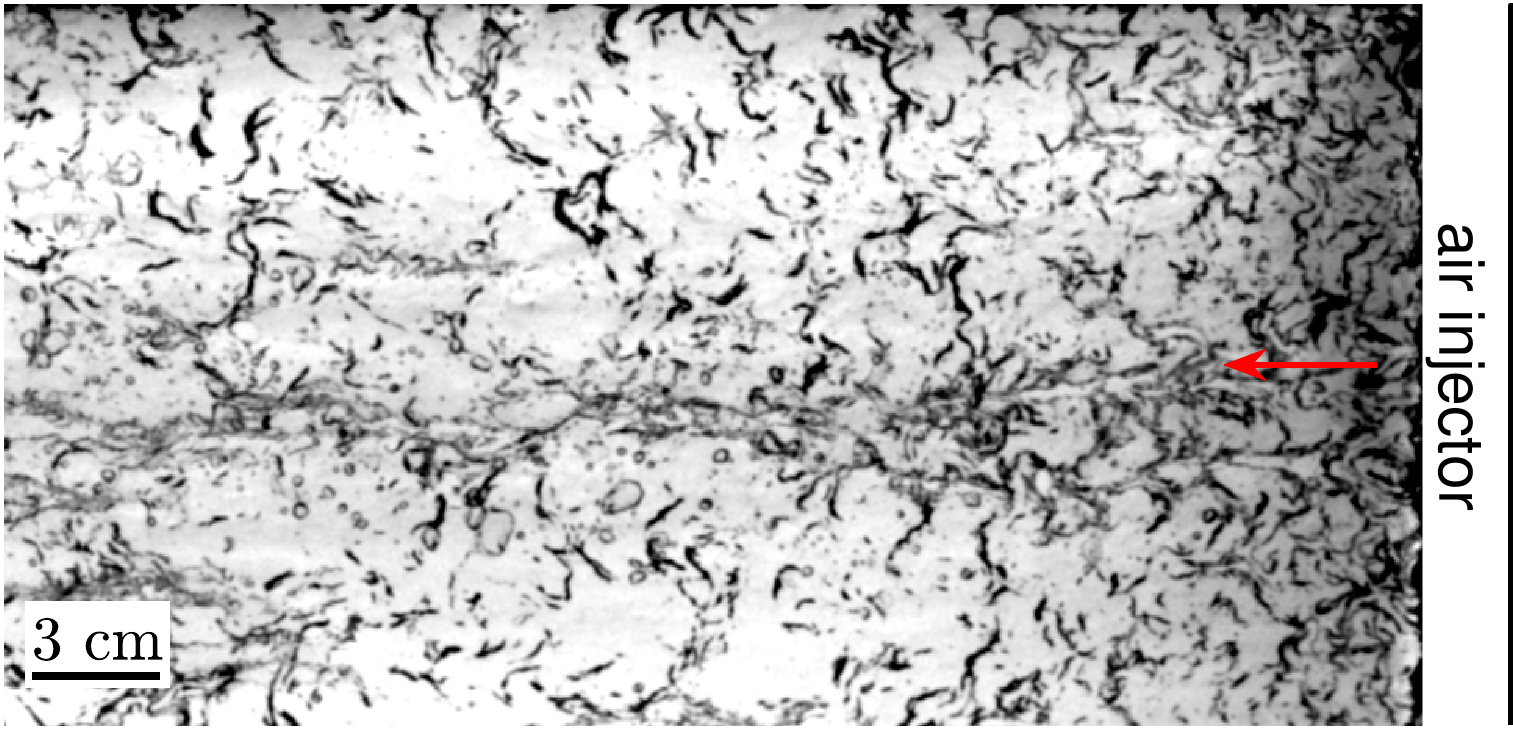}
  \caption{$Q_{air}=Q_{crit}=180$ l/min}
  \label{wet3}
  \end{subfigure}%
  \hspace{0.01cm}
\begin{subfigure}[t]{0.45\textwidth}
  \centering
\includegraphics[width=\textwidth]{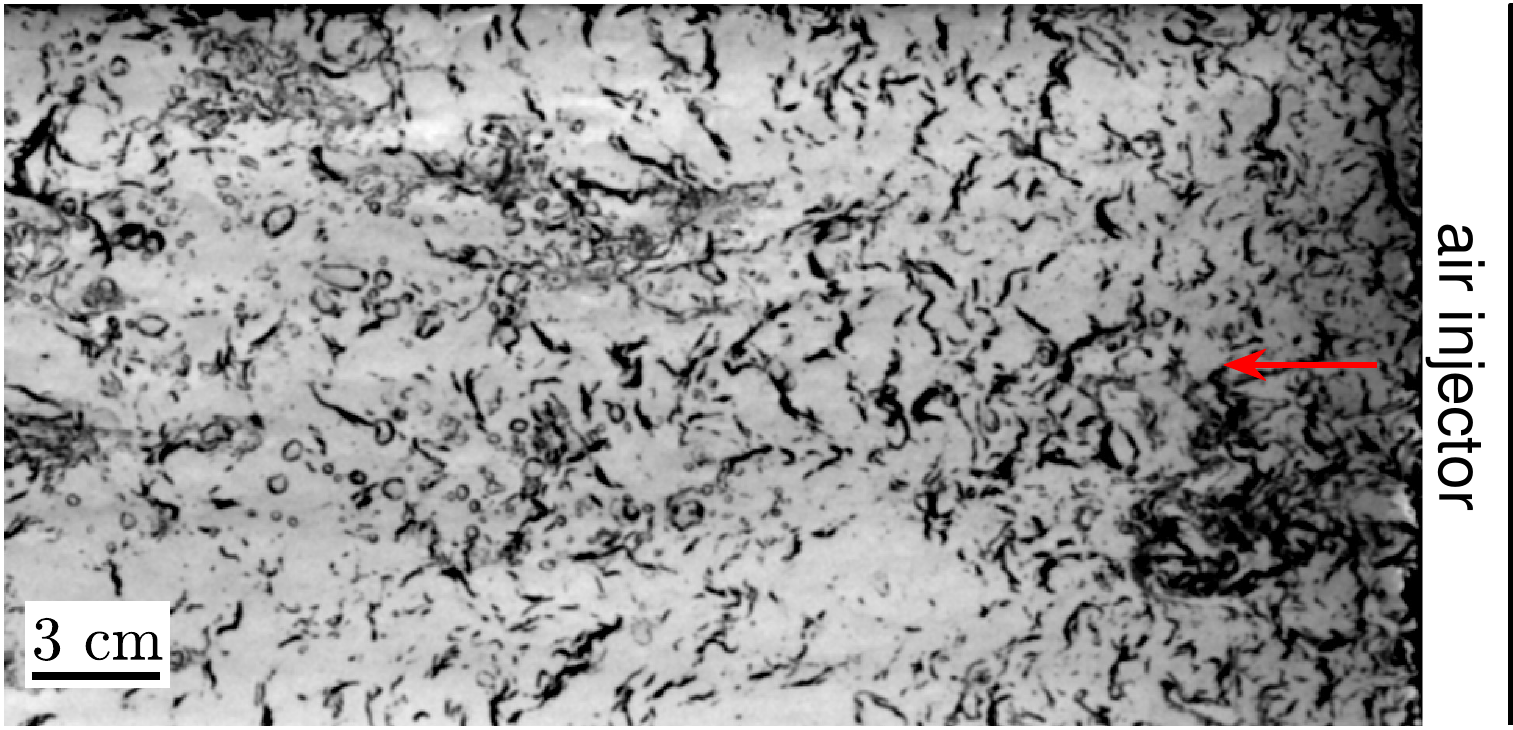}
  \caption{$Q_{air}=240$ l/min $>Q_{crit}$}
  \label{wet4}
  \end{subfigure}
  \caption{Characteristic images of wetted pockets near the injector at $Q_{air}=Q_{crit}$ and $Q_{air}>Q_{crit}$ for $U_{\infty}$ = 2.5 m/s (a) \& (b) and $U_{\infty}$ = 4 m/s (c) \& (d). A wetted patch is indicated in (a) for clarity.}
    \label{wettedpatches}
\end{center}
\end{figure}

Close to the injector, when looking at the evolution of both the thickness and the exit angle of the air phase, minimal variation is observed with increasing $Q_{air}$ (while only $U_{\infty}=2$ m/s is shown, this behavior is similar in all other velocities), especially in the ALR regime (Figure \ref{tinjthetaQ}). A slight decrease in exit angle and increase in $t_{inj}$ is seen for $Q_{air}<40$ l/min, associated with the slight increase in bubble diameter and bubble coalescence behavior as discussed in the previous section. Thus, while from a topological perspective, the robustness of the exit geometry of the air phase is of note, it cannot be meaningfully linked with the observed increase in DR for the low velocities. Moving to the estimates of the air layer thickness at the middle of the plate, $t_{air}$, the first thing to observe is that $t_{air}\approx 1.5t_{inj}$, indicating a growth in air layer thickness along the plate, similar for all $U_{\infty}$ (Figure \ref{tairQ}). The second thing to note is that for lower velocities, there is an increase in $t_{air}$ as $Q_{air}$ increases (we should repeat here that since the focus is on the ALR, we only look at $Q_{air}>Q_{crit}$). For $U_{\infty}$ = 2 m/s a relative increase of nearly 75\% is calculated, whereas higher velocities yield substantially smaller values. While this trend does mirror the one observed in DR, where an increase of about 15\% was observed, and an increase in ALR thickness would be associated with a DR increase (given the increase in non-wetted area in the side fences), the variation observed in $t_{air}$ is not significant enough to fully explain it. More specifically, the $t_{air}$ growth corresponds to a $3.4\%$ increase in terms of non-wetted area which is minimal considering the DR growth rate with $A_{nw}$ (Figure \ref{ANWvsDR}). Consequently, the observed increase in DR past $Q_{air}=Q_{crit}$ for low velocities is associated to an increase in air layer thickness (away from the injector), and can be partly be attributed to the resulting increase in non-wetted area, yet there must other factors responsible. 

One of these factors is the level of embedment of the air layer within the liquid TBL (the ratio $t_{air}/\delta$), which has been shown to have a significant effect on the air phase topology \citep{nikolaidou2024effect, zverkhovskyi2014ship} and is particularly relevant in full-scale applications where $t_{air}/\delta\ll1$. For the two lower velocities, this embedment decreases with $Q_{air}$ since the air layer thickness increases. Based on our single phase measurements (see section \ref{onephaseflow}), for $Q_{air}=90$ l/min we estimate $t_{inj}/\delta\approx0.5$ close to the injector and $t_{air}/\delta=0.165$ in the middle of the plate. This is assuming the same TBL growth with and without air, which is unlikely. However, it still provides a rough estimate of embedment. These ratios become 0.47 and 0.27 respectively, at the highest air flow rate. The higher $t_{air}/\delta$ ratio, indicates that the air/water interface is further away from the wall and thus encounters a much lower turbulent activity within the TBL. That is expected to act favorably in terms of drag reduction (see also next paragraph), aligning with the $DR$ increase measured for these velocities and air flow rates. This indicates a compound effect of the $t_{air}$ increase: larger non-wetted area and lower turbulent activity within the TBL disturbing the interface. For the higher velocities, the $t_{air}/\delta$ ratios are almost constant since the air layer thickness does not vary appreciably with $Q_{air}$. Thus no effect in $DR$ is expected, aligning again with the relatively unchanged measured $DR$ with increasing $Q_{air}$ (Figure \ref{DRcurvesall}). 

Finally, we turn our attention towards the wall-parallel imaging planes, where clear topological differences are observed between low ($U_{\infty}=2.5$ m/s, Figure \ref{wet1}, \ref{wet2}) and high velocities ($U_{\infty}=4$ m/s, Figure \ref{wet3} and \ref{wet4}). For the latter, the air layer interface is much more deformed, with signs of small scale features, a clear effect of higher Reynolds number and thus turbulent activity, not seen in the lower velocity cases. Yet, increasing $Q_{air}$ past $Q_{crit}$ does not lead to any observable change in that topology. On the other hand, for the lower velocities, at $Q_{air}=Q_{crit}$, while several small scale features persist deforming the air-water interface, these are significantly decreased for $Q_{air}>Q_{crit}$: as mentioned above, this can be attributed to a larger $t_{air}$, protruding further in outer region of the TBL, such that the interface is exposed to a lower liquid turbulent activity. 

In addition, for $Q_{air}=Q_{crit}$, several wetted pockets are observed close to the injector, locally breaking the spanwise coherence of the air layer (Figure \ref{wet1}). These liquid gaps, reflecting high speed streaks upstream of the injector (see \cite{laskari2025effects}), also shrink considerably in the lower velocities with increasing $Q_{air}$. This leads to a much smoother and undisturbed interface (Figure \ref{wet2}). Linking back to the DR behavior of these cases, a decrease in wetted pockets with increasing $Q_{air}$ and a smoother air–liquid interface accompany the increase in DR for the low velocities. This results in a higher maximum drag reduction ($DR_{max}\approx 75\%$) than that achieved for $U_{\infty}=5$ m/s ($DR_{max}\approx 65\%$), despite a similar maximum $t_{air}$ (Figure \ref{tairQ}). For the higher velocities, the air-liquid interface is significantly more deformed and at smaller scales than for lower $U_{\infty}$, owing to the much higher Re number. This topology that varies little with increasing $Q_{air}$ beyond the critical one, mirroring again the almost constant DR levels observed for that velocity. Thus, expanding on the discussion above regarding the air layer thickness, we can also safely assume here that the observed drag behavior for $DR>60\%$ as $U_{\infty}$ increases can also be partly attributed to the changes in the air layer spanwise integrity and local breakup.

\begin{figure*}
\begin{center}
\begin{subfigure}[t]{.6\textwidth}
  \centering   \includegraphics[width=\textwidth]{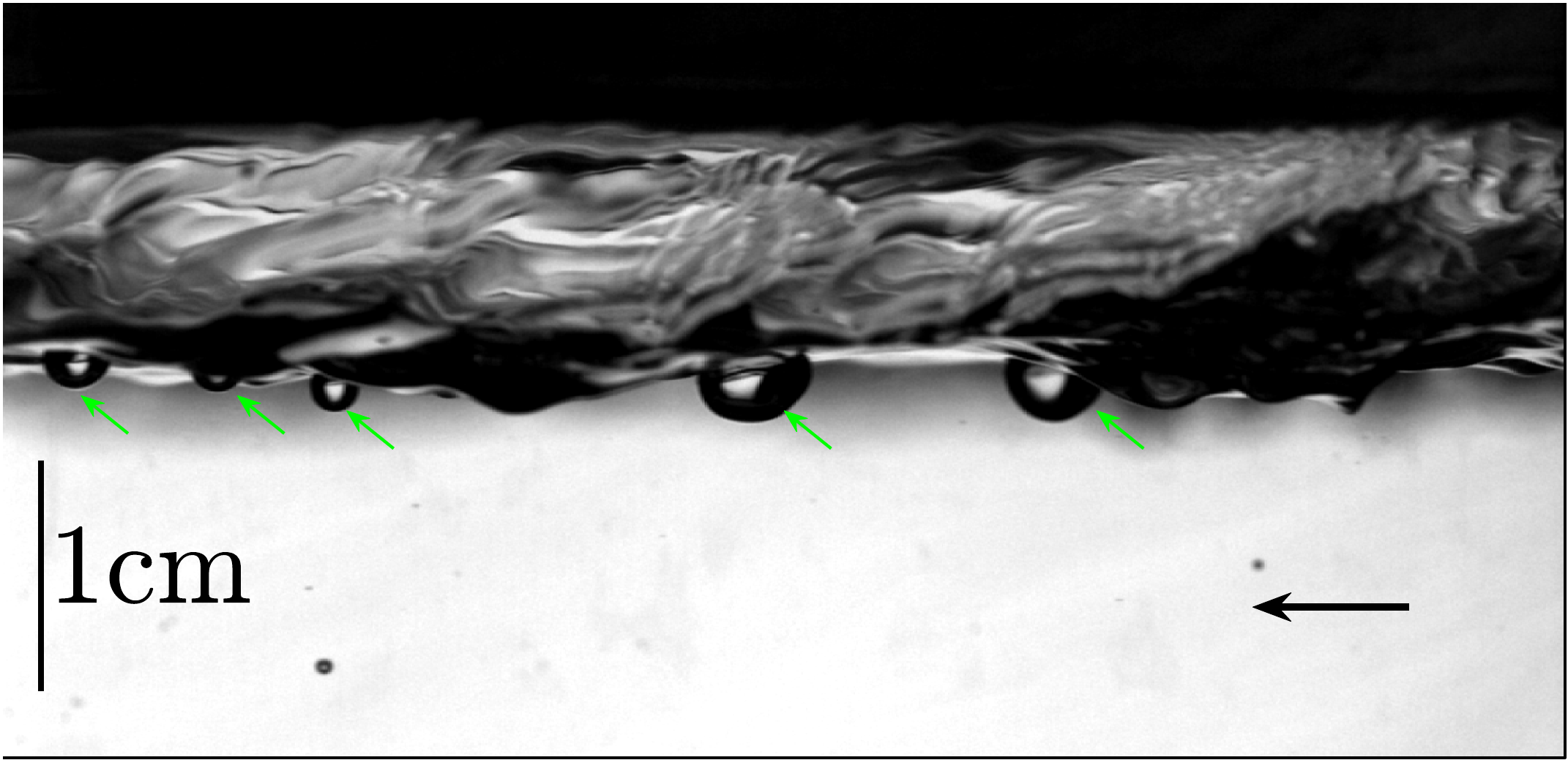}
  \caption{}
  \label{bubbles}
  \end{subfigure}%
\hspace{0.018\textwidth}
\begin{subfigure}[t]{.3\textwidth}
  \centering
\includegraphics[width=0.96\textwidth]{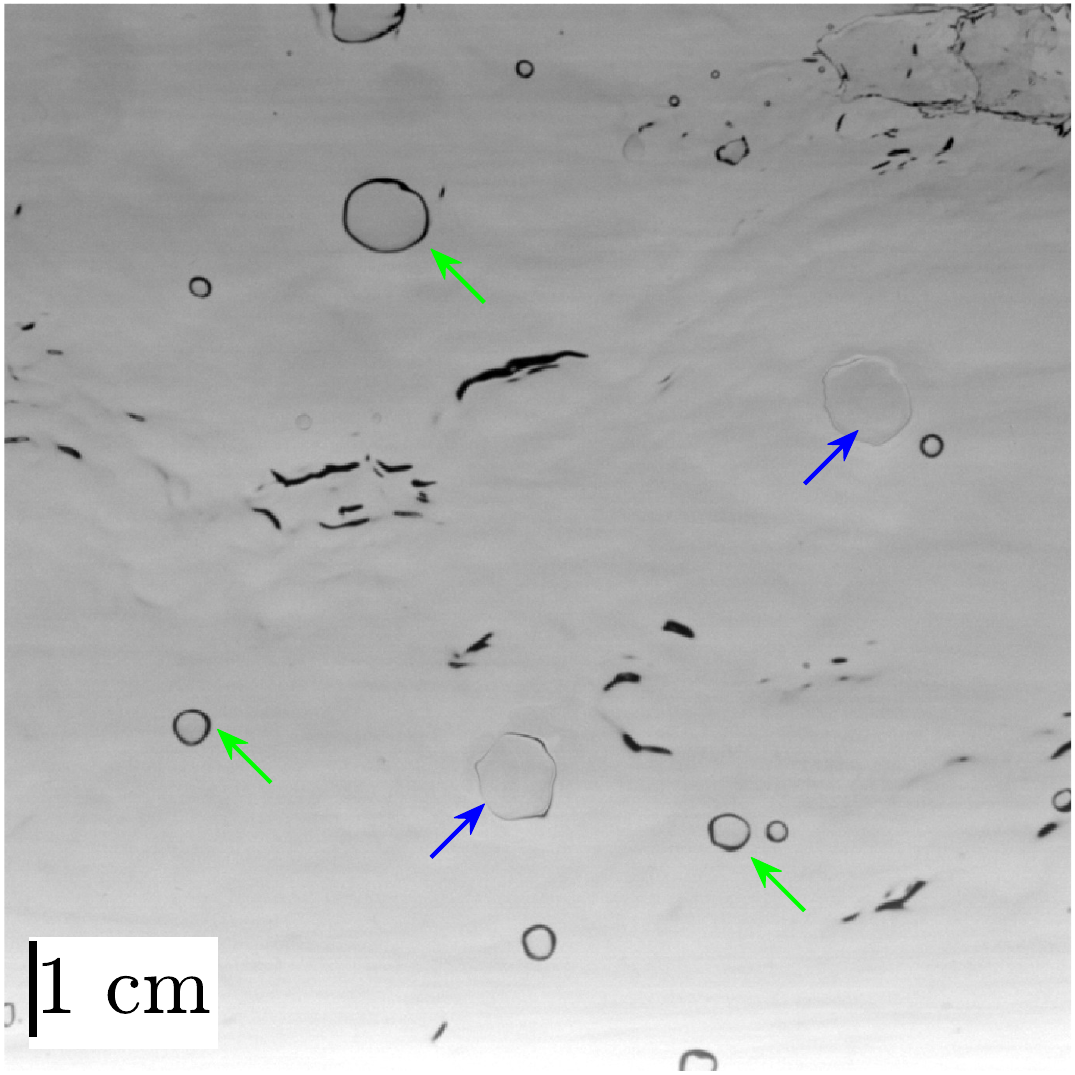}
  \caption{}
  \label{droplets}
  \end{subfigure}
  \caption{Morphological features of the air-water interface on the wall-normal plane (a) and the wall-parallel plane (b).
 Air bubbles are indicated in green and water droplets in blue. This is for $U_{\infty}=2$ m/s and $Q_{air}=100$ l/min.}
    \label{dropletsandbubbles}
\end{center}
\end{figure*}

To conclude this section, although a detailed analysis of the air layer instability mechanisms is beyond the scope of this work, some additional observations regarding the integrity of the air–water interface (Figure \ref{dropletsandbubbles}) will also be highlighted. Connecting these to $DR$ is more challenging than our discussion above, yet they have been identified as having the potential to promote air layer breakup and thus drag increase. Both normal and shear stresses originating from the liquid phase can contribute to interface deformation and eventual rupture: the former may increase 
interface curvature, while the latter could cause local wetting and disruption of the air layer. Wall-normal imaging revealed the existence of air bubbles near the air–water interface, likely originating from a partial layer breakage (green in Figure \ref{bubbles}). These bubbles, while consistently present, become more frequent with increasing $U_\infty$. The same features can be observed in the wall-parallel imaging (green in Figure \ref{droplets}). Additionally, lighter-perimeter features (blue in Figure \ref{droplets}) can be observed in 
the same wall parallel images. It is hypothesized that these are formed due to the wetted patches near the injector discussed earlier. When such a patch forms, the liquid is temporarily wetting the top wall, leaving behind small droplets. These droplets are then advected downstream 
by the air flow and eventually fall into the air–water interface, creating these lighter-perimeter features. While the exact contribution of these features to drag reduction cannot be quantified here, they are highlighted as potential areas of exploration in future studies.

\section{Air injection - Subcritical conditions}
\label{subcrtitical_results}
In the previous section, we discussed the topology of all the air phase regimes and the resulting drag reduction performance for freestream velocities corresponding to supercritical conditions ($Fr_d>1$). Here we extend the parameter space to subcritical conditions by lowering $U_{\infty}$ to reach $Fr_d\leq1$. The motivation stems from the different air phase topology observed in our earlier work \citep{nikolaidou2024effect} compared to the one presented in section \ref{supercritical_results}. More specifically, for the same air injector geometry, an air cavity of a limited length was observed for the highest air flow rates, instead of the unbounded air layer described in the current study. The hypothesis is that the governing parameter for this topology change is the $Fr_d$ number. Although many previous studies focused on one of the two topologies, the transition has not been explicitly showcased.

\begin{figure}[t!]
\centering
\includegraphics[width= 0.8\linewidth]{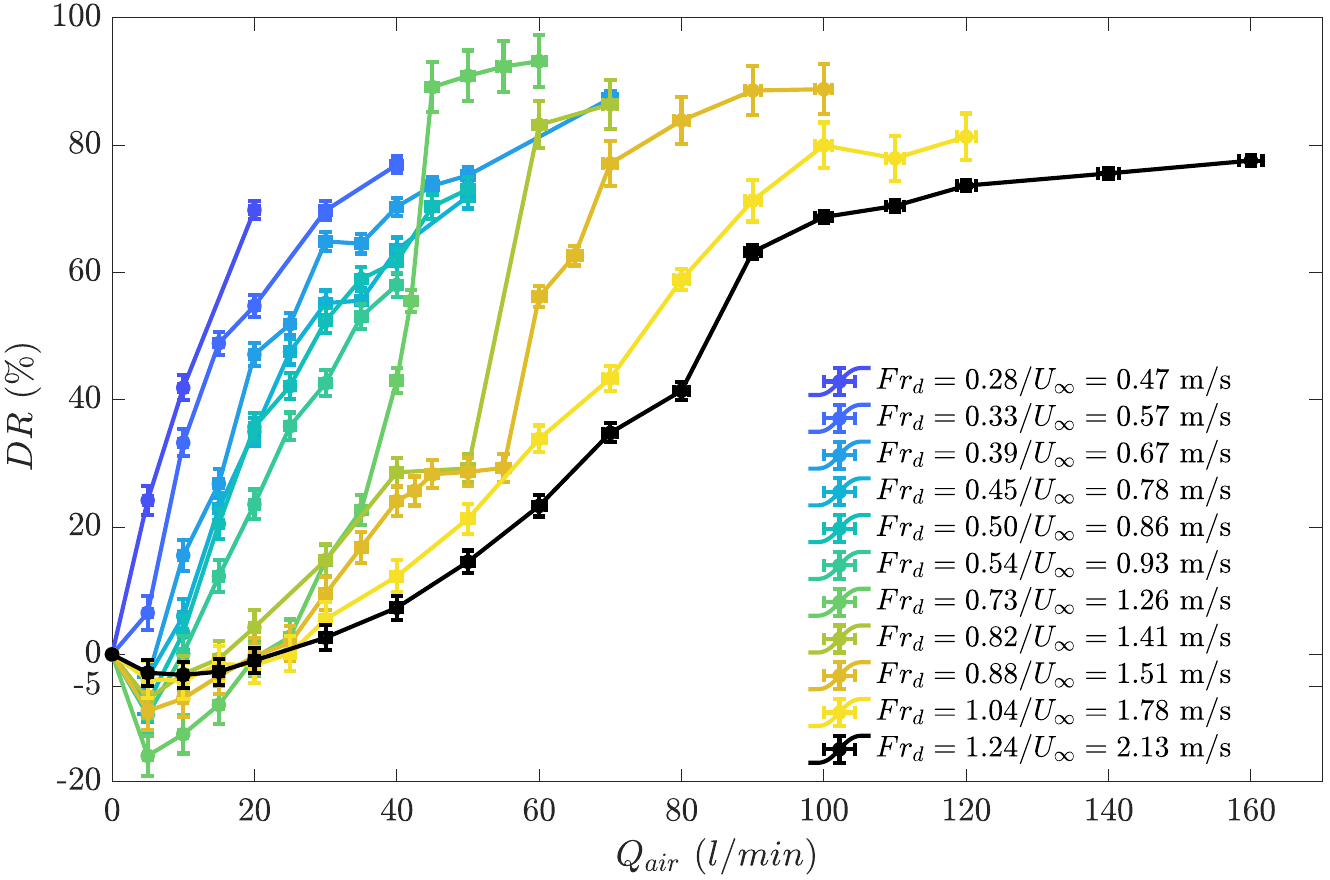}
\caption{Drag reduction curves, for different $U_{\infty}$. The black line ($Fr_d=1.24$) from Figure \ref{DRcurvesall} is included for comparison.}
\label{DR_orifice}
\end{figure}

To begin with, we look at drag reduction for $F_d\leq1$ (colored lines in Figure \ref{DR_orifice}), together with the lowest velocity from the supercritical regime for comparison (black line in Figure \ref{DR_orifice}). Based on the overall trends, two different families of drag reduction curves are observed: one for $Fr_d\geq0.73$ and one for $Fr_d\leq0.54$. We will discuss each one in what follows. For $Fr_d\geq0.73$, the curves follow a supercritical behavior similar to the one observed for $Fr_d>1$ in section \ref{supercritical_results}: relatively slower initial increase of DR with $Q_{air}$ (positive curvature) and a stabilization around a maximum DR. As also seen in section \ref{transReg}, this $DR_{max}$ increases monotonically with decreasing $U_{\infty}$, with values here reaching approximately 93\% for the lowest freestream velocity of this family ($U_{\infty} = 1.26 $m/s), as opposed to approximately 66\% for the highest one ($U_{\infty} = 5.5$ m/s in section \ref{transReg}) for $Fr_d>1$. Finally, in alignment with BDR observations from section \ref{bubbly}, drag increases ($DR<0$) at low air flow rates and the magnitude the minimum DR increases with decreasing $U_{\infty}$. Thus, the lowest velocity of this family ($U_{\infty} = 1.26~\mathrm{m/s}$), exhibits the best ALR performance ($DR=93\%$) and the worst BDR one ($DR=-16\%$). Since neither a better DR performance for ALR nor a higher drag increase for BDR are reached for lower $U_{\infty}$, this case corresponds to the best and worst performances respectively, across all cases. Shifting now our attention to the DR trends of the second family of curves ($Fr_d\leq0.54$ in Figure \ref{DR_orifice}), there are some distinct differences in behavior. Drag reduction still increases with $Q_{air}$ but at a steeper initial rate (negative curvature) for all $U_{\infty}$. For the highest $Q_{air}$ tested, there is also no DR plateau as observed for $Fr_d\geq0.73$ for the air flow rates tested. 
Finally, although there is still a drag increase measured for the lowest air flow rates (BDR), as also observed for $Fr_d\geq0.73$, its magnitude {\em{decreases}} as $U_{\infty}$ decreases, to eventually disappear for the lowest velocities tested.

\begin{figure}[!ht]
\begin{center}

  \begin{subfigure}[t]{0.75\textwidth}
  \centering
\includegraphics[width=\textwidth]{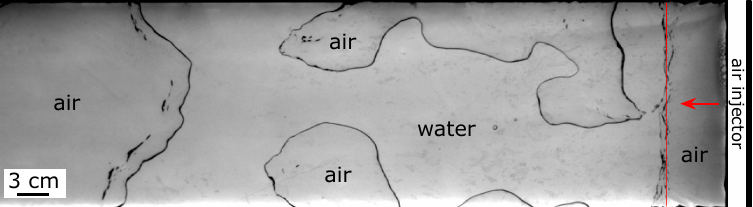}
  \caption{$Fr_d=0.28$}
  \label{cavity1}
  \end{subfigure}%
  \vspace{0.01cm}
  \begin{subfigure}[t]{0.75\textwidth}
  \centering
\includegraphics[width=\textwidth]{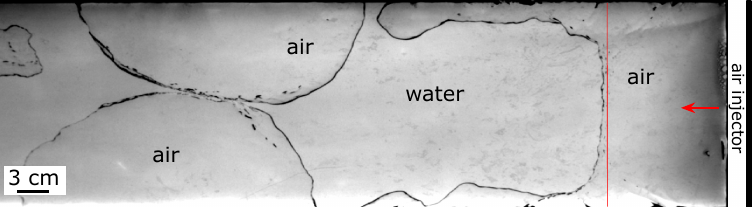}
  \caption{$Fr_d=0.33$}
    \label{cavity2}
  \end{subfigure}%
    \vspace{0.01cm}
    \begin{subfigure}[t]{0.75\textwidth}
  \centering
\includegraphics[width=\textwidth]{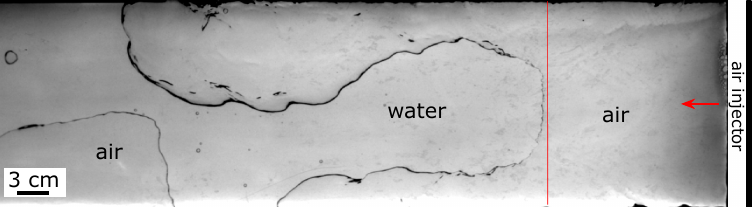}
  \caption{$Fr_d=0.39$}
   \label{cavity3}
  \end{subfigure}%
    \vspace{0.01cm}
    \begin{subfigure}[t]{0.75\textwidth}
  \centering
\includegraphics[width=\textwidth]{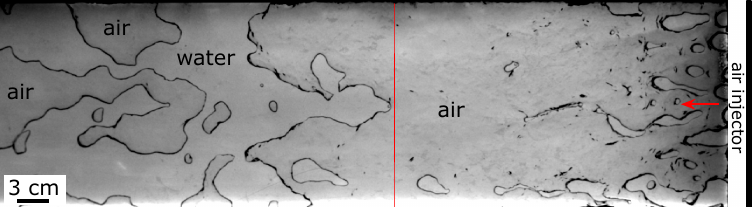}
  \caption{$Fr_d=0.54$}
      \label{brokenAirLayer}
  \end{subfigure}%
    \vspace{0.01cm}
    \begin{subfigure}[t]{0.75\textwidth}
  \centering
\includegraphics[width=\textwidth]{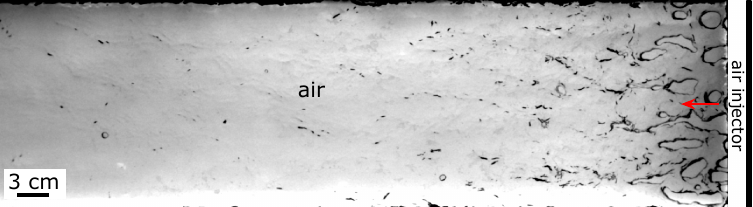}
  \caption{$Fr_d=0.73$}
  \label{smoothAirLayer}
  \end{subfigure}%
  \end{center}
    \caption{Characteristics images of the air layer regime for (a) $U_{\infty}=0.47$  m/s and $Q_{air}=20$ l/min, (b) $U_{\infty}=0.57$ m/s and $Q_{air}=40$ l/min, (c) $U_{\infty}=0.67$  m/s and $Q_{air}=40$ l/min, (d) $U_{\infty}=0.93$  m/s and $Q_{air}=40$ l/min and (e) $U_{\infty}=1.26$  m/s and $Q_{air}=60$ l/min. From top to bottom $Fr_d$ increases and there is a transition from air cavity to an air layer. The red vertical line indicates the air cavity closure in (a)--(d). In (e) the air layer is unbounded.}
    \label{airphasemap}
  \end{figure}

Based on the DR behavior alone, it becomes clear that a transition takes place between $0.54\leq Fr_{d,crit}\leq 0.73$, with the family of DR curves belonging to $Fr_d\geq0.73$ still closely following supercritical behavior. Using the dispersion relation for gravity waves, we estimate that $Fr_d\approx0.61$ demarcates deep from intermediate/shallow water conditions \citep[see][]{zverkhovskyi2014ship,nikolaidou2024effect}. This is right in the middle of the transition range extracted from the DR curves, indicating that the deviation from deep water conditions is associated with the different DR behavior. From a topological perspective, for deep water conditions and air injection behind a cavitator, it has been shown that the air layer length is limited, stable, and equal to one half of the gravity wavelength (\cite{matveev2003limiting} and \cite{butuzov1966limiting,butuzov1967artificial}). The same was also observed for vertical injection without a cavitator in our previous work \citep{nikolaidou2024effect} for $Fr_d<0.4$. Thus in the current study, a bounded air cavity of a certain length would be expected for $Fr_d<0.61$, while for larger $Fr_d$ deep water conditions are no longer satisfied and the theoretical cavity length would increase to approach infinity. As a result, an unbounded air layer would be expected similar to what was observed in section \ref{transReg} for $Fr_d>1$. In what follows, we will assess whether this indeed persists down to $Fr_d\approx0.61$ and whether a transition to a bounded cavity occurs for $Fr_{d}<0.61$, accompanying the change in DR behavior. 

To that end, instantaneous images of the air phase are subsequently assessed (Figure \ref{airphasemap}). It is immediately apparent that: a) for $Fr_d>0.61$ an unbounded air layer is formed, similar to the one observed for $Fr_d>1$ (Figure \ref{smoothAirLayer}) and b) for $Fr_d<0.61$ the air layer's streamwise continuity is broken throughout the entire width (Figure \ref{brokenAirLayer}), gradually transitioning to a stable, bounded cavity (Figures \ref{cavity1},\ref{cavity2}, \ref{cavity3})  as $Fr_d$ decreases. This confirms our DR-based classification of the $Fr_d$ regimes and we showcase this transition for the first time using both drag reduction and imaging data. At a second level, what can also be seen is that a steady, bounded cavity becomes clearer for $Fr_d<0.4$, with its length increasing with increasing $U_\infty$ (as predicted by the dispersion relation), matching our earlier results \citep{nikolaidou2024effect} remarkably well (different facility, same injection system). Further agreement with results from that study is seen with respect to shedding: an air pinch-off mechanism from two side branches is also observed here. Likely due to the narrower width of the current facility, these branches are also seen to intermittently coalesce here, leading to large non-wetted areas further downstream of the main cavity. Finally, the bubbly and transitional regimes (images not shown here) exhibit similar topological features across all cases and also match those of $Fr_d>1$ in section \ref{supercritical_results}. Within BDR, increasing $U_\infty$, leads to smaller size bubbles, as expected and seen also in the $Fr_d>1$ cases (Figure \ref{qair_db}). From side view imaging data, it is also confirmed that the bubbles are positioned within a single-bubble layer in the vertical direction, for all cases.

Linking the air phase topology with drag reduction can help us better understand the underlying mechanisms of DR. In particular, what is clearly highlighted from the available data, is the balance between air layer coverage (or non-wetted area) and interface integrity. By expanding our supercritical results here to a $Fr_d\geq0.73$ range, we confirmed that by further decreasing $U_\infty$, a monotonic increase of maximum DR is achieved (from a $DR_{max}=66\%$ for $U_\infty=5.5~\mathrm{m/s}$ to a $DR_{max}=93\%$ for $U_\infty=1.26~\mathrm{m/s}$), although the corresponding air coverage is mostly unchanged (and the thickness variation negligible). What does change is the Reynolds number across the cases, reflected as a marked variability of the interface smoothness (compare Figure \ref{smoothAirLayer} with Figure \ref{wettedpatches}): a lower Reynolds number TBL leads to a much smoother interface which is much less prone to local breakups, and is associated with higher DR. This is also supported by observations at the $Fr_d<0.4$ regime. In this case, an air cavity with spanwise coherence is formed, whose length extends only a few centimeters downstream of the air injector (Figure \ref{cavity3}). Beyond this point, the cavity breaks up into large air patches separated by liquid. These smooth patches however appear to still have a significant effect in reducing the drag, since higher levels of drag reduction are measured in this topology ($DR_{max} \approx80\%$ for $Q_{air}=70~\mathrm{l/min}$) compared to cases where a streamwise continuous but strongly deformed interface is present (eg. $DR_{max}<75\%$ for $Fr_d=1.24$ and $Q_{air}>Q_{crit}$). We should note here that, due to the global nature of the drag measurements employed, there are other factors (related to the areas of the plate which could not be covered by air) potentially contributing to the observed Reynolds number dependency. Thus, while our focus is on the air phase changes observed which can be meaningfully linked to the global drag reduction, reconstructing the full picture with all causal relationships involved, would require further (and more targeted) measurements. 

Finally, a comment is warranted regarding full-scale applications. While the low velocities used here to attain subcritical conditions are not representative of the ship scales, the resulting deep water conditions ($Fr_d<0.61$) are. However, due to facility limitations, most laboratory studies either satisfy deep water conditions with lower velocities \citep{nikolaidou2024effect,Qin2019,zverkhovskyi2014ship} or reach higher freestream velocities (and thus Reynolds numbers) but within intermediate or shallow water regimes \citep{elbing2008bubble,barbaca2019unsteady}. When both deep water conditions and high freestream velocities are present (as is the case in realistic ship conditions), based on the current analysis an air cavity would be expected, but the effect of high Reynolds number on it remains unclear. 

\section{Regime map}
\label{regimeMAPsec}

\begin{figure}[t!]
\centering
\includegraphics[width= 0.8\linewidth]{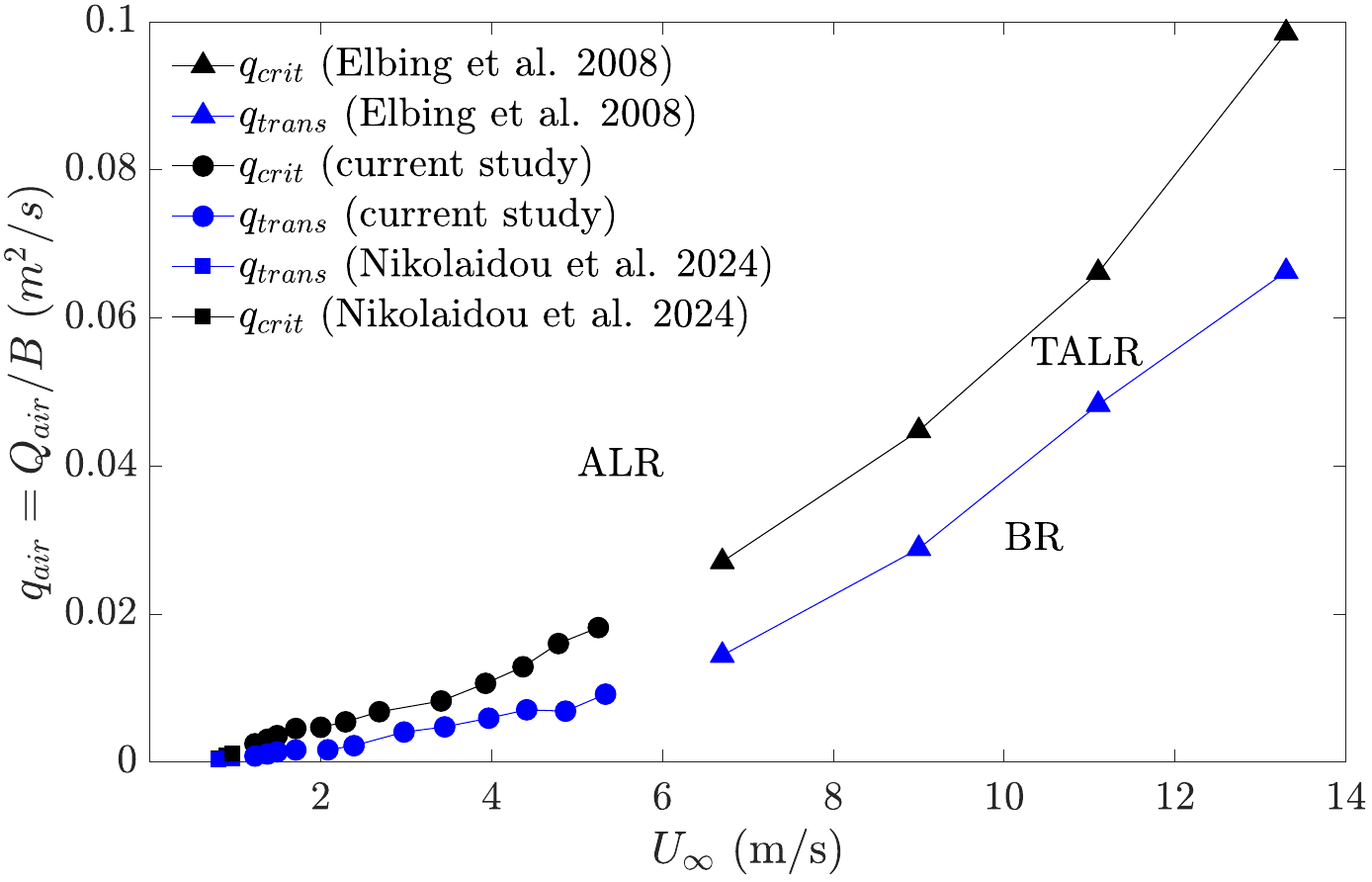}
\caption{Regime map. Transitional air flow rate $q_{trans}$ and critical air flow rate $q_{crit}$ demarcating the transition from bubbly regime (BR) to transitional regime (TALR) and from TALR to air layer regime (ALR) respectively. Measurements of the current study are shown along with measurements from \cite{elbing2008bubble} and \cite{nikolaidou2024effect}. Air flow rates are corrected accounting for the local pressure and density at the injection location and are given per meter width ($q_{air}=Q_{air}/B$).}
\label{regimeMAP}
\end{figure}

In the previous sections, results were presented for both subcritical and supercritical conditions, in which three air phase regimes were consistently observed: the bubbly regime (BR), the transitional air layer regime (TALR) and the air layer/air cavity regime (ALR). The air flow rates that demarcate the transition from BR to TALR ($q_{trans}$) and from the TALR to ALR ($q_{crit}$) are presented over a wide range of $U_{\infty}$ (Figure \ref{regimeMAP}; note that the air flow rate per meter width $q_{air}=Q_{air}/B$ is used here instead). In order to investigate any global trends, data from our study are shown together with those from \cite{elbing2008bubble} and \cite{nikolaidou2024effect}. While the experimental facilities used for these studies differ significantly in size and capabilities, in all of them a flat plate with a slot-type air injector was used. In addition, the definition of $q_{trans}$ and $q_{crit}$ also differs between them. \cite{elbing2008bubble} defined them based on a drag reduction level of $20\%$ and 80\% respectively. In \cite{nikolaidou2024effect} the classification was made based on the non-wetted area coverage. In the current study, $q_{trans}$ was defined qualitatively from the imaging data, while $q_{crit}$ was defined based on the drag reduction and air flow rate measurement trends ($DR$ curve asymptotically reaching a maximum level, see Figures \ref{DRcurvesall} and \ref{DR_orifice}). Despite all that, the transition points seem to follow the same trend across a wide range of $U_{\infty}$. The transitional and critical air flow rates roughly follow a quadratic relation with $U_{\infty}$. 

\subsection{Scaling of the critical air flow rate}

\begin{figure*}
\begin{center}
\includegraphics[width=0.8\linewidth]{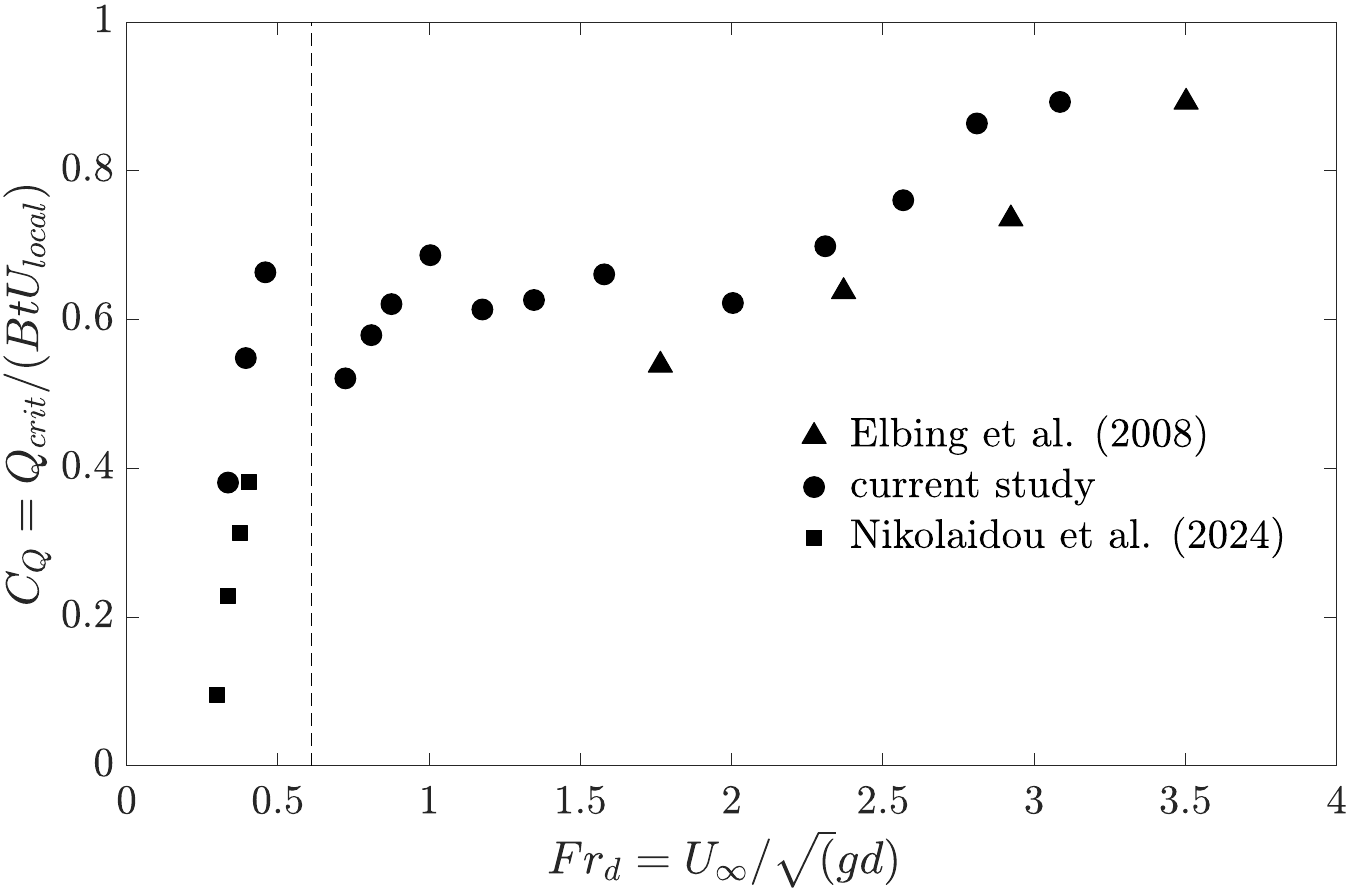}
  \caption{Froude-depth number versus non-dimensional air flow rate. The local velocity $U_{local}$ at the air-water interface is used for the velocity scale. The dashed vertical line demarcates the shallow/intermediate conditions (to the right) from deep water conditions (to the left).}
  \label{FrdQnd}
\end{center}
\end{figure*}

Finally, the focus is shifted to the critical air flow rate, $q_{crit}$. This air flow rate needed for transition to an air layer is important since the ALR is the most desired regime in terms of drag reduction. Thus, predicting $q_{crit}$ is crucial for industrial air lubrication systems and yet, despite a large number of air lubrication studies, there is still no universal scaling available, posing a significant challenge in applying laboratory scale findings to the real life application. In this section, a scaling is proposed based on the parameters that appear to play a significant role. 

For this purpose, the $U_{\infty}-q_{crit}$ data points from Figure \ref{regimeMAP}, are reused here, in a non-dimensional form. For the velocity scale, the $Fr_d$ number is used for normalization, which was previously shown to dictate the air phase topology for $q_{air}>q_{crit}$ (see section \ref{subcrtitical_results}). The critical air flow rate per meter width $q_{crit}$, is normalized with the slot width of the air injector, $t$ and the local velocity $U_{local}$. It should be noted that, the resulting coefficient, $C_Q=Q_{crit}/(BtU_{local})$, can be interpreted as the ratio of a mean air injection velocity with a local liquid velocity ($U_{local}$). Increasing $t$ enhances coalescence of the air phase and favors the transition to the ALR, resulting in a lower $q_{\mathrm{crit}}$ for larger $t$. This agrees well with visual observations during our experiments but further studies targeting the effect of the slot width are needed. $U_{local}$ is the liquid flow velocity at $y=t_{air}$. For small-scale experiments where $t_{air} \approx \delta$ it is reasonable to expect $U_{local} \approx U_{\infty}$, but for smaller $t_{air}/\delta$ ratios (which is also the case in full-scale ships), $U_{local}<<U_{\infty}$. Local flow velocities were also used in our previous study to explain incoming TBL differences: a large $U_{local}$ (closer to $U_{\infty}$) implies a smaller incoming TBL thickness which is shown to promote transition to the ALR \citep{nikolaidou2024effect}. As such, $U_{local}$ was also chosen here for normalization, to take into account the effect of local flow conditions rather than freestream ones. For the results in our previous study \citep{nikolaidou2024effect}, $U_{local}$ measurements were available, for the current ones, an approximation was made based on separate LDA measurements \citep{nikolaidou2026flowcharacterizationdelftmultiphase} and for  \cite{elbing2008bubble} an approximation was made based on the available velocity data at \cite{etter2005high}.

The resulting non-dimensional plot (Figure \ref{FrdQnd}) highlights a vastly different trend between shallow/intermediate and deep water conditions (separated by a dashed vertical line). A steeper slope is shown for the latter, while for shallow water conditions ($Fr_d>0.61$) only a very small increase of $C_Q$ with $Fr_d$ remains, with the normalization successfully bounding all data points to within $0.6<C_Q<0.9$. It is good to keep in mind that in absolute terms this corresponds to a very large range of air flow rates (44~l/min to 15662 ~l/min) and freestream velocities (0.57~m/s to 13.3~m/s). While the proposed scaling looks promising, further experiments should be undertaken with different flat plate widths and slot widths to test this further. In addition, the role of other parameters such as roughness should be also examined and potentially added to the scaling as proposed by \cite{peifer2020air}.

\section{Summary \& Conclusions}

Friction drag reduction via air lubrication of a flat plate TBL is studied via direct measurements of the drag force and synchronous imaging data. Measurements were performed over a large range of Reynolds numbers (by varying $U_{\infty}$) and air flow rates ($Q_{air}$) to assess the effect of freestream velocity. This provided insight in the physical mechanisms of drag reduction via linking the air phase topology and the resulting drag. 

Simultaneous measurements of the drag force of the flat plate and imaging of the total air coverage along the plate are performed for 2 m/s and across all three air phase regimes. It is found that drag reduction lags significantly behind the non-wetted area coverage at all cases and no simple correlation exists between the two parameters. 

A bubbly regime (BR) is identified for low $Q_{air}$. Bubble sizes are quantified via AI-based zero-shot image segmentation. The bubble size distribution becomes narrower and the mean bubble size decreases with $U_{\infty}$, an effect which relaxes for $U_{\infty}>3$m/s. For all cases the bubble size varies from millimeters to centimeters. Bubbles are found to increase the drag for $Q_{air}<20$ l/min and $0.67\mathrm{m/s}<U_{\infty}<2.5\mathrm{m/s}$, with a peak for $U_{\infty}=1.26 \mathrm{m/s}$. This drag increase is associated with a specific bubble organization over the wall-normal: bubbles residing in a single layer parallel to the wall. However, for $U_{\infty}>2.5$m/s bubbles disperse over the vertical and start to effectively decrease the drag. 

Further increasing the air flow rate results in the creation of air patches (TALR) and ultimately in the formation of an air layer (ALR) that extends beyond the test section length (for $Fr_d>1$ cases). The onset of this air layer at $Q_{air}=Q_{crit}$ is marked by a $60\%$ drag reduction, irrespectively of $U_{\infty}$. Further increasing the air flow rate within the ALR, results in an increase of drag reduction for the lower $U_{\infty}$. By examining both wall-parallel and wall-normal images it is clear that this is accompanied by an increase the air layer thickness and its surface continuity. Both of these are expected to contribute to the drag reduction increase. On the other hand, for higher $U_{\infty}$, increasing $Q_{air}$ within the ALR has a marginal effect on drag reduction accompanied by a marginal change in the air layer thickness and free surface continuity. In general, within the ALR, higher drag reduction levels are achieved for the lower velocities despite the similar non-wetted area coverage of all air layers. This behavior can be at least partially attributed to the lower susceptibility of the interface to breakup: a low Reynolds number TBL is associated with a much smoother air-liquid interface, with no deformations affecting its integrity.

The topology of the air phase regimes for $Fr_d>0.7$ is found to be similar to the supercritical ones ($Fr_d>1$). A big morphological difference is found for subcritical flow conditions pertaining to the deep water regime ($Fr_d<0.61$) however. In this regime the unbounded air layer transitions to an air cavity of a certain thickness and length. The Froude-depth number is then showcased to dictate this morphology transition. 

Finally, comparing the critical air flow rate to other ones from literature reveals a uniform trend across different scales, hinting at the absence of scaling effects. A new scaling of the critical air flow rate is suggested using the local velocity near the air layer, the slot width of the injector and the Froude-depth number. This scaling allows for a critical flow rate prediction across a wide range of scales.

\begin{bmhead}[Funding]
This work is part of the public–private research program “Water Quality in Maritime Hydrodynamics” (AQUA) project P17-07. The support by the Netherlands Organisation for Scientific Research (NWO) Domain Applied and Engineering Sciences, and project partners is gratefully acknowledged.
\end{bmhead}
\begin{bmhead}[Declaration of interests]
The authors report no conflict of interest.
\end{bmhead}

\begin{bmhead}[Acknowledgments]
{The authors gratefully acknowledge the technical staff of the Ship Hydromechanics Laboratory at Delft University of Technology (in particular Sebastian Schreier, Peter Poot and Pascal Taudin Chabot) for their assistance in the successful execution of the experiments.}
 \end{bmhead}
 
\begin{bmhead}
 [Data availability statement]
 {Selected data will be uploaded at the 4TU research data repository. The Machine Learning algorithm used for bubble segmentation is openly available at \url{https://github.com/AliRKhojasteh/Bubble_segmentation}.}
 \end{bmhead}

\begin{appen}

\section{Repeatability of air lubrication measurements}
\label{more_dr}

Since the repeatability/precision error was assessed based on single phase flow conditions, in order to estimate the repeatability of multiphase flow measurements, repeated measurements are performed for various freestream velocities and air flow rates (Figure \ref{repeatability}). It can be seen that for almost all conditions a good repeatability is achieved across all three regimes. Only in the case of 3 m/s and within the transitional regime, the repeated measurements lie outside the experimental uncertainty (due to the bias error). That is likely due to the highly dynamic nature of this air phase topology introducing additional effects, rather than the measurement technique.

\begin{figure}[!ht]
\centering
\begin{subfigure}[t]{0.45\linewidth}
    \centering
    \includegraphics[width=\linewidth]{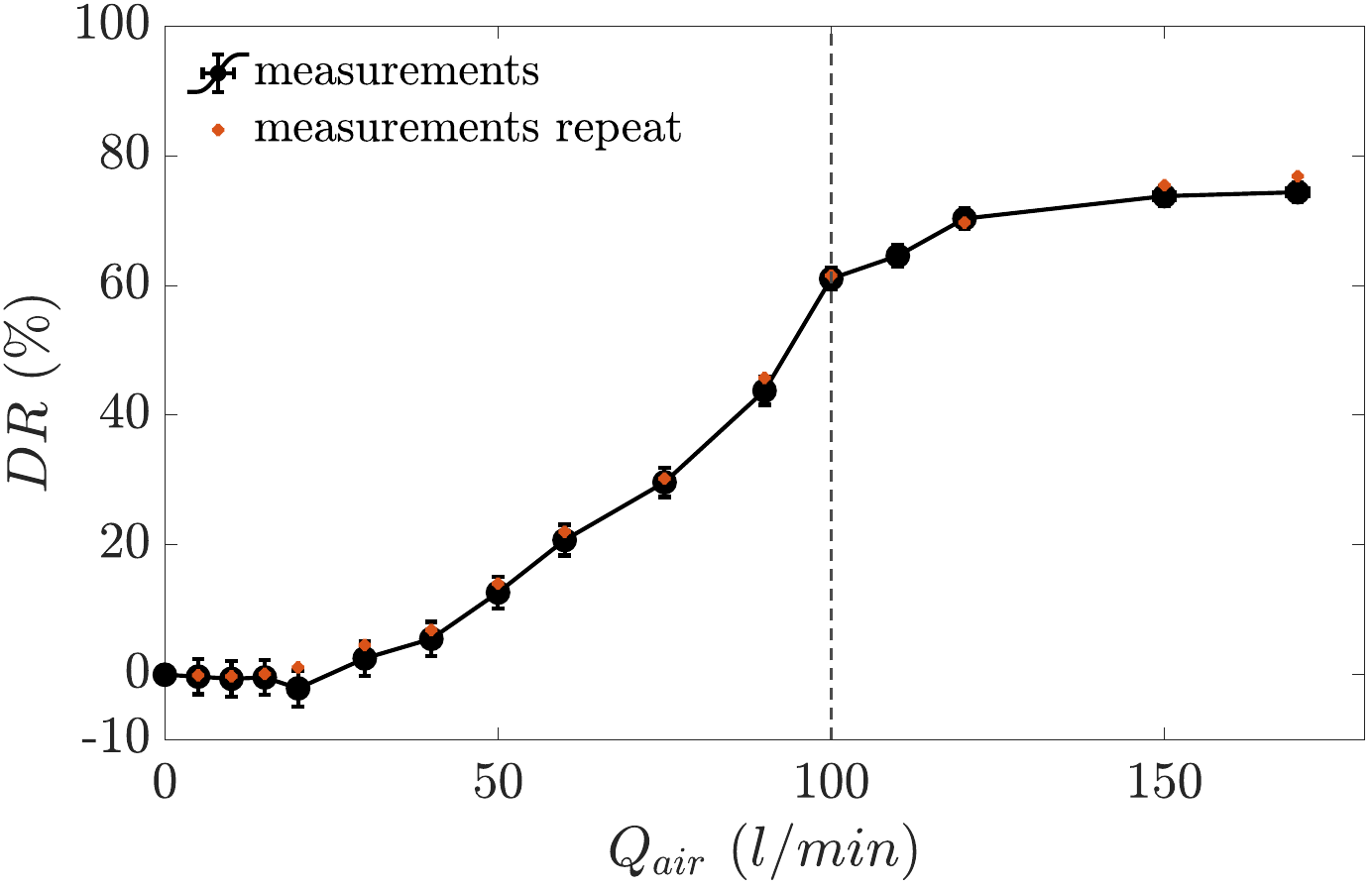}
    \caption{$U_{\infty}=2.5$ m/s}
\end{subfigure}
\begin{subfigure}[t]{0.45\linewidth}
    \centering
    \includegraphics[width=\linewidth]{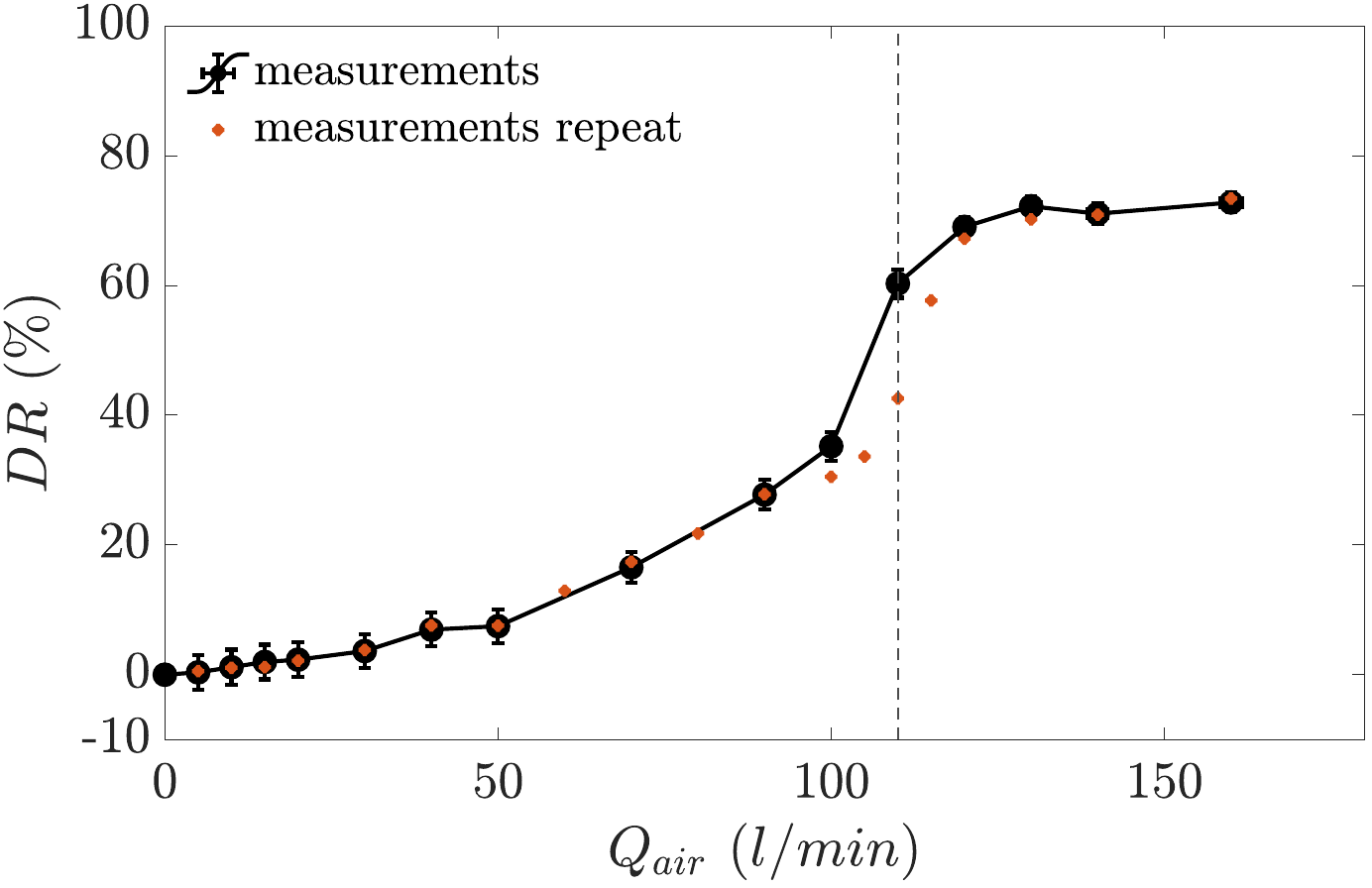}
    \caption{$U_{\infty}=3$ m/s}
\end{subfigure}
\begin{subfigure}[t]{0.45\linewidth}
    \centering
    \includegraphics[width=\linewidth]{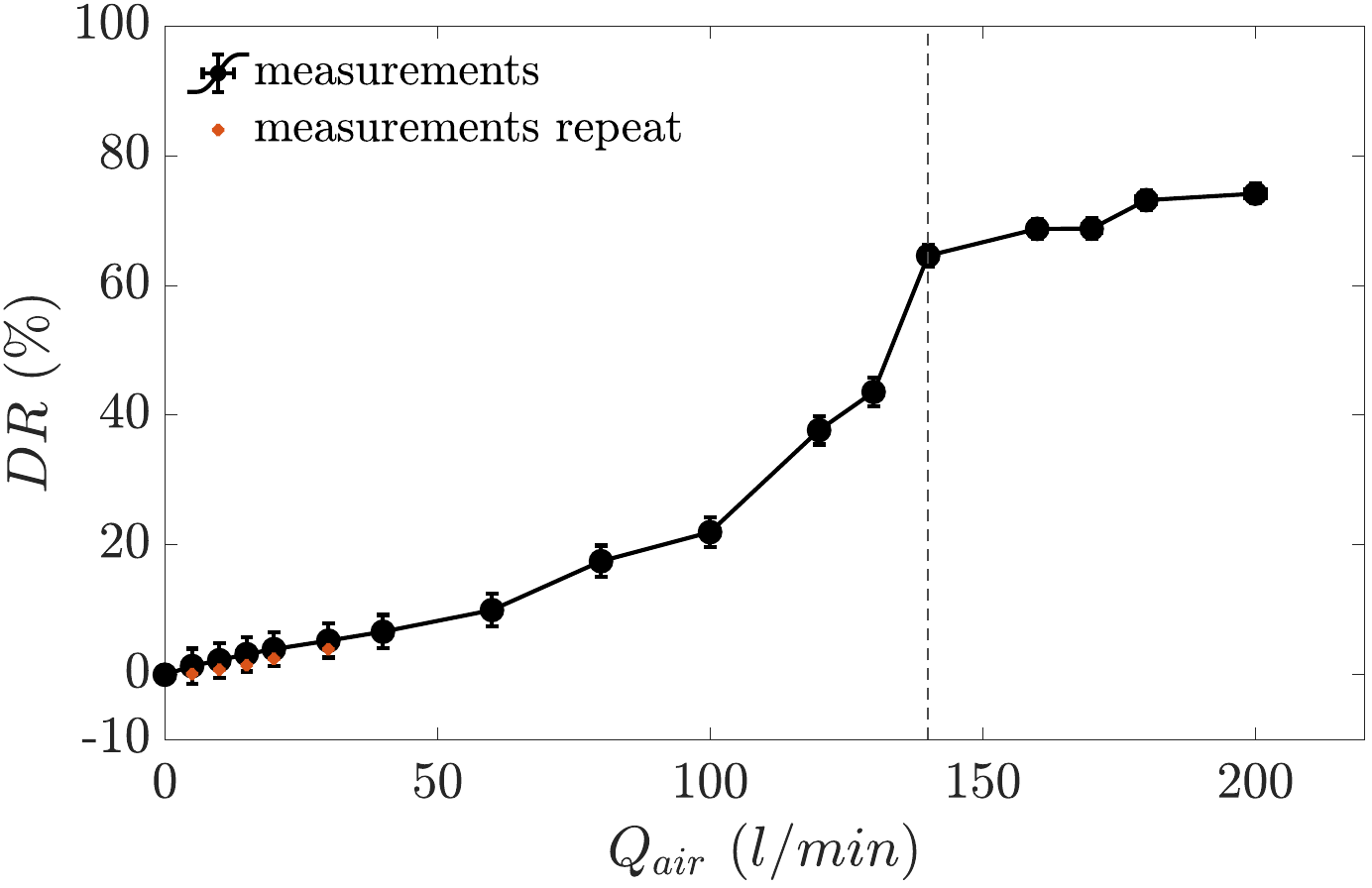}
        \caption{$U_{\infty}=3.5$ m/s}
\end{subfigure}
\begin{subfigure}[t]{0.45\linewidth}
    \centering
    \includegraphics[width=\linewidth]{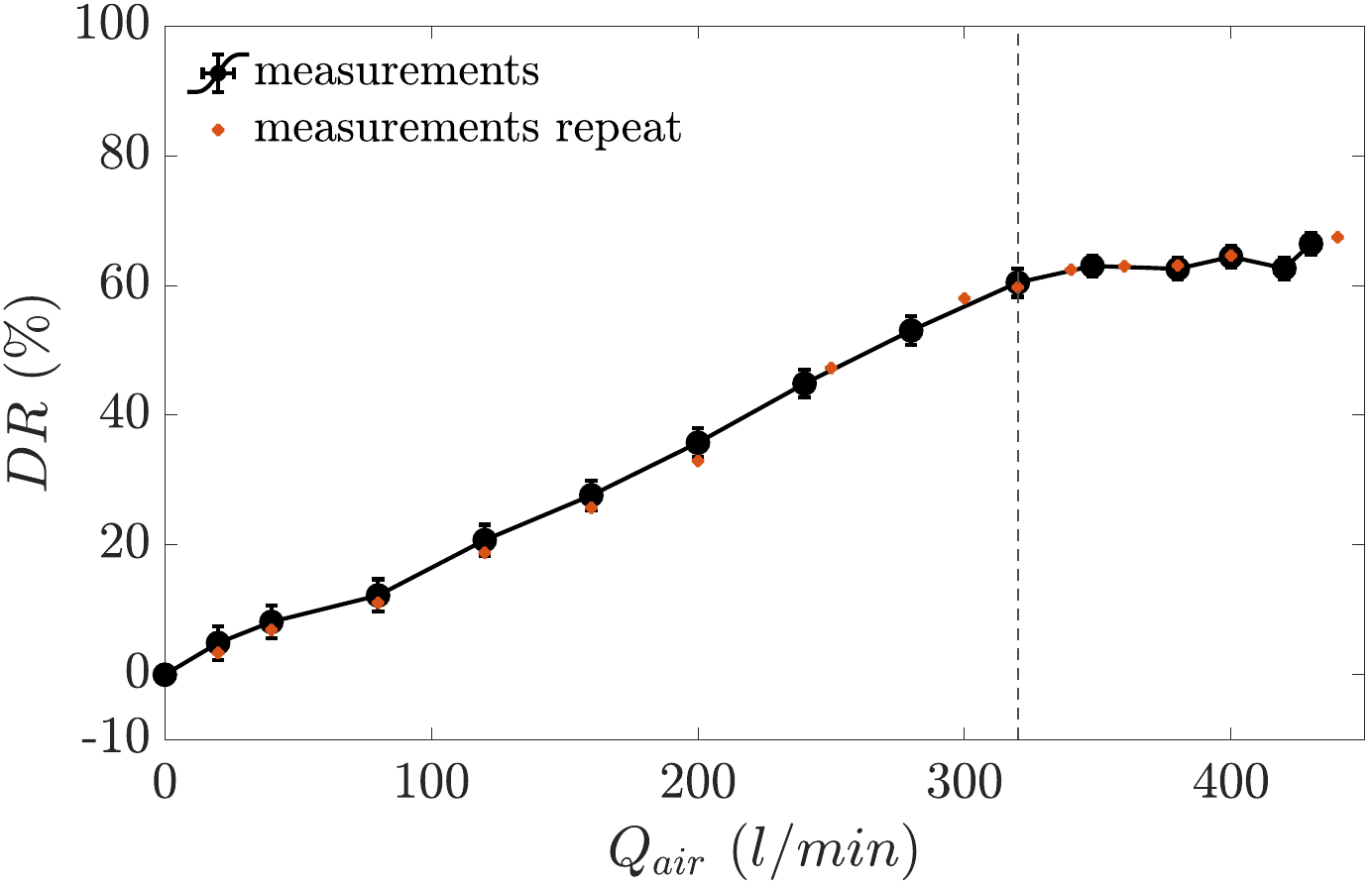}
    \caption{$U_{\infty}=5.5$ m/s}
\end{subfigure}
\caption{Repeatability of drag force measurements for various air flow rates for (a) $U_{\infty}$=2.5 m/s, (b) $U_{\infty}$=3 m/s, (c) $U_{\infty}$=3.5 m/s and (d) $U_{\infty}$=5.5 m/s.}
\label{repeatability}
\end{figure}

\section{Additional bubble size distributions}
\label{ML}

Additional bubble size distributions obtained at higher $U_{\infty}$ than the one shown in Figure \ref{2mps_PDF} reveal a consistent trend: both the bubble size distribution and the mean values of $d_b$ decrease with increasing $U_{\infty}$ (Figure \ref{bubble_stats_5_35}). This indicates that the bubble sizes become more uniform. Furthermore, the mode is again mostly insensitive to $Q_{air}$ ($\approx3-4$ mm) and approximately equal to the injector slot width ($t=4$ mm).

\begin{figure*}[!ht]
\begin{center}
\begin{subfigure}[t]{0.45\textwidth}
  \centering
    \includegraphics[width=\textwidth]{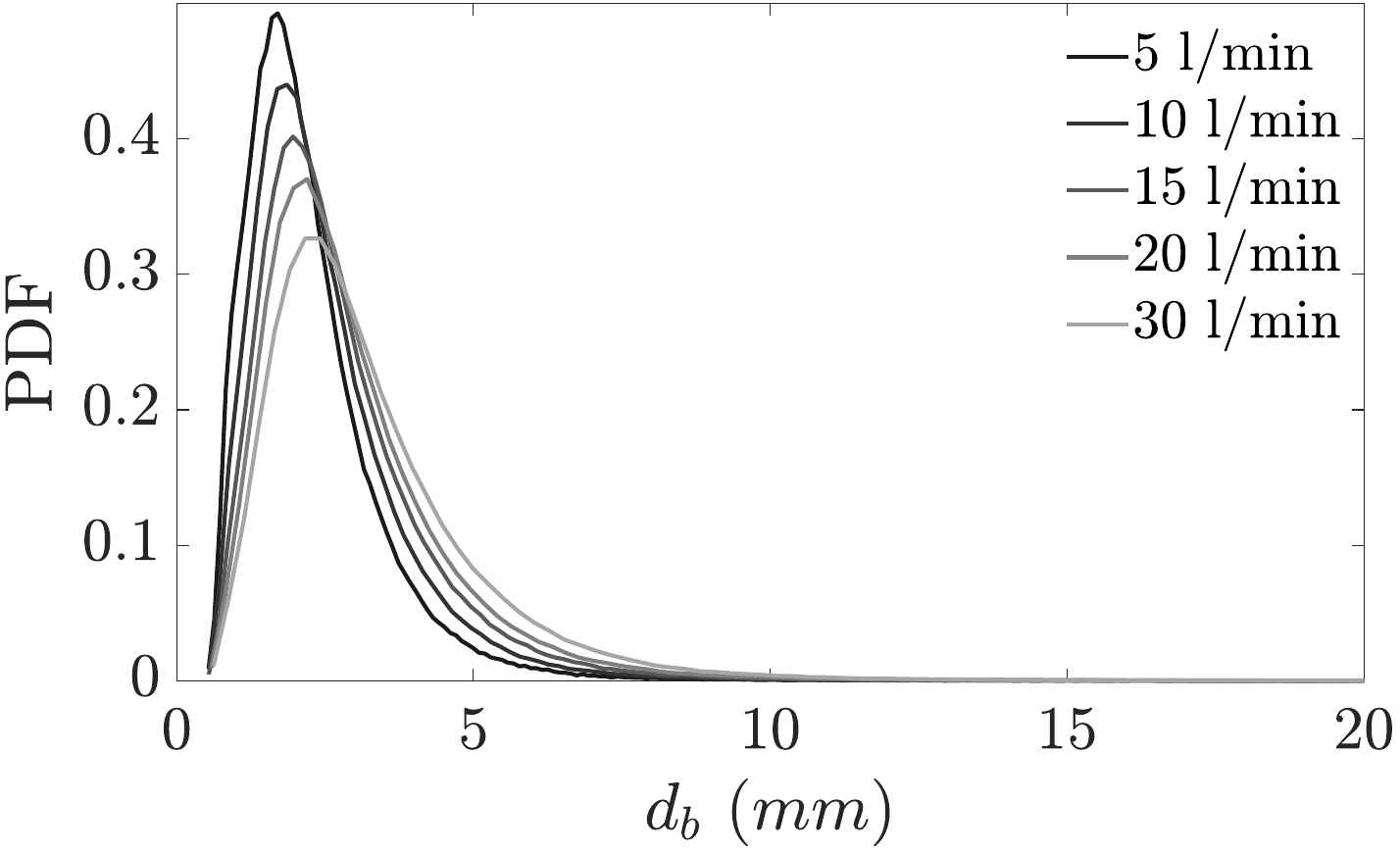}
  \caption{$U_{\infty}=3.5$ m/s}
  \end{subfigure}%
  \hspace{0.01cm}
\begin{subfigure}[t]{0.45\textwidth}
  \centering
\includegraphics[width=\textwidth]{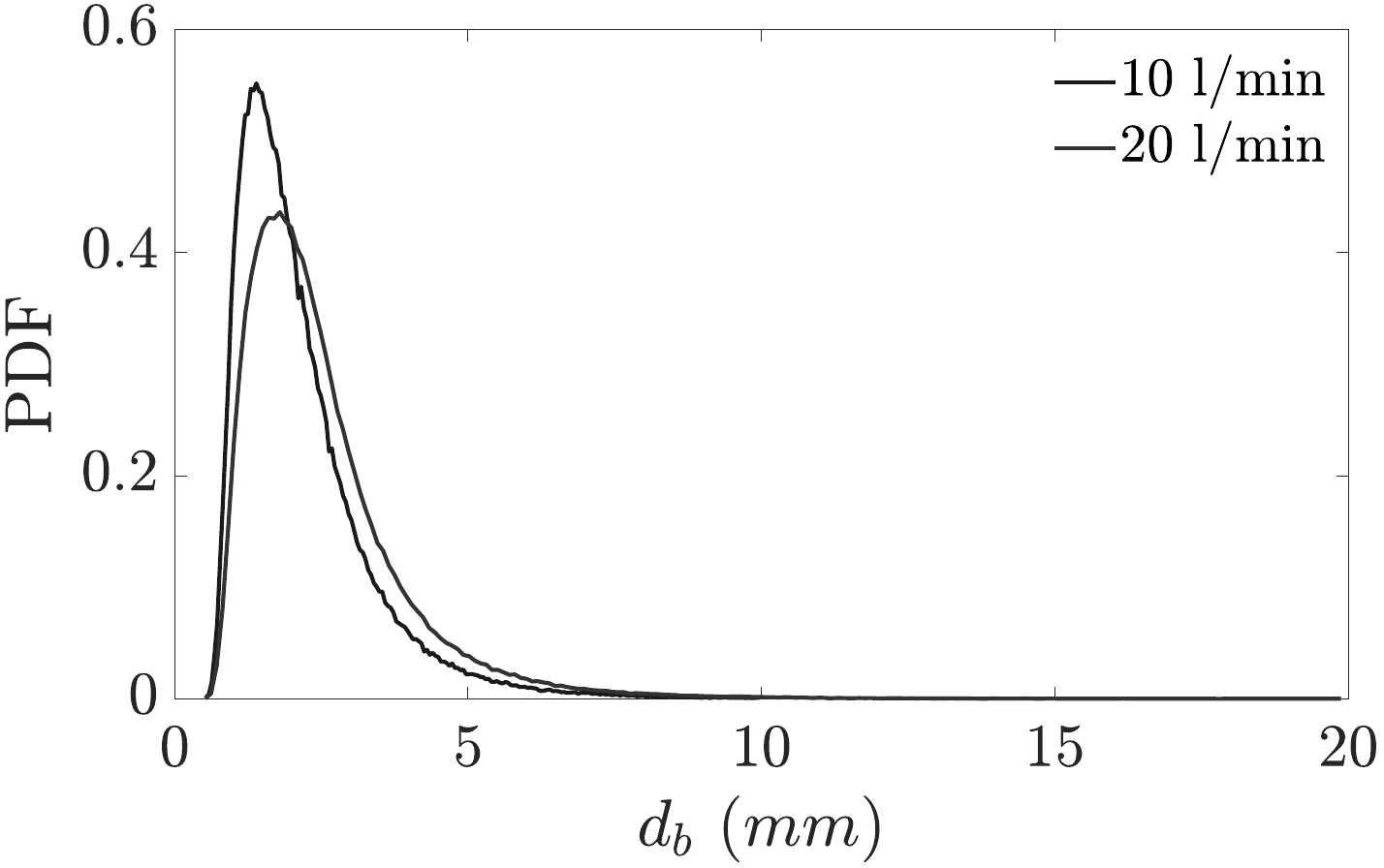}
  \caption{$U_{\infty}=5$ m/s}
  \end{subfigure}
  \caption{Probability distribution of the bubble diameter (\( d_b \)) for various air flow rates within the bubbly regime.}
    \label{bubble_stats_5_35}
\end{center}
\end{figure*}

\end{appen}\clearpage
\bibliographystyle{jfm}
\bibliography{jfm}
\end{document}